\documentclass[12pt,preprint]{aastex}

\usepackage{lscape}
\begin{document}

\title{On the Evolutionary Status of Class I Stars and Herbig-Haro Energy 
  Sources in Taurus-Auriga\footnote{The data presented herein were obtained
  at the W.M. Keck Observatory, which is operated as a scientific
  partnership among the California Institute of Technology, the University
  of California and the National Aeronautics and Space Administration. The
  Observatory was made possible by the generous financial support of the
  W.M. Keck Foundation.}}

\author{Russel J. White and Lynne A. Hillenbrand}

\affil{Department of Astronomy, California Institute of Technology, MS 
  105-24, Pasadena, CA 91125}

\begin{abstract}

We present high resolution (R $\sim$ 34,000) optical (6330 - 8750 \AA)
spectra obtained with the HIRES spectrograph on the W. M. Keck I telescope
of stars in Taurus-Auriga whose circumstellar environment suggests that they
are less evolved than optically revealed T Tauri stars.  Many of the stars
are seen only via scattered light.  The sample includes 15 Class I stars
and all Class II stars that power Herbig-Haro flows in this region.  For 
28 of the 36 stars observed, our measurements are the first high dispersion
optical spectra ever obtained.  Photospheric features are observed in all 
stars with detected continuum, 11/15 Class I stars (42\% of known Taurus
Class I stars) and 21/21 Class II stars; strong emission lines
(e.g. H$\alpha$) are detected in the spectra of all stars.  These spectra,
in combination with previous measurements, are used to search for
differences between stars which power Herbig-Haro flows and stars which do
not, and to reassess the evolutionary state of so-called protostars (Class
I stars) relative to optically revealed T Tauri stars (Class II stars).

The stellar mass distribution of Class I stars is
similar to that of Class II stars and includes 3 spectroscopically
confirmed Class I brown dwarfs.  Class I stars (and brown dwarfs) in Taurus
are slowly rotating ($v$sin$i$ $<$ 35 km/s); the angular momentum of a
young star appears to dissipate prior to the optically revealed T Tauri
phase.  The amount of optical veiling and the inferred mass
accretion rates of Class I stars are surprisingly indistinguishable from
Class II stars.  Class I stars do not have accretion dominated
luminosities; the accretion luminosity accounts for $\sim 25$\% of the
bolometric luminosity.  The median mass accretion rate of Class I and
Class II stars of K7-M1 spectral type is $4 \times 10^{-8}$ M$_\odot$/yr
and the median mass outflow rate is 5\% of the mass accretion rate.
The large ranges in mass accretion rate ($\sim 2$ orders of magnitude),
mass outflow rate ($\sim 3$ orders of magnitude) and ratio of these
quantities ($\sim 2$ orders of magnitude) represent real dispersions in
young accreting stars of similar mass.  We confirm 
previous results that find larger forbidden-line emission associated with
Class I stars than Class II stars.  We suggest that this is caused by an
orientation bias that allows a more direct view of the somewhat extended
forbidden emission line regions than of the obscured stellar photospheres,
rather than because of larger mass outflow rates.  Overall, the
similar masses, luminosities, rotation rates, mass accretion rates, mass
outflow rates, and millimeter flux densities of Class I stars and Class II
stars are best explained by a scenario in which most Class I stars are no
longer in the main accretion phase and are much older than traditionally
assumed.  Similarly, although stars which power Herbig-Haro flows
appear to have larger mass outflow rates, their
stellar and circumstellar properties are generally indistinguishable from
those of similar mass stars that do not power these flows.

\end{abstract}


\section{Introduction}

Optical spectroscopy both initiated and continues to drive theories of low
mass star formation.  Strong HI and CaII emission superimposed upon a
late-type stellar absorption spectrum is one of the original defining
characteristics of the T Tauri variable star class \citep{joy45, joy49}.
Subsequent observations of these emission features has fueled theories for
active chromospheres, mass accretion from disks, and mass loss in winds and
jets \citep[e.g.][]{calvet84, lp74, d81}.  High resolution optical
spectroscopy is still the most accurate tool for characterizing the stellar
properties (e.g. T$_{eff}$, log$g$, $v$sin$i$, [Fe/H]) and investigating
mass accretion and stellar jet/wind processes of young stars
\citep{johns-krull99, sp03, heg95, muzerolle98}.  For many young accreting 
stars, high resolution spectroscopy is often the only tool capable of
extracting the stellar photospheric properties from the optically veiled
spectrum \citep[e.g.][]{wb03}.  
However, because of high extinction, optical observers have been inhibited
from pushing to the earliest stages of low mass star formation.  These
objects, often called protostars or Class I stars \citep{lada87},
consist of a young star embedded within, but perhaps beginning to emerge
from the collapsing envelope of material from which it is forming.  Infall
models with infall rates of a few$\times 10^{-6}$ M$_\odot$/yr are
consistent with spectral energy distributions (SEDs) and scattered light
images of Class I stars \citep[e.g.][assuming a T Tauri-like mass and
luminosity] {kenyon93a, kenyon93b, whitney97}.  This is believed to be the
main accretion phase when the majority of stellar mass is acquired.  How
and how much of the 
infalling envelope material is transferred onto the central star, and the
properties of this central star are largely unknown.  Measuring stellar
properties and the processes at the star-disk interface generally requires
spectroscopic observations at short wavelengths ($\lesssim 2 \mu$m) where
these self-embedded stars are the faintest.  As a consequence, the current
populations of protostars are all \textit{assumed} to be younger than T
Tauri stars because of their less evolved circumstellar environment; there
is currently no photospheric evidence that demonstrates that these
obscured stars are younger than T Tauri stars.

The available (but very limited) spectroscopic observations of some
embedded stars suggest that they are T Tauri-like (i.e. Class II-like), but
with heavily veiled spectra and strong emission lines, implying high mass 
accretion rates and powerful stellar jets \citep{ce96, gl96, kenyon98}.  
The onset of a stellar jet or wind is generally believed to be the
mechanism which clears the surrounding envelope, revealing the central star
\citep[e.g.][]{shu87}.  At least $\sim$ 60\% of all Class I objects are
associated with Herbig-Haro (HH) objects \citep{gomez97, kenyon98,
reipurth99}, which are regions of shocked gas believed to be powered by an
outflow or jet \citep{rb01}.  In comparison, only about 10\% of Class II
stars appear to power optical outflows, though this sub-sample has many of
the highest accretion rates among T Tauri stars \citep[e.g. DR Tau, DG
Tau;][]{gullbring00}.  Class II HH energy sources are often associated
with spatially extended nebulosity \citep[e.g. HL Tau, FS
  Tau;][]{stapelfeldt95, krist98}. Their semblance to younger protostars 
suggests that these stars could be transitioning between the Class I and
Class II stages.

A handful of studies over the last 2 decades have demonstrated using low
resolution \textit{optical} spectra that photospheric features can be seen
in the scattered light continua of some Class I stars and HH energy
sources \citep{mundt85, cohen86, kenyon98}.  Although the advances in
infrared techniques over this same time period led to a widespread shift
from optical to infrared observations of the youngest stars
\citep[e.g.][]{gl00}, optical spectroscopy still retains 2 advantages over 
near-infrared spectroscopy for characterizing the stellar and accretion
properties.  First, unlike near-infrared light which can be dominated by
thermal emission from warm dust and gas, optical light is dominated by
emission from the photosphere and the high temperature accretion shocks.
It therefore offers the most direct view of the stellar properties and the
accretion luminosity.  Second, for small dust grains ($\lesssim 1 \mu$m),
optical light scatters more efficiently than near-infrared light.  Thus,
even if we can not see the stellar photosphere directly because of very
high circumstellar extinction, the cavities commonly seen in the envelopes 
surrounding Class I stars \citep[e.g.][]{padgett99} may allow us to observe
the photosphere and inner accretion processes through scattered light.
This of course is only feasible in low column density star forming
environments like Taurus-Auriga where the young stars are not deeply 
embedded within the large-scale molecular cloud.  The protostars in Taurus
present an opportunity to observe at optical wavelengths young stars at an 
unprecedentedly early age - potentially up to a factor of 10 younger than
more optically revealed T Tauri stars if the statistical ages of embedded
stars \citep{myers87, kenyon90} are correct.

In this paper we present a high spectral resolution optical survey of
roughly 1/2 of all objects classified as Class I stars (the remaining 1/2
are currently inaccessible due to faintness) and nearly all objects
(including both Class I and Class II stars) powering HH flows in Taurus.
Our goal is to understand how stars evolve from embedded protostars to the
optically revealed T Tauri stars.  This requires determining if Class
I stars are in fact younger than T Tauri stars, and if they are in the 
main accretion phase.  Our approach is one which we hope will
begin to bridge the observational segregation of T Tauri star and
protostellar studies; although both the stellar and circumstellar
properties of T Tauri stars are regularly investigated, the overwhelming
majority of protostellar observations are directed towards understanding
the circumstellar properties without regard for their stellar properties.

\section{Sample Definition, Observations and Data Analysis}

\subsection{Sample}

The observational goal of this project was to observe a sample of stars 
in Taurus that are younger than T Tauri stars.  Since the ages of young
stars are often difficult to determine, especially at the earliest
observable stages, we relied on 2 phenomenological indicators of youth to
construct an initial target list.  The first was based on measurements of
the spectral energy distribution.  We identified all stars that are Class
I-like based on either their mid-infrared spectral index
\citep{myers87} or their bolometric temperature \citep{ml93}.  The 
Infrared Astronomy Satellite (IRAS) provided an unbiased and complete (but
flux and confusion limited) 
survey for such objects.  The spectral indices, defined as $\alpha =
d$log($\lambda F_\lambda$)/($d$log$\lambda$), were determined over the
wavelength interval 2 to 25 $\mu$m using the K-[25] colors listed 
in \citet{kh95}.  We included stars with $\alpha > 0$, the generally
accepted value used to distinguish Class I stars from Class II stars
\citep{kh95}.  The bolometric temperatures, defined as the temperature
of a blackbody with the same mean frequency as the observed SED, were taken
from the compilation of \citet{chen95}.  We included stars with T$_{bol}
\le 650$ K, the value proposed by \citet{chen95} to distinguish Class I
stars from Class II stars.  This sample consists of 32 stars.  We caution
that the spectral indices and the bolometric temperatures are in
approximately half of the cases (57\%) determined from SEDs that include
flux contributions from multiple stars.  As an extreme example, HH30 IRS,
HL Tau, XZ Tau and LkH$\alpha$ 358 only have a combined, spatially
unresolved 25 $\mu$m measurement.  In these cases the evolutionary classes
of the components are assigned that of the integrated system.  We note that
if the 25 $\mu$m flux is equally divided among the components and
individual indices are calculated using spatially resolved K-band
magnitudes, these indices generally agree with the system's index
(typically $\pm 0.2$ dex).  However, there are remarkable exceptions.  If
the 25 $\mu$m flux of CoKu Tau 1 is divided in this way from its optically
bright neighbor Hubble 4 (a weak-lined T Tauri star), its spectral index
increase from $-0.48$ to $+0.81$, suggesting that it is a Class I star
instead of a Class II star.

The second indicator of youth we used is the presence of a spatially
resolved optical jet.  As discussed in the Introduction, stars which power
HH flows are almost all either Class I stars or nebulous Class II stars.
Adopting this selection criterion helped to identify a less SED-dependent
sample; several apparently very young stars (based on spatially extended
circumstellar emission) have insufficient photometric measurements to
construct a SED (e.g. CoKu Tau 1, HV Tau C), and would have been excluded
otherwise.  Specifically, we included all stars that that are ``Suspected
Sources'' of HH jets/flows in Taurus-Auriga as listed in the General
Catalogue of HH Objects \citep{reipurth99}.  We also included the young
star ZZ Tau IRS, which is the suspected energy source of HH 393
\citep{gomez97}, and has never been observed spectroscopically.  We refer
to stars that power a HH flow as HH stars, and those which do not as non-HH
stars.  This second selection criterion added 12 additional stars, for a
total sample of 44 stars.

In light of the circumstellar selection criteria used to identify this
sample, we will refer to stars in this sample as \textit{environmentally
young}.  We distinguish a strict sub-sample of bona fide Class I stars, by
requiring that both $\alpha > 0.0$ and $70 <$ T$_{bol} < 650^\circ$ K (if
only one of these SED measurements is available, the Class classification
is based solely on that quantity).  Stars with T$_{bol} < 70^\circ$ K are
considered Class 0 stars.  Following this criterion, 26 of the 44
environmentally young stars are Class I stars, and 1 is a Class 0 star
(IRAS 04368+2557).  The remaining 17 stars are classified as Class II
stars.

This combined sample of 44 stars is listed in Table \ref{tab_sample} and
is presumed to include the youngest stars in Taurus; the list does not
include stars not detected at 2 $\mu$m \citep[e.g. IRAM 04191+1522;
  starless cores][]{ma01, onishi02}.  The components of
binary systems \citep[e.g.][]{ghez93, leinert93, duchene04} are indicated
with the suffix A or B while the spatially unresolved multiples are
indicated with the suffix AB or ABC.  Stars newly 
observed in this study are distinguished (new) in the third column from
stars for which we use previous high dispersion measurements (prev).
Coordinates from the 2MASS database are given and in most cases are a
substantial improvement to the available astrometry, especially for stars
within dark cloud cores.  We also list the 2 to 25 $\mu$m spectral indices,
bolometric temperatures from \citet{chen95}, our evolutionary
classification (I or II), whether the source powers a spatially resolved
optical jet \citep{gomez97, reipurth99}, if 
the star is visible in the POSS-II Red plates (inspection by eye), $I_c$
magnitudes and references, and 2MASS $J$-band magnitudes.  31 of the 44
stars power an optical jet; 15 of the 27 Class I/0 stars have optical
jets.  28 of these 44 stars are visible on the red POSS-II plates, and thus
emit some observable light short-ward of 1 $\mu$m; the 16 stars not visible
consist of 15 Class I stars and 1 (borderline) Class II star, IRAS
04154+2823.

Figure 1 shows the distributions of $\alpha$ and T$_{bol}$ for
the stars spectroscopically observed here (Section 2.3), and for the larger 
sample of Taurus T Tauri stars in \citet{kh95} which have both measurements
available.  For illustrative purposes, the 3 Class I stars without
$T_{bol}$ estimates (IRAS 04260+2642, IRAS 04325+2402, Haro 6-28) are
plotted assuming a value of 500$^\circ$ K, and the 1 Class II star without
a $T_{bol}$ estimate (CoKu Tau 1) is plotted assuming a value of
2000$^\circ$ K.

\subsection{$I_c$-band Imaging, Reduction, and Photometry}

Using the W. M. Keck II telescope and the Low Resolution Imaging
Spectrograph \citep[LRIS;][]{oke95} in its imaging mode, we obtained
$I_c$-band images of 17 of the 44 environmentally young stars (mostly Class
I stars) listed in Table 1 on 1998 Oct 30-31 and 1999 Dec 13.  In all
cases, the integration time was set to 300 seconds which yields a sky count
level of $>$ 1/2 full-well.  The images are 6 x 8 arcmin$^2$ in size and
were processed by subtracting a median-averaged set of bias frames and
dividing by a median-filtered set of flat field frames constructed from
observation of the twilight sky.  Equatorial ``selected area'' standards
\citep{landolt92} were observed for photometric calibration, assuming
typical Mauna Kea extinction; all images were obtained at an airmass of $<
1.1$.  Magnitudes were extracted using an aperture 6\farcs3 in diameter and
a sky annulus extending from 26-30 arcseconds.  These values are listed in
Table 1.  Errors are $<$0.1 magnitude unless marked with a colon.  The
measurements errors for stars listed with a colon are larger and more
difficult to quantify because their emission is primarily spatially
extended.

\subsection{High Resolution Optical Spectroscopy}

The W. M. Keck I telescope and High-Resolution Echelle Spectrometer
\citep[HIRES;][]{vogt94} were used by us to conduct observations on 3
separate observing runs: 1999 Dec 5-6, 2002 Dec 13, and 2003 Feb 17-18.
Variable and sometimes heavy cirrus conditions dominated the 1999 Dec run,
while the other 2 runs had apparently photometric conditions.  HIRES was
used with the red collimator and the RG-610 filter.  The D1 decker (1.15" x
14") was used, which has a projected slit-width of 4 pixels yielding a
resolving power of $\sim$ 34,000.  The cross-disperser and echelle angles
were set (at approximately 1.41 deg and -0.28 deg, respectively) to achieve
a wavelength coverage of $\sim$ 6330 to 8750 \AA, spanning 16 orders with 
wavelength gaps between the orders.  This wavelength range includes several
regions with temperature sensitive photospheric features and many
of the strongest permitted emission lines (H$\alpha$, OI 8446, Ca II 8498,
Ca II 8662) and forbidden emission lines ([OI] 6364, [NII] 6583, [SII]
6716, [SII] 6731) seen in T Tauri star spectra.  Calibration frames
obtained included those of an internal Quartz lamp for flat-fielding and
a ThAr lamp for wavelength calibration.


From the environmentally young star sample listed in Table 1, observational
priority was given to targets which showed optical emission in the POSS-II
Red plates (28/44) and for which no previous high resolution spectra were
available.  We observed 23 of the 28 optically visible stars.  The
remaining five stars have been observed previously with either high  
dispersion optical spectroscopy \citep[CW Tau, GK Tau, DO Tau, UY
Aur;][]{heg95} or spatially resolved medium dispersion optical spectroscopy 
\citep[XZ Tau A\&B;][]{hk03}.  We include the available results of these
previous measurements in our analysis and discussion (Sections 3 and 4).
Finally, we observed 3 of the 16 environmentally young stars that are not
visible on POSS-II plates, but are seen in our LRIS images (IRAS
04154+2823, IRAS 04295+2251, IRAS 04489+3042), yielding a total of 26 of
the 44 stars listed in Table 1, 31 including previous spectroscopy.

In addition to these primary targets, we obtained spectra of 6 additional
Taurus T Tauri star systems.  IRAS 04278+2253 A \& B, Haro 6-13, IRAS
04303+2240 all have bolometric temperatures less than 1000$^\circ$ K and
have never been observed at high dispersion.  HK Tau A \& B have been
classified as transitional Class I/II stars \citep{kh95}; HK Tau B is also
an edge-on disk system \citep{stapelfeldt98, koresko98} and has never 
been observed optically at high dispersion.  We observed this pair of
stars, even though the system's $\alpha$ and T$_{bol}$
do not corroborate the claim that it is a transitional Class
I/II star ($\alpha = -0.51$, T$_{bol} = 2148^\circ$ K).  The lightly veiled
T Tauri stars DN Tau and V836 Tau were observed as well.  The properties of
these 6 star systems are also listed in Table 1, separately.  They are
included in the analysis, but distinguished from the environmentally young
star sample.

The spectroscopic observations spatially resolved the components of the
binary systems MHO 1/2 IRAS 04278+2253, HK Tau, and RW Aur.  For the HV Tau 
system, only a spectrum of HV Tau C was obtained.  In the case of L1551 IRS
5, we obtained spectra of the point-like optically visible object at the
position of IRS 5, at the position of the HH object $\sim$ 12 arcseconds
east of IRS 5, and at a position centered on the brightest portion of the 
optical reflection nebula that is 2\farcm5 directly west of IRS 5
\citep[see][]{stocke88}.

For 28 of the 36 stars observed (in 32 systems), these measurements are
the first high dispersion optical spectra obtained.  Exposure times and
epochs of observations are listed in Table 2.  To assist in the
spectroscopic analysis, we observed 4 slowly rotating weak-lined T Tauri
stars in the TW Hydrae association (TWA 8 A \& B, TWA 9 A \& B) and
numerous dwarf and giant spectral type standards, 9 of which have
accurately known radial velocities \citep{nidever02}.  The observed 
spectroscopic standards were chosen to span a range of G, K and M spectral
types at both dwarf and giant surface gravities.


\subsection{Spectroscopic Reduction}

The HIRES data were reduced using the facility ``makee'' reduction script
written by Tom Barlow.  The reductions included bias subtraction,
flat-fielding, spectral extraction, sky subtraction, wavelength
calibration, and heliocentric radial velocity corrections.  The projected
spectra from target stars have typical full-widths at half-maximum of 1-2
arcseconds, though 5 stars (IRAS 04016+2610, IRAS 04302+2247, L1551 IRS 5,
DG Tau B, Haro 6-5 B) have much broader profiles because of spatially
extended emission.  In these cases, a broader extraction aperture was used 
to include this extended emission; the sky level was determined outside
these apertures.

The 2 binary stars RW Aur and IRAS 04278+2253 were observed with the
components aligned along the slit.  The spectra of the components were
extracted with a choice of apertures that minimize contamination from the
companion.  In the case of RW Aur, in which the secondary is separated by
only 1\farcs4 from the primary and is nearly 10 times fainter, the
extracted spectrum of the secondary star is still contaminated by that of
the primary; the secondary's spectrum showed a broad, but very distinct low 
level component to the Ca II emission lines ($\lambda$ 8498 \AA, $\lambda$
8662 \AA), as well as very weak OI circumstellar absorption, both of which
we attribute to contamination RW Aur A.  These artifact features disappear
cleanly if a spectrum of RW Aur A, scaled to $20\pm5$\% of RW Aur B's
continuum level, is subtracted from the spectrum of RW Aur B.  Our final
reduced spectrum of RW Aur B has this 20\% subtraction.

Figure 2 shows portions of the HIRES spectra for the 28 stars with a
continuum signal-to-noise ratio (S/N) greater than 2.  Table \ref{tab_obs}
lists the continuum S/N per pixel at 6700 \AA\, and 8400 \AA\, for all
observations.  For the 6 stars with S/N $<$ 2 (2 Class II stars: DG Tau B
and IRAS 04154+2823, and 4 Class I stars: MHO 1, L1551 IRS 5, IRAS
04295+2251, and IRAS 04302+2247) the continuum detected (if real) is
insufficient to measure any photospheric lines.  These 6 stars are excluded
from our photospheric analyses.  Li I (6708\AA) is detected in the spectra
of 24 of the 28 stars with a genuine continuum.  H$\alpha$ emission is
detected in all 36 stars observed.

\section{Analysis}

Figure 1 illustrates how this survey has extended the sample of stars with
spectroscopic detections of the photosphere well into the Class I regime.
Of the 26 Class I stars in Taurus (Table 1), 11 now have spectroscopic 
measurements
of their photospheres and 15 have optical emission-line measurements.  In
addition to the remaining 15 Class I stars with no spectroscopic detection
of their photosphere, there are 7 Class II stars from \citet{kh95}
with no spectroscopically measured photospheres:
IRAS 04370+2559,
IRAS 04414+2506,
IRAS 04301+2608,
FZ Tau,
FT Tau,
IRAS 04200+2759, and
T Tau South. 
Note that only 3 of these Class II stars have both $\alpha$ and 
$T_{bol}$ measurements and thus are shown in Figure 1.

Though not complete, the considerable expansion of the available
spectroscopy towards environmentally younger stars enables for the first 
time an investigation of the stellar and circumstellar properties between
evolutionary classes.  We wish to accomplish this without relying too
strongly on any one evolutionary diagnostics ($T_{bol}$, $\alpha$, HH
properties), which may be flawed and/or have unaccounted for biases.
Therefore we take the approach of comparing all measured and inferred
properties with both the bolometric temperatures and spectral indices (when
available), and distinguishing HH stars from non-HH stars.  Possible
differences between the properties of Class I stars and Class II stars
(defined by $T_{bol}$ and $\alpha$ criteria; Section 2.1) and HH stars and
non-HH stars are investigated by comparing median values and by conducting
K-S tests on these distributions \citep{press97}.  K-S probabilities are
reported in most cases; we consider distributions to be different if the
K-S probability is less than 0.056 (ie. 2$\sigma$ difference).

To increase the significance of these comparisons we identify and include
an additional sample of Class II T Tauri stars.  This sample includes the 6
additional Taurus T 
Tauri stars observed here (all optically veiled), and the 26 optically
veiled (at $\sim 6000$ \AA) Taurus T Tauri stars from the high spectral
resolution studies of \citet{bb90}, \citet{heg95} and \citet{wb03}.  Our
comparisons thus include only stars that are either environmentally young
(Table 1) or are classical T Tauri stars defined by their optically veiled
spectra.  No weak-lined T Tauri stars (Class III stars) are included.

\subsection{Measured Spectral Properties}

\subsubsection{Radial and Rotational Velocities}

Radial and projected rotational velocities are determined via a 
cross-correlation analysis \citep[e.g.][]{hartmann86}.  First, we identify
up to 10 spectral regions of length 20-30 \AA\ with sufficient S/N ($> 2$)
and without prominent telluric absorption features \citep[e.g.][]{tr98},
strong gravity sensitive features (K I 7665 \& 7699 \AA, Na I 8183 \& 8195
\AA, Ca II 8498 \AA), or emission lines.  These regions are
cross-correlated with at least 3 spectral standards of spectral type
similar to the target star (see Section 3.1.2) and with radial velocities 
accurate to 0.3 - 0.4  km/s \citep{nidever02}.

To determine the final radial velocities of the stars, the velocity offsets
from the standards must be corrected for barycentric velocity corrections
and possible errors in the wavelength solution of either the standard or
target spectrum.  The former is determined using the \textit{rvcorrect}
task in IRAF.  The latter correction may only be relevant when combining
data from multiple nights and/or multiple runs, as is the case here.  This
correction is determined by cross-correlating the telluric A-band of the
standard and the target star.   Failure
to correct for this effect leads to radial velocity errors of 3-4 km/s.
Correcting for this effect yields consistent radial velocities of the
standards to better than 0.5 km/s, consistent with the accuracy of their
radial velocities.  Uncertainties are estimated by adding, in quadrature,
the standard deviation of the multiple radial velocity estimates (3-7
standards are used in each case) and the uncertainty in each radial
velocity estimate, which is assumed to be the error in the mean of all
spectral regions used per star.

Rotational velocities ($v$sin$i$) for the observed targets are
determined from the width of the peak in their cross-correlation with a
slowly rotating standard.  The width is estimated by fitting a Gaussian plus
quadratic function to the cross-correlation peak.  The width versus
$v$sin$i$ relation is determined empirically by cross-correlating a
``spun-up'' spectrum of a slowly rotating standard.  The rotationally
broadened spectra are constructed using the profiles given in Gray (1992;
$\epsilon = 0.6$).  As has been noted in previous work
\citep[e.g.][]{wb03}, the presence of a continuum excess (see Section
3.1.3) has little effect on the inferred projected rotational velocities,
as long as the features are sufficiently well measured.  Uncertainties in
$v$sin$i$ are determined following the same procedure used for the radial
velocities.  We find that all templates within a few spectral subclasses of
the target star generally give consistent $v$sin$i$ values; an exact match
in spectral type is not critical in determining $v$sin$i$.  However,
differences in spectral type 
and luminosity class can affect $v$sin$i$ measurements at low rotational
velocities \citep[less than $\sim$ 10 km/s; see][]{wb03}.  To avoid
introducing this bias, we adopt the conservative approach in determining
$v$sin$i$ upper limits.  If the uncertainty is less than 1 standard
deviation from the theoretical $v$sin$i$ detection limit (7 - 15 km/s, set
by S/N), the measurement is assigned an upper limit.

The described procedure worked well for determining radial and
projected rotational velocities for 24 of the 28 stars with a continuum
detection.  Because of low S/N and, in some cases, high veiling (Section
3.1.2),
the cross-correlation analysis for the remaining 4 stars (IRAS 04016+2610,
IRAS 04264+2433, IC 2087, and IRAS 04489+3042) was successful in only a few
segments (2-3) of their spectra.  The correlation results for these 4
stars imply no measurable rotation, but do yield radial velocities that are
consistent with that of Taurus.  In the cases of IRAS 04016+2610 and IC 
2087, we are especially cautious about the results.  The photospheric-like
features in their spectra are quite weak and/or noisy, and possibly could
be spurious noise events.  The results for IRAS 04264+2433 and IRAS
04489+3042 are slightly more robust, however, since both stars show TiO
absorption bands that can be more confidently identified (see Figures 1 \&
2).  The radial velocities and $v$sin$i$ detection limits (15 km/s) for
these stars are listed in Table 2 followed by colons to indicate possible
larger systematic uncertainties.

Although for the majority of these stars their measurements are the first
epoch of high resolution spectroscopy, no definitive spectroscopic binaries
are identified.  With the exception of HV Tau C, all stars have radial
velocities within 2$\sigma$ of the mean of Taurus-Auriga ($17.4 \pm 2.1$;
Hartmann et al. 1986), and thus are consistent with being Taurus
members.  The radial velocity of HV Tau C (23.1$\pm$0.6) is 2.6$\sigma$
from the mean of 
Taurus, and thus the star could be a single-lined spectroscopic binary.
Unfortunately we do not have a spectrum of its companions HV Tau AB for a
more strict comparison.  We also note that radial velocities of HK Tau A
and HK Tau B are significantly different ($21.6\pm0.5$ and $17.4\pm0.3$),
several times that expected from relative orbital motion alone.  We suggest
that either one of these stars is a single-lined spectroscopic binary or
that the pair is not physically associated.  The latter case would explain
their apparently non-coplanar disks \citep{duchene03}.

The projected rotational velocities of the environmentally young star
sample are relatively slow, ranging from 10 - 34 km/s.  The measured
$v$sin$i$ values of this sample and the additional sample of Class II stars
are plotted in Figure \ref{rot_age} versus the evolutionary diagnostics
$T_{bol}$ and $\alpha$.  The environmentally young stars have measured
$v$sin$i$ values (excluding upper limits) that are generally quite similar
to the values of environmentally older stars.  Treating $v$sin$i$ upper
limits as detections, the distributions of $v$sin$i$ values for Class I and 
Class II stars are similar (K-S probability of 0.079), though the median 
$v$sin$i$ of Class I stars is slightly larger than that of Class II stars
(median Class I $v$sin$i = 18.0$ km/s, $\sigma =$ 5.1; median Class II
$v$sin$i = 14.4$ km/s, $\sigma =$ 10.8).  Comparing HH stars and non-HH
stars, the distributions of $v$sin$i$ are more discrepant (K-S probability
of 0.0034; a $2.9\sigma$ difference); HH stars rotate slighly faster than
non-HH stars (median $v$sin$i = 19.4$ km/s, $\sigma =$ 5.9; median $v$sin$i
= 12.1$ km/s, $\sigma =$ 11.5).  Table \ref{tab_median} summarizes the
median statistics, excluding brown dwarfs.

\subsubsection{Spectral Types and Continuum Excesses}

Spectral types and continuum excesses are determined by matching the
line/band ratios and line/band depths, respectively, to those of
rotationally broadened and artificially veiled dwarf spectral standards.
The procedure works as follows for a given star.  First, all of the
comparison dwarf spectral standards are rotationally broadened to have the
same $v$sin$i$ as the young star.  For each rotationally broadened
standard, a set of ``veiled'' templates are then generated by adding a range
of continuum excess values, all assumed to be constant over each $\sim$ 20
\AA\, spectral region.  This set of rotationally broadened and veiled
templates are then compared to the young star sample to determine which
spectral type gives the best match.  Quantitatively, the best match is 
determined by minimizing the root-mean-squared values over 3 temperature
sensitive regions: 6490 - 6502 \AA, 7120 - 7128 \AA, and 8431 - 8436 \AA\,
(see Figures 2-4).  In most cases, the assigned spectral type is that of the
spectral standard which gives the best fit.  In some cases, however,
emission lines or anomalous low S/N features yield clearly discrepant
spectral types.  Consequently, all spectral types are confirmed by visual
inspection.  Uncertainties in the spectral type are determined by the
range of spectral types which can reasonably match the spectrum; the
uncertainties range from 0.5 to 3 spectral subclasses.  More robust
techniques for determining temperatures such as synthetic modeling of the
spectra are currently unwarranted given the low S/N of the data.
The low S/N of IRAS 04016+2610 and the high optical veiling of 
IC 2087 prevent us from determining their spectral types.  However, both
show Ti I (8435 \AA, 8436 \AA) absorption that is stronger than Nb I (8439
\AA) absorption, suggesting a spectral type cooler than G.  The absence of
any clear TiO bands suggests a spectral type hotter than M.  We classify
these two stars as K stars.  The inferred spectral types for all stars are
listed in Table 1.

The inferred spectral types range from K0 to M6 for the environmentally
young star sample.  Class I stars and Class II stars have a similar
distributions of stellar spectral types.  HH stars, on the other hand, 
tend to be slightly hotter than non-HH stars.  Of the 46 stars studied here
with spectral types hotter than M3.5 (M $> 0.24$ M$_\odot$), 20 are HH 
stars (43\%).  In contrast, of the 9 stars with spectral types of M3.5 or
cooler, only 2 are HH stars (22\%).

Once the spectral type of the star is determined, the continuum excess,
called $r$ and defined as $r = F_{excess}/F_{photosphere}$, is determined
by comparing the depths of photospheric features to those of a rotationally
broadened and artificially veiled dwarf spectrum of the same spectral
type.  The continuum excess added to the standard is allowed to be negative;
this results in a comparison standard with deeper absorption features.
Consequently, stars with no continuum excess should have $r$'s that scatter
about zero, and this scatter can be used to estimate the uncertainty in
$r$.  The continuum excess is determined in 2 spectral region, 6500 \AA\,
and 8400 \AA, by averaging the results within 50 \AA\, of these wavelengths.
The uncertainties in the continuum excess are estimated by changing the
rotationally broadened and veiled template by the spectral type uncertainty
and redetermining the continuum excess; the spectral type uncertainty
typically dominates the uncertainty in the continuum excess.  If the
uncertainty in the continuum excess is larger than the measured value,
the uncertainty is assigned as an upper limit to detection.  As a check of
our procedure, we note that the 4 weak-lined T Tauri stars TWA 8a, TWA 8b,
TWA 9a, and TWA 9b, with spectral types ranging from K7 to M5.5, can be
successfully modeled using a rotationally broadened dwarf spectrum without
any continuum excess (e.g. $<r_{6500}>$ $= -0.04\pm0.11$).  For the K stars
IRAS 04016+2610 and IC 2087, we use a K4 dwarf as the template from which
the veiling is determined.  The inferred continuum excess values at
6500 \AA\, ($r_{6500}$) and at 8400 \AA\, ($r_{8400}$) are listed in Table
2.  For 5 stars, the S/N is insufficient to determine the continuum excess
at 6500 \AA\, but is sufficient to measure the excess at
8400 \AA.

Figure \ref{veil_ratio} shows the ratio of the continuum veiling at 8400
\AA\, to the continuum veiling at 6500 \AA\, as a function of stellar
effective temperature.  Values plotted include measurements from this study
(Table \ref{tab_obs}) and from \citet{wb03}.  In addition, we show the
ratio of optical veiling values 
from \citet{hk03}, determined from similar but slightly shorter wavelength
intervals ($r_{6100}$/$r_{8115}$), using low resolution spectra; only
veiling values above 0.1 are included from this data set to avoid
spuriously large ratios.  For early- to mid-K stars, the continuum excesses
at 8400 \AA\, are in many cases larger than the values determined at 6400
\AA.  This is somewhat surprising, since the optical excesses are generally
attributed to featureless continuum emitted by a $\sim 10^4$ K hot spot(s)
\citep[e.g.][]{kenyon94, cg98}, and therefore should be significantly less at
8400 \AA.  We note that this effect is consistent with previous veiling
studies which find the veiling to be ``almost flat'' in the red
\citep{bb90}; this previous work primarily 
focused on K5-M0 type stars for which an excess of constant flux would
yield nearly a constant veiling value ($r_{8400}$/$r_{6500} \sim 1.0$;
Figure \ref{veil_ratio}).  We postpone discussion of this interesting
phenomenon until Section
3.3.1, but use these empirical ratios to estimate the continuum excess at
6500 \AA\, for the 5 stars with a $r_{8400}$ measurement, but no $r_{6500}$ 
measurement.  For the 2 K stars, $r_{6500}$ is estimated to be
$r_{8400}$/1.2; for the M1 star, $r_{6500}$ is estimated to be to be 
$r_{8400}$/0.70; for the M5.5 and M6 stars, $r_{6500}$ is estimated to be
$r_{8400}$/0.43.  These estimated $r_{6500}$ values are listed in Table 2
followed by colons.

For IRAS 04303+2240, which was observed both on 1999 Dec 6 and 2003 Feb 17,
no photospheric features were detected during the first epoch, but 
early-M spectral features were clearly identified during the second (Figure
1).  The star was also much brighter during the first epoch than the second
epoch; the relative counts in the spectra suggest a brightness decrease by
factors of 3.25 at 6500 \AA\, and 2.47 at 8400 \AA\, (see S/N in Table 2).
Although no continuum excesses could be measured in the first epoch since
the photosphere is not detected, we estimate the excesses during this epoch
by assuming that the change in brightness between the two observations is
caused only by a change in the continuum excess.  These larger first epoch
values are listed in Table 2 and followed by colons.

The distribution of continuum excesses at 6500 \AA\,
($r_{6500}$) is shown in Figure \ref{veil_age} as a function of $T_{bol}$
and $\alpha$.  Values for the additional sample of Class
II stars are also shown, with continuum excess measurements from
\citet{bb90, heg95} and \citet{wb03}.  Environmentally young stars have
optical veiling values that are similar to those of environmentally older
stars.  K-S tests show no differences in the distributions, but slight
differences are seen in the median values.  The continuum excesses of the
11 Class I stars (median $r_{6500} = 1.0$, $\sigma = 0.6$) are marginally
larger than 
the continuum excesses of the 50 Class II stars (median $r_{6500} = 0.6$,
$\sigma =$ 1.3).  Similarly, the continuum excesses of the 24 HH stars
(median $r_{6500} = 1.1$, $\sigma = 1.3$) are marginally larger than the 
continuum excesses of the 36 non-HH stars (median $r_{6500} = 0.6$, $\sigma
= 1.0$).  Table \ref{tab_median} summarizes the median statistics,
excluding brown dwarfs.

\subsubsection{Surface Gravity Indicators}

Pressure sensitive photospheric features offer a means of determining if
the surface gravities of Class I stars are less than 
those of Class II stars, for a given mass.  This would indicate that they
have larger radii and hence younger ages.  Some of the more distinguishing
pressure sensitive features for K and M spectral types include CaH (6382
\AA, 6389 \AA, 6903 - 6946 \AA), K I (7665 \AA, 7699 \AA), Na I (8183 \AA,
8195 \AA), Ca II (8498 \AA, 8542 \AA, 8662 \AA) and Fe I lines
\citep[see e.g.][]{kirkpatrick91}.  The molecular absorption bands of TiO
and VO present in M dwarfs are also known to be pressure sensitive
\citep{tdw93, wb03}.  Our spectra include K I (7699 \AA), Ca II (8498 \AA,
8662 \AA), several Fe I lines and CaH and TiO absorption bands.  Of these,
Ca II is not a useful diagnostic of surface gravity, since it often
shows either strong emission from accretion activity, or core emission from
chromospheric activity.

The spectra shown in Figures \ref{spec1} - \ref{spec3} include the K
I at 7699 \AA.  This feature is one of the strongest and most easily
identifiable absorption features in our modest S/N optically veiled
spectra.  However, many stars show non-photospheric components to this
feature, including both broad emission profiles with narrow absorption
superimposed (e.g. DG Tau, HL Tau, T Tau) and broad asymmetric absorption
with possible emission (e.g. IRAS 04303+2240, IC 2087, RW Aur A).
For objects with no obvious non-photospheric emission, however, the wing
strengths of K I for both Class I and Class II stars are intermediate
between those of narrow giants and those of pressure broadened dwarfs.
This is a well known characteristic of T Tauri stars.

For early- and mid-M spectral types, we find that dwarfs, as opposed to
giants or some combination of the two, provide a better match to the
pressure sensitive Fe I lines and TiO bands of both Class I and Class II
stars \citet[see also][]{wb03}.  In addition, CaH bands at 6382 \AA\, and
6389 \AA\, are clearly seen in many of the Class I and Class II stars
observed here.  These features are strongest in the spectra of early M
dwarfs, and absent in spectra of similar spectral type giants.

The combined results of these comparisons suggest that the surface
gravities of both Class I and Class II stars are intermediate between those
of dwarfs and giants, but somewhat more similar to dwarfs.  This is
consistent with the predicted surface gravities for young stars
\citep[e.g.][]{siess00}, whose radii should be only a few times 
their main sequence values, as opposed to $\sim$ 100 times that of their
main-sequence values in the case of giant stars.  Theoretical predictions
suggest Class I stars should have radii less than 2 times those of T Tauri
stars \citep[e.g.][]{stahler88}, which translates into differences in
surface gravity of less than a few tenths of a dex.  Given the relatively
low S/N of the Class I spectra, measurements of these small log(g)
differences are not possible from line profile analysis.  We are also
cautious of such an approach, given the non-photospheric emission
associated with gravity sensitive features (K I 7699 \AA; Section 3.1.4).
Analyses which focus on determining surface gravities from molecular-band
strengths \citep[e.g.][]{mohanty04} offer a more promising approach.
With the current data at hand, we can neither prove nor disprove that Class
I stars have lower surface gravities, as would be expected if they are
younger than Class II stars.

\subsubsection{Prominent Emission Lines}

The emission-lines of young stars are powerful probes of the mass accretion
and mass outflow processes.  Their ratios characterize temperatures and
densities of the emitting regions and their intensities and velocity
profiles trace the flow of circumstellar material \citep[e.g.][]{edwards87,
hp92, hamann94, heg95}.  In Figures \ref{emission1}-\ref{emission4} are
shown emission-line profiles of the observed Taurus stars in Table 2.  The
plotted profiles include the 2 permitted lines H$\alpha$ 6563 \AA\, and
CaII 8498 \AA, and the 3 forbidden emission-lines [OI] 6364 \AA, [NII] 6583
\AA, and [SII] 6731 \AA.  Fe II 6456 \AA, K I 7699 \AA\, and OI 8446 \AA\,
emission can be seen in the spectra plotted in Figures \ref{spec1} -
\ref{spec3}, when present.

Equivalent widths (EWs) of H$\alpha$, CaII 8498 \AA, CaII 8662 \AA, [OI]
6364 \AA, [NII] 6583 \AA, [SII] 6716 \AA, and [SII] 6731 \AA\, are listed
in Table 3.  These EWs include emission at all velocities which, for
several forbidden lines, means summing distinct velocity components.  In
cases where the CaII emission appears in the center of the photospheric 
absorption feature (CoKu Tau 1, Haro 6-28, Haro 6-33, RW
Aur B, HK Tau A \& B, DN Tau, V836 Tau), the EWs are estimated from the
approximate bottom of the stellar absorption profile; no corrections for
the stellar absorption are made for stronger emission-line stars.
Detection limits are estimated based on the S/N of the spectra 
and/or the quality of the sky subtraction for telluric sensitive features
(e.g. [OI] 6364 \AA).  In cases where an emission-line is present, but the
S/N in the continuum is less than 1.0, the equivalent width is assumed to
be a lower limit (up to 999 \AA).  Lower limit EW measurements of [SII]
6716 \AA\, and 6731 \AA\, use the same continuum value to preserve the
ratio of these neighboring features.  For stars with S/N $< 1.0$
and no emission-lines, no limits are given.  Uncertainties in the EWs are 
generally 10\% of the listed value, and usually limited by the difficulty
in determining the continuum level of a cool type photosphere underneath
an emission-line with broad wings.  The EWs of [OI] 6364 \AA\, and to a
lesser extent [SII] 6716 \AA\, have larger uncertainties (10-25\%) because
of the uncertain removal of sky emission and stellar absorption features at
or near those wavelengths.  In several cases, poor sky subtraction yielded
lower limits on the EW measurements.

\subsubsubsection{H$\alpha$ Emission}

Broad H$\alpha$ emission is one of the most ubiquitous characteristics of
mass accretion.  Measurements of the line profiles potentially offer a
means to characterize the accretion velocity and mass accretion rate
\citep[e.g.][]{muzerolle98}.  H$\alpha$ emission is detected from all stars
observed, albeit marginally for IRAS 04154+2823 and IRAS 04295+2251.
Figure \ref{ha_age} shows the distributions of H$\alpha$ EWs versus
$T_{bol}$ and $\alpha$ for the environmentally young star sample and the
additional Class II star sample.  For illustrative clarity, no lower limits
are plotted (typical of stars with no continuum detection); since all stars
show H$\alpha$ emission, there are no upper limits.  Environmentally
young and old stars have indistinguishable H$\alpha$ EW distributions,
according to K-S tests.  The median EW[H$\alpha$]s of the 16 Class I stars
and the 49 Class II stars are both $-59$ \AA.  The median EW[H$\alpha$]s of
the 25 HH stars is $-63$ \AA\, and is similar to that of the 39 non-HH
stars, $-58$ \AA.  Table \ref{tab_median} summarizes the median statistics,
excluding brown dwarfs.

Several environmentally young stars show evidence of a blue-shifted
absorption component superimposed upon the H$\alpha$ emission profile.
This feature is often seen in the profiles of accreting TTSs and 
is usually attributed to absorption from an outflowing jet or wind 
\citep[e.g.][]{edwards87, ab00}.  

The full widths at 10\% of the peak flux levels (called 10\%-widths,
hereafter) of the H$\alpha$ profiles are listed in Table 4.  This profile
measurement has been proposed as a useful diagnostic of accretion
\citep[e.g.][]{wb03}.  For the environmentally young star sample, these
values range from 139 km/s to 610 km/s.  The majority of the H$\alpha$
profiles are broad (25/28 have 10\%-widths $\gtrsim 250$ km/s), consistent
with the predictions of a magnetically channeled accretion flow
\citep[e.g.][]{hartmann94}.  Nevertheless, several optically veiled and 
presumably accreting stars show relatively narrow 10\%-widths (e.g. IRAS
04260+2642 and HK Tau B).  
The distribution of H$\alpha$ 10\%-widths of Class I stars are marginally
different for that of non-HH stars (K-S probability = 0.062; a 1.9$\sigma$
difference); Class I stars have narrower 10\%-widths, in the median, than
Class II star (356 km/s versus 429 km/s).  In contrast, distributions of
10\%-widths for HH stars and non-HH stars are indistinguishable (K-S
probability = 0.77), and the median values are essentially the same (410
km/s and 413 km/s, respectively).  Table \ref{tab_median} lists the median
values, excluding brown dwarfs.

\subsubsubsection{Forbidden-line Emission}

The optically thin forbidden emission-lines are believed to originate in
an outflowing jet or wind.  Their intensity is therefore expected to be
directly proportional to the amount of material being funneled along the
jet, as viewed through the slit of the spectrograph.  The strongest
forbidden emission-lines in our spectra are [OI] 6364 \AA\, and [SII] 6731
\AA, both of which are detected in 21 of the 28 environmentally young stars
observed.  Figure \ref{sii_age} shows the distributions of EW[SII]s
versus $T_{bol}$ and $\alpha$.  No lower limits are plotted, but upper
limits are.  Additional EW measurements of Taurus T Tauri stars are
included using the values measured by \citet[][all velocities]{heg95}. 

The environmentally young stars have EW[SII] within the range
of values measured for environmentally older stars, but the distributions
are different.  The median EW[SII] of Class I stars is larger than that of
Class II stars (median Class I EW $= -3.1$ \AA; median Class II EW $=
-0.15$ \AA) and the distributions are statistically different (K-S
probability of $1.9\times10^{-5}$).  Similarly, the median EW[SII] of 
HH stars is larger than that of non-HH stars (median HH EW $= -1.8$
\AA; median non-HH EW $= -0.14$ \AA) and the distributions are different,
though less significantly (K-S probability of 0.0086).  Table
\ref{tab_median} lists the median values, excluding brown dwarfs.
These comparisons may be somewhat biased by the large fraction of
[SII] non-detections (31\% for Class I stars, 34\% for Class II stars, 15\%
for HH stars, 47\% for non-HH stars), which are treated as detections.
Nevertheless, comparing only systems with detected emission, the EW[SII]s
of environmentally young stars are still systematically larger than those
of environmentally older T Tauri stars.

The forbidden emission-lines of T Tauri stars are often blue-shifted
and occasionally show double-peaked profiles, with one peak near the stellar
velocity and the second offset blue-ward by 50 - 300 km/s
\citep[e.g.][]{appenzeller84, mundt87, edwards87}.  Only 3 of the 21
environmentally young stars with forbidden-line emission shows a distinct
double peaked emission profile (DG Tau B, L1551 IRS 5, HL Tau).  One of the 
additional TTS stars observed also shows a double-peak (Haro 6-13).  In the
2 cases for which the systemic velocity is known (HL Tau and Haro 6-13),
the blue-shifted peaks are offset by approximately -200 km/s and -100 km/s,
respectively.  About a 
third of the 21 environmentally young stars with [OI] emission have
forbidden emission peaks close to the systemic velocity, but with wings
extending blue-ward by $\gtrsim 50$ km/s.  Half of the stars observed,
however, show only forbidden emission at or very near the systemic
velocity.  This fraction is slightly higher if only Class I stars are
considered.  Of the 13 Class I stars with [OI] emission, only 5 show
evidence of a blue-shifted component (MHO-2, GV Tau A, L1551 IRS 5, HL Tau,
IRAS 04489+3042).  This effect could be explained if the jet emission in
the majority of Class I stars observed propigates perpendicular to our line
of sight.

\subsubsubsection{The Effect of Orientation on the Emission-line Profiles}

The orientation of each star/disk/jet system relative to our
line-of-sight may have effects on both the measured EWs and emission-line
profiles.  One of the most useful qualities of an EW is that, by
definition, it offers a direct measure of the emission-line flux relative
to the continuum, independent of extinction.  Thus, if the spectra are not
flux calibrated and the extinction is not accurately known, as is the
typical case here, the total emission-line flux may still be determined if
the intrinsic flux from the stellar continuum can be estimated (e.g. by
assuming a flux density distribution based on age and 
distance). However, this assumption is only valid if the continuum and
emission-line fluxes originate from the same region, and thus have the same
extinction.  This is most likely true for the permitted
emission-lines of young stars, which in general appear to be optically
thick and originate in high density regions close to the star.  However,
this assumption may not be true for the forbidden emission lines, which
often originate in regions that are spatially extended
from the star \citep[e.g.][]{reipurth99}.  In such a case, the
emission-line region may be more directly observable than the young
partially embedded central star is.  The preferentially attenuated
continuum flux will consequently produce artificially large EW values.

The three systems with well defined edge-on disks based on direct imaging,
HH 30, HV Tau C, and HK Tau B, likely suffer from this bias to some degree.
The forbidden emission line profiles of the edge-on disk systems are narrow
and centered on the systemic velocity, consistent with a bipolar flow
perpendicular to our line-of-sight.  High spatial resolution Hubble Space
Telescope images of both HH 30 and HV Tau C show spatially extended [OI]
and [SII] emission \citep{burrows96, bacciotti99, stapelfeldt03}; this
emission is more directly observable than the continuum stellar emission
which is scattered into our line of sight.  Preferential attenuation of the
continuum likely explains why the EWs of [SII] 6731 \AA\, for HH 30 and HV
Tau C are more that 2 orders of magnitude larger than those of any Taurus T
Tauri star in the survey of \citet{heg95}.  The possibly biased EWs of
edge-on disk systems are therefore distinguished in Figure \ref{sii_age}.
A preferentially attenuated continuum may also explain the large forbidden
emission-line EW lower limits of the 2 stars with no detected continuum (DG
Tau B and L1551 IRS5).  Without spatial information on the origin of these
emission-line features and the efficiency of scattered continuum emission,
this is not something we can correct for.  We highlight this effect,
however, as a potential bias of the forbidden emission-line EWs and the
mass outflow rates inferred from them (Section 3.3.2).

In contrast to the forbidden emission-line EWs, the H$\alpha$ EWs of the 3
edge-on type disk systems are consistent with the EW distribution of
the larger, presumably randomly oriented sample.  This agrees with the
expectation that the majority of the H$\alpha$ emission originates very
close to the star and therefore experiences the same attenuation as the
stellar continuum.  However, they nevertheless have relatively narrow
H$\alpha$ 10\%-widths (293 km/s, 261 km/s, and 196 km/s).  On average these
values are more consistent with those of non-accreting T Tauri stars, with
typical 10\%-widths $< 270$ km/s \citep{wb03}. 
Assuming that the accretion occurs via the same magnetically chanelled high
velocity flow for these 3 stars, then either the edge-on orientation
prevents a direct view of the high velocity emission or the projected
velocities along our line of sight are low.  The absence of strong
blue-shifted absorption component superimposed on the H$\alpha$ emission
is also consistent with an outflow perpendicular to our line of sight.

In light of the emission-line properties of the known edge-on disk systems
HH 30, HV Tau C, and HK Tau B, we suggest that 3 additional stars with
similar properties, CoKu Tau 1, IRAS 04260+2642, and ZZ Tau  
IRS, may also have nearly edge-on orientations.  This prediction is based on
(1) the relatively narrow H$\alpha$ 10\%-widths (285 km/s, 139 km/s, 268
km/s), (2) the absence of any superimposed absorption on the H$\alpha$
emission profiles, and (3) the unusually large strengths of their forbidden
emission-lines, which are both narrow and very close to the systemic
velocity (see Figures \ref{emission1} - \ref{emission3} and Table 3).  The 
emission-lines of IRAS 04158+2805 and IRAS 04248+2612 are also marginal
consistent with an edge-on orientation, but the narrow profiles of these
two stars may simply result from the lower infall/outflow velocities expected
in substellar objects \citep[e.g.][]{jayawardhana03}.  In a few
of these cases, however, we note that the narrow forbidden emission lines
are centered $\sim 10-40$ km/s blueward of the systemic velocity (e.g. IRAS
04158+2805, IRAS 04260+2642, ZZ Tau IRS), suggesting that these systems are
close to, but not completely edge-on.

\subsubsubsection{Emission-line Variability}

L1551 IRS 5 and IRAS 04303+2240 show significant changes in emission-line
properties over a time scale of $\sim$ 3 years. 
During the first epoch observations of IRAS 04303+2240, when the star is
much more optically veiled and presumably accreting at a higher rate, the
permitted emission-lines (H$\alpha$, Ca II) show evidence of absorption
from a blue-shifted outflow, but the forbidden emission-lines are very weak
or absent.  During the second epoch, when the star is less optically
veiled, the permitted line profiles are slightly narrower, more symmetric,
and show no evidence a superimposed absorption.  The forbidden
emission-lines, however, are considerably stronger.  Even after accounting
for the diminished continuum excess, the EW fluxes of [OI], [NII] and [SII]
are all an order of magnitude larger during the second epoch.

No continuum emission is detected in the first observation of L1551 IRS 5,
but is marginally detected in the second (S/N $= 1.4$ at 8450 \AA).  Like
IRAS 04303+2240, the emission-lines profiles of L1551 IRS 5 changed
significantly over the same 3-year time-scale.
The changes included broader emission profiles for both H$\alpha$ and the
forbidden emission lines, the disappearance of the Ca II emission, and
possibly weaker line-emission overall (lower upper-limits).  It is
interesting that the emission-line profiles of the first epoch of IRS 5 are
quite similar to the emission-line profiles of the HH knot obtained during
the second epoch.  The similarities include both the approximate velocity
profile of the H$\alpha$ emission and the 3-peaked profiles of the
forbidden-line emission.

The spectrum of the L1551 nebula is the only spectrum from this system that
shows H$\alpha$ emission with a peak close to the expected systemic
velocity (i.e. the mean of Taurus).  This may actually be scattered
emission from the star.  Low resolution spectra of this nebula have
identified stellar absorption features of a G-type photosphere
\citep{mundt85}.  No photospheric features were detected in our spectrum.
Observations of this scattered light nebula may be the only way to
observed the stellar emission from L1551 IRS 5 at optical wavelengths.

\subsection{Inferred Stellar Properties}

\subsubsection{Stellar and Bolometric Luminosities}

One of the main motivations for determining the stellar luminosities of the 
stars identified as environmentally young in Table 1 is to see if they are
in fact younger than more optically revealed T Tauri stars.  Since low 
mass stars contract primarily along Hayashi tracks, maintaining
roughly constant temperature, younger stars should have larger radii and be
more luminous.  The luminosity of a young star is typically calculated from
a reddening corrected photometric measurement (e.g. $I_c$ or $J$) and a 
bolometric correction corresponding to its spectral type.  Although most of
the stars observed here have both measured photometry and now a spectral
type, ascertaining the intrinsic stellar energy distribution and the proper
reddening correction necessary to estimate their stellar luminosities is
inhibited by the extensive non-photospheric emission associated with these
stars.  At optical wavelengths, scattered light emission can increase the
apparent stellar flux measured in a fixed aperture and can cause
underestimates of extinction because of preferential blue scattering.  At
near-infrared wavelengths, the general trend of increasing thermal emission
from the disk toward longer wavelengths can lead to over-estimates of the
extinction.  Despite these possible systematic uncertainties, we explore
several traditional means for calculating stellar luminosities.  We
consider de-reddening the stars using colors which are as uncontaminated,
that is as close to photospheric as possible, namely $I_c-J$, $J-H$, and
$H-K_s$.  $I_c$ magnitudes are listed in Table 1; $J$, $H$, and $K_s$
magnitudes are from the 2MASS database.  The underlying photospheric colors
are adopted according to our spectral types and a standard interstellar
reddening law \citep{cohen81}.  For comparison purposes, we apply
bolometric corrections to 3 de-reddened magnitudes: $I_c$, $J$, and $K_s$.
For the fraction of our sample where all three methods can be applied,
comparison of the results reveals systematic effects.  Luminosities
calculated from $K_s$ and $H-K$ are systematically higher than those
computed from $J$ and $J-H$ (by 0.35 dex) or from $I_c$ and $I_c-J$ (0.45
dex).  This is likely due to the systematically larger excess emission at
$K_s$ caused by thermal emission from hot dust and gas.  However, the
differences in luminosities calculated from $J$, $J-H$ and I$_c$, $I_c-J$  
are scattered about unity though with large dispersion.  This suggests that
either may provide an adequate measure of the stellar luminosities.
Since we have $J$ and $H$ magnitudes for the majority of stars, we adopt
the stellar luminosities calculated from $J-H$ extinctions and a bolometric
correction to $J$-band.   The calculated visual extinction and stellar
luminosities are listed in the fourth and fifth columns of Table
\ref{tab_prop}.  The uncertainties in these estimates are dominated by
systematic effects, which lead to errors of 1-2 magnitudes in $A_V$ and 
factors of $2-3$ in $L_{star}$, with even larger errors for edge-on disk
systems (e.g. HH 30).

The average visual extinction of the 11 Class I stars with stellar spectral
types is $10.5$ mag, with a standard deviation $4.2$ mag.  This is only a
factor of a few larger than typical values for Class II stars in Taurus, 
and is a factor of a few less than is often suggested by the rising energy
distributions \citep{gl96} and scattered light images
\citep[e.g.][]{kenyon93b, whitney97} of most Class I stars.  As described in
Section 2.1, our Class I sample is biased towards those which are more
optically revealed.

Figure \ref{sdf00} shows the distribution of stellar luminosities for
environmentally young stars in Taurus as a function of effective
temperature.  The temperatures are estimated from the new spectral types
listed in Table 2, assuming the dwarf temperature scale adopted by
\citet{hw04}.  The 6 environmentally young stars which are likely edge-on
disk systems (Section 3.1.4) are marked since this orientation
likely leads to underestimates of their stellar luminosity.
The 3 panels separate stars with T$_{bol} \le 650$, stars
with $\alpha > 0.0$, and stars which power HH flows.   Also shown for
comparison are the mean stellar luminosities of Class II and Class III T
Tauri 
stars in Taurus (environmentally young stars are excluded) measured in 6
temperature bins.  The T Tauri stellar luminosities are taken from
\citet{kh95}, and are determined from a bolometric correction to the
$J$-band, as was done here\footnote{The mean T Tauri luminosities does
not include the stars HBC 358, 359, 360, 361, 362, 372, and 392, which have
insufficient Lithium to be confidently considered T Tauri stars, and does
not include the binaries HBC 354/355 or HBC 356/357, whose primaries lie
below the main-sequence assuming a distance of 140 pc.}.  Figure
\ref{sdf00} also shows the pre-main sequence evolutionary models of
\citet{siess00}.  In comparison to these models, T Tauri stars in Taurus
have a mean age of a few $\times$ 10$^6$ years, but with systematically
younger ages ($< 10^6$ years) at the lowest masses and larger ages ($\sim
10^7$ years) at the highest masses.  The apparent mass dependent age
inferred here is often seen in comparisons of cluster isochrones with
pre-main sequence evolutionary models, and may be caused by incorrect
assumptions in evolutionary models or the assumed temperature scale used
for T Tauri stars.  Finally, the temperatures and luminosities
corresponding to the ``stellar birthline'' for spherical accretion
\citep{stahler88} are shown, using the mass-radius relationship defined in
\citet{fs94} as applied to the \citet{siess00} evolutionary models.

The stellar luminosity estimates of the environmentally young stars are
consistent, on average, with the mean T Tauri star luminosities.  At masses
below $\sim 1$ M$_\odot$, the stellar birthline is systematically
over-luminous by $0.3 - 0.5$ dex compared to the mean T Tauri luminosities
and the environmentally young star luminosities, though statistically
consistent with both of these distributions.  At masses above $\sim 1$
M$_\odot$, however, the stellar birthline is systematically larger by $0.5
- 1.0$ dex than the mean T Tauri luminosities, and only marginally
consistent with the environmentally young luminosities.  We note that the
$10^6$ yr 
isochrone reasonably matches the mean luminosities of environmentally young
stars over all masses.  For direct comparison (and further use in Section
3.3 below), we list the $10^6$ yr luminosity corresponding to the stellar
temperature for all stars with temperature estimates in Table 4
($L_{star}^{1Myr}$).  Independent of selection criteria, environmentally
young stars appear to be coeval with populations of Class II and III stars
in Taurus.

\subsubsection{Stellar Masses and Angular Momentum}

The stellar luminosity and temperature estimates of the environmentally
young stars allow direct comparison with pre-main sequence evolutionary
models to estimate stellar masses.  However, we do not favor computing
masses via direct comparison given the large spread in stellar
luminosities, likely caused by biases introduced from circumstellar
material.  The cool temperatures of these K and M type young stars suggest
that most should reside on the convective and predominantly vertical
(constant temperature) portion of their pre-main sequence evolution.
Their temperature principally will determine their stellar mass.  Therefore 
we use the temperature - mass relation defined by the \citet{siess00}
1 Myr isochrone 
to estimate masses for the stars in our sample with spectral type
estimates.  As noted above, the $10^6$ yr isochrone reasonably represents
the average apparent age of these stars.  Masses for objects cooler than
the lowest mass track (0.1 M$_\odot$) are estimated to be 0.07 M$_\odot$ at
spectral type M5.5 and 0.05 M$_\odot$ at spectral type M6, based on both a
modest extrapolation of the \citet{siess00} models and comparisons with
other evolutionary models \citep[e.g.][]{baraffe98}.

The inferred stellar/substellar masses are listed in Table 3, and span
nearly 2 orders of magnitude: 0.05 M$_\odot$ to 3.5 M$_\odot$.  The
distribution of masses is shown in Figure \ref{mass_age} as a function of
the two evolutionary diagnostics $T_{bol}$ and $\alpha$.
For comparison, the additional sample of Class II stars is shown; their
masses are determined from their assigned spectral types and the same
temperature - mass relationship used above.  The similar spectral type
distribution of Class I and Class II stars (Section 3.1.2) translates into
a similar mass distribution.  Likewise, the slight trend towards hotter
spectral types for HH stars translates into a slight difference in the
stellar mass distributions; HH stars are slightly more massive.  
It should be realized, however, that the comparison Class II population
shown in Figure \ref{mass_age} is not representative of the true
distribution of stellar (and substellar) masses in Taurus.  The now well
established populations of low mass stars and brown dwarfs in Taurus
\citep[e.g.]{briceno98, briceno02, martin01, luhman03} are in general too
faint to have been detected in previous mid- and far-infrared surveys
(e.g. IRAS), from which spectral indices and bolometric 
temperatures can be determined.  Thus without these evolutionary diagnostic
measurements, they can not be plotted in Figure \ref{mass_age}.
Nevertheless, the 3 Class I stars IRAS 04158+2805, IRAS 04248+2612, and
IRAS 04489+3042 are of particular interest.  With spectral types of M5.5 or
cooler, they have substellar masses and are the first spectroscopically
confirmed Class I brown dwarfs.

Figure \ref{break_age} shows the $v$sin$i$ values measured in Section 3.1.1
normalized by their break-up velocity, defined as $v_{br} = \sqrt{GM/R}$.
Stellar mass and radius estimates are from the adopted \citet{siess00}
$10^6$ yr isochrone described above.  The normalized velocities range from
less than 5\% to 20\% of the break-up velocities.  K-S comparisons of Class
I stars with Class II stars, and HH stars with non-HH stars show no 
differences in distributions (K-S probabilities of 0.079 and 0.096,
respectively).  The difference in the $v$sin$i$ distributions of HH stars
and non-HH stars (Section 3.1.1) disappears when these values are
normalized by the break-up velocity, possibly because of the slight mass
difference between HH and non-HH stars.  There is no evidence for a change
in the angular momentum between the evironmentally young and the T Tauri 
samples.

\subsection{Inferred Circumstellar Properties}

In this section mass accretion rates are estimated from the observed
continuum excesses and mass outflow rates are estimated from the strengths
of forbidden-line emission, under the assumption of a bipolar flow.  We
outline the steps of these procedures in some detail to highlight the
substantial uncertainties in current estimates.  The ratio of these
quantities are used to constrain the origin and energetics of these
apparently physically related processes.

\subsubsection{Mass Accretion Rates}

The continuum excesses observed in T Tauri star spectra are attributed to
the high-temperature regions generated as accreting material shock
decelerates at the stellar surface.  Measurements of this excess can
therefore be used to estimate the mass accretion rate under the assumption
that the liberated energy is gravitational potential energy.  The steps
involved in estimating mass accretion rates are illustrated in the 5 panels
of Figure \ref{macc_mass}.  Figure \ref{macc_mass} contains only a subset
of our data; it is restricted to a narrower mass range for direct
comparison with the stellar and accretion properties of the 17 Taurus T
Tauri stars studies by \citet{gullbring98}.  The mean and range from this
study are indicated in each panel.

The top panel of Figure \ref{macc_mass} shows the distribution of continuum
excesses at 6500 \AA\, ($r_{6500}$) as a function of stellar mass for stars
observed here and the additional sample of Class II stars\footnote{The
continuum excess measurements for the additional Class II stars are from
\citet{bb90, heg95} and \citet{wb03}.  Although the values measured by
\citet{bb90} and \citet{wb03} are determined near 6500 \AA, as are those
newly presented here, the values measured by \citet{heg95} are determined
at a slightly shorter wavelength, 5700 \AA.  However, direct comparison of
the veiling values for stars observed by both \citet{heg95} and
\citet{bb90} shows no difference (16 star overlap; median $r_{6500}$ -
$r_{5700}$ = +0.03).  We use the $r_{5700}$ values as $r_{6500}$ values.}
(Section 3).  The distribution of continuum excesses is indistinguishable
from that measured by \citet{gullbring98} at bluer 
wavelengths (3200 - 5300 \AA).  As noted by \citep{bb90} and further
supported by the results in Section 3.1.2, the veiling is relatively
constant redward of $\sim$ 5000 \AA.

Since the continuum excess measurements are determined relative to the
photosphere ($r = F_{excess}/F_{photosphere}$), the photospheric flux needs
to be known in order to determine the continuum excess flux.  Since current
estimates of the photospheric flux are subject to very large uncertainties
(Section 3.2.1), we assign stellar flux values based on the predictions of
the \citet{siess00} evolutionary models for 1 Myr aged stars as observed 
through the $R_c$ passband.  An $R_c$ zero point of $2.32\times10^{-9}$ erg
s$^{-1}$ cm$^{-1}$ \AA$^{-1}$ \citep{hayes85} is assumed.  The second panel 
of Figure \ref{macc_mass} shows the distribution of stellar luminosities
corresponding to the adopted photospheric flux levels.  We show stellar
luminosities, rather than $R_c$-band fluxes, for direct comparison with the 
\citet{gullbring98} values.  Our luminosities are slightly larger than
those determined by \citet{gullbring98}, but by less than a factor of 2 
in the mean.  The third panel in Figure \ref{macc_mass} shows the
distribution of excess fluxes measured over the 6000 - 6500 \AA\,
wavelength interval.  Surprisingly, the distribution is indistinguishable
from that measured over the 3200 - 5300 \AA\, interval by
\citet{gullbring98}; standard hot spot models predict the excess flux 
should be only one-tenth of that seen in the bluer passband \citep{hk03}.
Up to a factor of 2 of this discrepancy may be caused by larger stellar
luminosities, but the observed red excesses are still larger by a
factor of $\sim 5$ compared with standard hot spot model predictions.

In order to convert the excess measurements to the total accretion
luminosity, a bolometric correction is needed.  Based on the calculations
of \citet{hartigan91} who model the emission from the accretion shock as a
slab of pure hydrogen gas of constant temperature and density in LTE,
\citet{hk03} estimate that the bolometric correction for an excess
measured over 6000 - 6500 \AA\, is $\sim 35$.  However, this
bolometric correction, which is 10 times larger than the bolometric
correction predicted for the 3200 - 5300 \AA\,
passband, will yield accretion luminosities that are 10 times larger than 
that measured by \citet{gullbring98} since the two passbands have similar
flux excesses.  Consequently, we suspect that the mass accretion rates
determined by \citet{hk03} are most likely over-estimates, as their best
matched model would predict very large blue wavelength excesses, in
contrast to what is observed.  Similarly, we suspect that inferred mass
accretion rates determined by \citet{gullbring98} are most likely
under-estimates, as their best matched model predict negligible red
excesses, in contrast to what is observed.  Without a better model for the 
total continuum excess flux, we adopt bolometric correction of 11,
corresponding to the logarithm average of the two extremes (3.5 and 35).
The resulting accretion luminosities are shown in the fourth panel of
Figure \ref{macc_mass} and are roughly a factor of 3 larger than those
estimated by \citet{gullbring98}.

Finally, the accretion luminosities are converted to mass accretion rates
(Figure \ref{macc_mass}; bottom panel) by assuming that the accretion shock 
luminosity equals the gravitational energy released per second as
material free-falls along magnetic field lines from an inner disk radius of
3$R_{star}$ \citep[e.g.][]{gullbring98, hk03}.  This assumption will
under-estimate the accretion rate if a non-negligible fraction of the
gravitational energy is used to power a wind or jet (the accretion rate
will increase by a factor of 2 in the case where 1/2 the energy is used in 
mass loss processes).  Comparison of the mechanical luminosity of the
outflow to the accretion luminosity suggests this is unlikely (Section
3.3.3).  The stellar radii used in these calculations are determined from
the temperature - radii relation of the \citet{siess00} $10^6$ yr
isochrone.  Our geometric assumptions differ from those of
\citet{gullbring98} only in the choice of inner disk radius.  This small 
difference, in combination with the difference in accretion luminosities,
leads to mass accretion rates that are, on average, a factor of 4 larger
than those of \citet{gullbring98}.  The fifth panel of Figure
\ref{macc_mass} shows the distribution of inferred mass accretion rates.

The mass accretion rates for stars observed here are listed in Table 4 and
span over 3 orders of magnitude, from $4 \times 10^{-10}$ M$_\odot$/yr to
$7 \times 10^{-7}$ M$_\odot$/yr.  The distribution of mass accretion
rates determined for environmentally young stars and for the additional
sample of Class II stars are shown in Figure \ref{macc_age} as a function
of the evolutionary diagnostics $T_{bol}$ and $\alpha$.
The median mass accretion rate of Class I stars is slightly less than that
of Class II stars (median Class I log($\dot{M}_{Acc}$) = -8.1; median Class
II log($\dot{M}_{Acc}$) = -7.3).  This is primarily because of the strong
mass dependence of the accretion rate \citep[e.g.][]{wb03, muzerolle03a};
unlike the Class I sample, the Class II sample does not extend into the
substellar regime because of insufficient data for determining 
bolometric temperatures and spectral indices.  Excluding the 3 Class I
brown dwarfs, the distributions of mass accretion rates for Class I stars
and Class II stars are indistinguishable (K-S probability of 0.60) and have 
more similar median values (median stellar Class I log($\dot{M}_{Acc}$) =
-7.1).  The distributions of mass accretion rates for HH stars and non-HH
stars are also similar (K-S probability of 0.11) though HH stars have mass
accretion rates that are larger by a factor of 2.5 in the median (median HH
log($\dot{M}_{Acc}$) = -7.0; median non-HH log($\dot{M}_{Acc}$) = -7.4).
Table \ref{tab_median} summarizes the median statistics, excluding brown
dwarfs.  Combining all types, the typical (median) mass accretion rate for
an optically veiled K7-M1 star is $4 \times 10^{-8}$ M$_\odot$/yr, with a
range of 2 orders of magnitude.

The inconsistencies between the predictions of standard hot spot models and
the large red excesses observed merit some discussion.  Previous work has
demonstrated that the variations in the veiling at blue wavelength are
simultaneously correlated with variations at red wavelengths \citep[see
Figure 4 in][]{bb90}, suggesting a physically related origin.  Emission
from the inner disk is unlikely to contribute much optical excess flux if
the temperature is limited to the dust destruction temperature of
silicate-type grains ($\sim 1500$ K), as some observations suggest
\citep{muzerolle03b}.  The interpretation that we favor is that the this
emission originates in a cooler component of the accretion generated shock.
If this cooler component dominates the excess at red wavelengths, then the
observed constant flux over the wavelength range 6500 \AA - 8400 \AA\,
(Figure \ref{veil_ratio}) requires a surprisingly cool temperature of only 
$\sim 4000$ K.  Thus for a K7 star of similar temperature, the excess 
emission should be constant with wavelength, as observed \citep[Figure 
\ref{veil_ratio};][]{bb90}.  This cooler component 
must be featureless, suggesting that it is heated from above.  The cooler 
excess component must also cover a large faction of the star in order to be
observed.  For example, in the case where the star and the excess have the
same effective temperature (e.g. for a K7 star), the veiling, or the ratio
of excess flux to photospheric flux, will be equal the ratio of the area of
these emitting regions ($r = F_{ex} / F_{phot} = A_{ex} / A_{phot}$).  Thus
for $r = 1.0$, the cool featureless emission will cover 1/2 of the stellar 
surface.  A more detailed investigation of the accretion generated excess
will require broader wavelength coverage that what is presented here 
\citep[e.g.][]{sp03} and is consequently beyond the scope of this study.

\subsubsection{Mass Loss Rates}

Optically thin forbidden-line emission can be used to estimate the mass
loss rate from young accreting stars.  Forbidden emission lines such as
[OI] 6300 \AA\, and [SII] 6731 \AA\, in the spectra of accreting T Tauri
stars nearly always show emission at the systemic velocity and often show a
separate blue-shifted peak or blue-shifted asymmetry \citep[corresponding
  to velocities of a few$\times10$ to a few$\times100$ km/s;
  see][]{appenzeller84, mundt87, edwards87, cabrit90, hamann94, heg95}.
The blue shifted emission is attributed to a
high velocity wind or jet on the near side of the star; the presence of a
circumstellar disk is believed to prevent direct observation of the
redshifted component of the presumed bipolar flow.  High spatial resolution
imaging studies of young stars with strong forbidden-line emission have
shown that the high and low velocity components likely originate in
separate regions.  The low velocity emission is usually spatially
coincident with the stellar position, while the high velocity emission
appears spatially extended, often in the form of a well collimated jet
\citep[e.g.][]{mundt87, lavalley97, bacciotti00}.  The presence of a high
velocity component of course requires the flow to be directed nearly along
our line of sight; several systems with edge-on disk orientation power
optical jets but have no high velocity forbidden line emission, as expected
(Section 3.1.4).  Interpretation of the low velocity forbidden
line-emission is still uncertain \citep[see e.g.][]{heg95}; it too may
originate from outflowing material at slower velocities, possibly a disk
wind \citep{kt95}.

We estimate mass outflow rates using the prescription given by
\citet{heg95} for [SII] 6731 \AA\, emission.  In our spectra, this
forbidden emission-line is typically the strongest and 
the least contaminated by telluric or stellar features.  Since the [SII]
emission is optically thin, the observed [SII] luminosity can be used to
estimate the total mass of the emission region.  \citet{heg95} predict the
mass of the emitting region to be $1.43\times10^{-3} (L_{6731}/L_\odot)$
M$_\odot$, where $L_{6731}$ is the [SII] 6731 \AA\, luminosity in solar units.
This mass estimate assumes the electron density is in the high density limit
($N_e > 2\times10^{4}$ cm$^{-3}$; consistent with [SII] line ratios), that
all the Sulfur atoms are singly ionized, and a cosmic sulfur 
abundance.  With a mass $M$ of the emitting region, the mass loss rate can
then be estimated from the flow speed $V$ and the length scale $L$ over
which the emission is observed ($\dot{M}_{Out} = MV/L$).

The EW[SII]s are converted to fluxes by assuming that the 
underlying continuum flux level is that of the star in the $R_c$ passband,
as predicted by the \citet{siess00} 1 Myr isochrone, multiplied
by (1+$r_{6500}$) to account for veiling.  This is the
same continuum assumption used to estimate the mass accretion rates.
The majority of stars observed do not have well-separated high- and 
low-velocity forbidden-line emission, which would allow us to directly
measure the velocity of the flow and the fraction of high velocity
flux relative to the total line flux.  Therefore for all sources we
assume a flow speed of 150 km/s and a high velocity flux equal to
40\% of the total [SII] 6731 \AA\, line flux.  This flow speed is the same
as that used by \citet{heg95} and is consistent with the
de-projected velocities of propagating HH knots
\citep[e.g.][]{krist99, bacciotti99}.  The adopted percentage of high
velocity flux is the mean percentage of high velocity emission measured by 
\citet[][compare their Tables 3 and 4]{heg95}.  The scale length
is assumed to be the slit-width of our observations (1\farcs15),
projected to the distance of Taurus (140 pc).  We note that the slit-width
is of similar size to the seeing disk.  Thus even if the jet was projected
along the length of the slit, the characteristic length scale would still
be appropriate since the width of the aperture extraction is defined by the
seeing-limited width of the continuum, and is relatively unaffected by
spatially extended emission lines.

Following this simple prescription, mass outflow rates are determined for
all stars with [SII] 6731 \AA\, emission, and upper limits to the mass
outflow rate are determined in cases where only detection limits are
available.  We note that for edge-on systems, the apparent bias towards
larger forbidden emission-line EWs will correspondingly bias the inferred
mass outflow rates.  The determined mass outflow rates are listed in Table
4 and range from $2 \times 10^{-10}$ M$_\odot$/yr to $3 \times 10^{-6}$
M$_\odot$/yr.  Ignoring edge-on systems, the largest mass outflow rate is
then $6 \times 10^{-7}$ M$_\odot$/yr (DG Tau).  The calculated mass
outflow rates are shown in Figure \ref{mout_age} as a function of
$T_{bol}$ and $\alpha$.  Likely edge-on systems are
indicated.  [SII] 6731 \AA\, measurements for the additional Class II
star sample are taken from \citet{heg95}; the outflow rates are determined
similarly.  We note that the inferred measured mass outflow rates are
slightly larger than, by a factor of 3, the values determined from
high spatial  
resolution spectroscopy \citep[e.g.][]{bacciotti02, woitas02}.  The mass
outflow rates are also roughly a factor of 2 larger than those measured by
\citet{heg95} using the high velocity component of [OI] 6300 \AA.  The
latter difference is primarily due to the slightly higher stellar
luminosities of our adopted isochrone.  Both samples show a very large
dispersion in the mass outflow rate ($\sim 3$ orders of magnitude for K7-M1
spectral types).  Thus, even though we use a weaker forbidden emission line
to measure outflow rates, the similar dispersions suggest that the large
range of values are dominated by stochastic differences between systems.

The median mass outflow rates of Class I stars and Class II stars are
similar (median Class I log($\dot{M_{Out}}$) = -8.0; median Class II
log($\dot{M_{Out}}$) = -8.2).  Although the [SII] 6731 \AA\, emission
lines, from which the mass outflow rates are calculated, are systematically
larger for Class I stars than Class II stars, this dependence is less
significant than the stellar mass  dependence of the mass outflow rate.  If
the 3 Class I brown dwarfs are excluded, the median mass outflow rate of
Class I stars is 31 times greater than that of Class II stars (median
stellar Class I log($\dot{M_{Out}}$) = -7.0).  The distributions are
nevertheless statistically indistinguishable (K-S probability of 0.29;
edge-on disk systems have been excluded).  The outflow rates of HH stars
are, in general, greater than those of non-HH stars (median HH star
log($\dot{M_{Out}}$) = -7.4; median non-HH star log($\dot{M_{Out}}$) =
-8.7).  These distributions show a more significant difference (K-S
probability of 0.00049).  Table \ref{tab_median} summarizes the median
statistics, excluding brown dwarfs.  Combining all types except edge-on
systems, the typical (median) mass outflow rate for an optically veiled
K7-M1 star is $2 \times 10^{-9}$ M$_\odot$/yr.

\subsubsection{Comparison of Mass Accretion and Mass Loss Rates}

The measured mass outflow rates are correlated with the measured mass
accretion rates over the 3 orders of magnitude in $\dot{M_{Acc}}$ inferred 
here.  This correlation has been demonstrated previously
\citep[e.g.][]{heg95} and suggests a physically related origin, though it
in part stems from the mass dependence of the accretion and outflow rates. 
This correlation is not uniquely one-to-one, however.  There are 14
optically veiled stars which show no [SII] emission, and 3 stars with
optical veiling upper limits which show strong [SII] emission. \citet{hk03}
note similar disparities using the stronger [OI] 6300 \AA\, emission line.

Although both the inferred mass accretion and outflow rates depend directly
on the photospheric luminosities, assumed in our analysis to correspond to
1 Myr aged stars, the ratio of these rates is independent of the
photospheric luminosity.  These mass flow ratios are listed following the
mass accretion and mass outflow rates in Table \ref{tab_prop}.  In cases
where [SII] 6731 \AA\, emission is measured, but only veiling upper limits
are available, the ratio is a lower limit.  In cases where the star is
optically veiled, but no [SII] 6731 \AA\, emission is detected, the ratios
are upper limits.  These values are plotted versus the evolutionary
diagnostics $T_{bol}$ and $\alpha$ in Figure \ref{mflow_age}.  The ratio
$\dot{M_{Out}}/\dot{M_{Acc}}$ for 
5 of the 6 edge-on disk systems are $\sim 10$.  Since the majority of disk
material is expected to accrete onto the star, as opposed to being ejected
in an outflow, the large ratios for these systems support the 
hypothesis that the forbidden emission-lines are biased towards
artificially large values.  Excluding edge-on disk systems, the measured
ratios range from $3.2 \times 10^{-3}$ to $1.7$.  The median
log ratio (including limits) is -1.2, corresponding to a ratio of 0.05.
This value is similar to, but larger by a factor of $\sim 5$, the
average value measured by \citet{heg95}. 

Although the distributions of mass flow ratios for Class I stars and Class
II stars span similar ranges, the mass flow ratios of Class I stars are 
larger, in the median, than those of Class II stars (median Class I
log($\dot{M_{Out}}$/$\dot{M_{Acc}}$) = -0.04; median Class II
log($\dot{M_{Out}}$/$\dot{M_{Acc}}$) = -1.2).  The mass flow rates
of HH stars are similarly larger than the values for non-HH stars (median
HH star log($\dot{M_{Out}}$/$\dot{M_{Acc}}$) = -0.13; median non-HH star 
log($\dot{M_{Out}}$/$\dot{M_{Acc}}$) = -1.4).  These differences are
correlated however, since many Class I stars are also HH stars.  Excluding
Class I stars, the median HH star log($\dot{M_{Out}}$/$\dot{M_{Acc}}$)
decreases to -0.81.  If many of the Class I stars observed here have 
edge-on orientations (Section 4.5), the effect of this on the observed
forbidden-line EWs (Section 3.1.4) could explain their higher ratios of
$\dot{M_{Out}}$/$\dot{M_{Acc}}$.

The ratio of $\dot{M_{Out}}$ / $\dot{M_{Acc}}$ appears to be independent of
the mass accretion rate.  Over the $\dot{M_{Acc}}$ range $10^{-8} -
10^{-6}$ M$_\odot$/yr, the mass flow ratios are uniformly distributed.  We 
note that the large range in $\dot{M_{Out}}$ / $\dot{M_{Acc}}$ is not
because of observational error, but a consequence of significant dispersion
in the relative strengths of the forbidden emission-lines and the continuum
excesses.  The two observations of IRAS 04303+2240, which measure
log($\dot{M_{Out}}$/$\dot{M_{Acc}}$) at -2.0 and -0.5, suggest this ratio
changes for a given star-disk system (thereby excluding random orientation
effects).  These changes may arise from either actual changes in the mass
accretion and/or mass outflow rates, or from changes in the properties of
the emission regions (e.g. temperature and density).  The highly variable
mass accretion rate will cause an additional source of variation for any 
given epoch.  The previously ejected, spatially extended forbidden emission 
should be more correlated with a time averaged mass accretion rate, as
opposed to the instantaneous measured mass accretion rates we present here.

Theoretical estimates for the ratio of mass outflow to mass accretion rate
range from $\sim 1/3$ \citep[e.g.][]{shu94} for a collimated jet powered by
a magnetospheric accretion flow to $\sim 10^{-4}$ for a slow disk
wind from a magnetically threaded accretion disk \citep[e.g.]{wk93}.
Although the high velocity components and the current ranges of
log($\dot{M_{Out}}$/$\dot{M_{Acc}}$) appear to exclude low velocity disk
winds as a dominant mass loss mechanism, the large range in mass flow
ratios currently offer little constraint on proposed models for powering
the observed jets.

Finally, the ratio $\dot{M_{Out}}$/$\dot{M_{Acc}}$ offers a useful
measure of the mechanical luminosity of the outflow
(1/2$\dot{M_{Out}}$V$^2$) relative to the energy generated by accretion
($L_{Acc}$).  The right hand ordinate of Figure \ref{mflow_age}
shows this ratio, assuming accretion onto a 0.5 M$_\odot$ star of radius
2.0 R$_\odot$, and a mass loss velocity ($V$) of 150 km/s.  The median
luminosity ratio is 0.076.  Thus the mechanical luminosity of the outflow
is approximately 8\% of the observed accretion luminosity.

\section{Discussion}

The insights gained from our spectroscopic observations address various 
long-standing assumptions regarding the earliest optically visible stages 
of star formation.  In order to place our new results in context, we
present a brief review of the standard paradigm of low mass star formation 
and the taxonomy of young stars.

The formation of a low mass star (M $\lesssim$ 3 M$_\odot$) is thought to 
begin with the dynamical inside-out collapse of a dense molecular cloud
core \citep[e.g.][]{shu87}.  A 3 stage description of evolution from this
point towards the main sequence was proposed to explain 3 relatively
distinct types of young stars as distinguished by their optical through
millimeter SEDs \citep[e.g.][]{lada87, adams87, wilking89}.  Young stars
with SEDs that peak at far-infrared wavelengths ($30-100 \mu$m) 
are presumed to be in the initial main accretion phase.  These stars are
often called Class I stars or protostars. The majority of their luminosity
escapes at far infrared wavelengths as reprocessed radiation in the
form of thermal dust emission; extinction to the star is often high.
Young stars with SEDs that peak at near infrared wavelengths
are Class II stars or classical T Tauri stars.  Their SEDs typically
resemble a revealed stellar photosphere with significant infrared and
millimeter excess emission attributed to thermal emission from a
circumstellar disk.  Young stars with SEDs that peak in the optical, with
little or no evidence for an infrared excess are called Class III stars or
weak-lined T Tauri stars.  These stars are presumed to have dispersed the
majority of their circumstellar disk material.  Subsequent to this
3 Class classification scenario, \citet{andre93} proposed a new class of
less evolved objects, called Class 0 stars.  These are stars that are so
deeply embedded that they are only observable at far-infrared and
millimeter wavelengths.  Distinction between Class 0 and Class I stars
is often observationally ambiguous, however \citep[e.g.][]{young03}; both
are believed to be in the main accretion phase, though the envelope is
presumed to be more mass than the star during the Class 0 stage.

The association of strong outflow/jet signatures, such as HH flows, with
many protostars suggest that this process plays an important role in
clearing envelope material.  Within the Class 0/I/II/III classification
scheme, HH flows first appear associated with Class 0 stars
\citep[e.g. IRAS 04368+2557][]{eiroa94}, are most common among Class I
stars, are associated with the about 10\% of Class II stars
\citep{gomez97}, and are not associated with any Class III stars.
HH stars are therefore expected to be, on average, younger than most Class
II or Class III T Tauri stars.

Since this classification scheme describing the evolution of circumstellar
material around solar-type stars was first proposed, a wealth of direct
observational evidence has been obtained in support of it.  
However, a correlation with stellar age has never been established.  In
this section we use the stellar properties (mass, rotation, age), and
circumstellar properties (accretion rate, outflow rate, disk orientation,
disk mass) of the 11 Class I stars and 43 Class II stars studied here
to address issues related to the evolution from the protostellar stage to
the T Tauri stage.  We use these same properties to search for differences
between HH stars and non-stars.

\subsection{The Masses of Class I Stars}


If the inside-out collapse of a singular isothermal sphere proceeds under
the control of slow ambipolar diffusion processes \citep[e.g.][]{shu77,
shu87}, the resulting infall rates are set by the local isothermal sound
speed and therefore should remain roughly constant in time throughout the
cloud.  One implication is that more massive stars require more time to
form than less massive stars.  Thus, assuming coeval star formation, 
these conditions would yield Class I stars that are more massive, on
average, than Class II stars.  However, recent observational, theoretical
and numerical work suggests that supersonic turbulent flows rather than
static magnetic fields control star formation \citep[see the review
by][]{mk04}, affecting both the infall rate and the formation time scale.
As an example, numerical models of gravoturbulent fragmentation suggest
that the initial infall rates may be time dependent (initially large),
scale with the final stellar mass, and yield a main accretion timescale
that is nearly independent of the final stellar mass \citep{sk04}.

Our results are more consistent with the proposal that supersonic turbulent
flows control star formation, rather than static magnetic fields.  As shown
in Figure \ref{mass_age}, the distributions of stellar mass for Class I
and Class II stars are statistically indistinguishable.  Although the
sample of Class I stars with mass estimates determined here is small, there
are 3 brown dwarfs, 6 sub-solar mass stars (0.2 - 0.9 M$_\odot$), and 2
super-solar mass stars.  Previous work has determined the masses of 2 Class
I stars based on disk kinematics under the assumption of Keplerian rotation
\citep[0.2-0.4 M$_\odot$ for IRAS 04381+2540 and 0.35-0.7 M$_\odot$ for
IRAS 04365+2535;][]{bc99}.  This distribution of Class I star 
masses agrees very well with the initial mass function measured from T
Tauri stars in Taurus \citep{briceno02}.  The implication is that the known
population of optically revealed T Tauri stars is likely to be a good
representation of the final initial mass function produced by the cloud.
The current and presumably subsequent generations of Class I stars are
unlikely to skew the mass function.

The cool spectral types of 3 Class I stars
suggest that they are Class I brown dwarfs (IRAS 04248+2612, IRAS
04158+2805, and IRAS 04489+3042).  With the exception of the most massive
of these 3, IRAS 04248+2612, the 1.3 millimeter continuum fluxes suggest
circumstellar disk+envelope masses of only a few hundreths of a solar mass
\citep{ma01}.  Thus, even if the majority of circumstellar material were to
accrete onto these brown dwarfs, both IRAS 04158+2805 and IRAS 04489+3042
will remain substellar.  If all of the 0.25 M$_\odot$ envelope accretes onto
IRAS 04248+2612, it will be a low mass star.  IRAS 04248+2612 is also of
interest as it is the lowest mass object known to power a molecular outflow
(Moriarty-Schieven et al. 1992) and a HH object \citep[HH
31;][]{gomez97}.  We suggest that the lowest mass Class I stars/brown
dwarfs are low mass because they formed from lower mass cloud cores, and
not because they are at an earlier evolutionary stage and have not yet
accreted the majority of their final mass.  The implication is that the
formation process of substellar mass objects, down to $\sim 0.05$
M$_\odot$, is simply a scaled-down version of that for solar mass 
stars.  Brown dwarfs likely form via the dynamical collapse of a cloud
core, experience an embedded Class I phase of evolution, and are capable of
powering both molecular and HH flows.  The presence of circumstellar
material and accretion argues against low mass star/brown dwarf formation
scenarios involving early ejection from a competitively accreting cluster
\citep[e.g.][]{rc01}.

Finally we note that without stellar temperature estimates, previous
attempts to estimate the stellar masses of protostars have proceeded by 
assuming accretion dominated luminosities; the stellar mass can be
estimated by calculating the gravitational potential well required to
liberate the observed luminosity.  The low luminosities of Class I stars in
Taurus consequently lead to very low substellar masses \cite[e.g. Haro 6-33
  $=$ IRAS 04385+2550;][estimate M$= 0.01$ M$_\odot$]{young03}.  These
masses are generally inconsistent with the values determined more directly
here (e.g. Table 4; for Haro 6-33, M$ = 0.46$ M$_\odot$).  Mass estimates
assuming accretion dominated luminosities appear to be inappropriate for
many Taurus protostars.

\subsection{The Ages of Class I Stars}


Unlike the self-embedded Class I stars, the optically revealed nature of
Class II and Class III stars have allowed more direct measurements of their
stellar temperatures and luminosities for comparison with the predictions
of pre-main sequence evolutionary models.  These comparisons imply T Tauri
ages of a few$\times 10^{6}$ yr, with no statistically significant
difference between the HR diagram locations of Class II and Class III stars
\citep{kh95, wg01, briceno02}.  Class II and III stars also have
similar lithium abundances \citep[e.g.][]{strom89, martin94} supporting
similar ages.  Given that Class I stars in Taurus are approximately
one-tenth as numerous as T Tauri stars (both Class II and Class III stars 
combined), it has been postulated that the Class I phase lasts for roughly
10\% of the T Tauri phase, under the assumption of a constant star
formation rate \citep[e.g.][]{myers87, kenyon90}.  The implied statistical
age of Class I stars is then a few$\times 10^5$ yr.

With stellar properties now determined for 42\% of the Class I stars in
Taurus, their stellar ages can be assessed more directly.  As illustrated
in Figure \ref{sdf00}, environmentally young stars have ages that scatter
uniformly about those of Class II and Class III stars, independent of the
selection criteria used to identify them ($T_{bol}$, $\alpha$, HH
properties).  The mean stellar luminosities of Class I stars,
specifically, are also more consistent with those of optically revealed
T Tauri stars than with the stellar birthline, though the difference is
only signficant above 1.0 M$_\odot$.  Class I stars appear to be coeval
with Class II and Class III stars.

\subsection{The Early Evolution of Angular Momentum}

The accretion of high angular momentum material from a rotating, collapsing
protostellar envelope during the earliest stages of star formation is
expected to produce very rapidly rotating stars \citep[e.g.][]{durisen89},
with rotational velocities comparable to the break-up velocity ($v_{br} =
\sqrt{GM/R} \sim 300$ km/s).  Thus, it came as a surprise to discover that
most T Tauri stars, at least in the Taurus star forming region, are slowly
rotating.  Nearly all of these stars have rotational velocities
less than one-tenth $v_{br}$ \citep[e.g.][]{hartmann86, bouvier90,
bouvier95}.  Proposed explanations for the slow rotation rates usually
involve ``magnetic braking'', a mechanism in which the star is
magnetically coupled to the accretion disk and transfers angular
momentum to the more slowly rotating parts of the outer disk \citep{bp82,
pn86}, the inner disk \citep{konigl91, shu94} or possibly a stellar
wind \citep{kt88}.  After the disk dissipates, substantial angular momentum
loss must still occur, likely through a magnetically coupled stellar wind,
in order to explain the slow rotation of somewhat older stars such as those
in the Pleiades and Hyades \citep{stauffer97, krishnamurthi98}.

Thusfar, the rotational properties of stars during the main accretion phase
have been difficult to ascertain.  High dispersion near-infrared spectra of
stars in the $\rho$ Ophiuchus dark cloud have found that at least some
environmentally young stars rotate moderately rapidly \citep[$v$sin$i$ =
  30-50 km/s;][]{gl97, gl00}, with tentative evidence that stars with 
large spectral indices rotate the fastest \citep{doppmann03}, though this
is based on only 3 stars with spectral indices $> 0.0$ (i.e. Class I-like).
Here we have demonstrated that Class I stars in Taurus are, as a group,
slowly rotating, with a $v$sin$i$ distribution indistinguishable from that
of the more environmentally evolved Class II stars (Figure \ref{rot_age}).
None of the observed Class I stars in Taurus rotate with $v$sin$i$ $>$30
km/s.  Further, since the Class I luminosities and hence radii are similar
to those of the Class II stars, there is no evidence for an change in
angular momentum from the Class I to the Class II stages (e.g. Figure
\ref{break_age}).  The youngest optically visible stars have already been 
slowed to well below ($<$5-15\%) break-up velocity.  Stars in Taurus 
are known to be systematically slower rotators than stars in Orion
\citep{cb00, wb03}.  Similar differences in the cloud properties or
possibly stellar ages (see below) may explain apparent differences in the
rotation rates of Taurus Class I stars and $\rho$ Ophiuchus Class I stars,
once confirmed with a larger $\rho$ Ophiuchus sample.

It has been proposed that T Tauri stars in Taurus rotate more slowly than
stars in other star forming regions (e.g. $\rho$ Ophiuchus, Orion) because
Taurus stars are older (by a factor of $\sim 3$), having ages larger than
the expected disk braking timescale \citep{hartmann02}.  In this scenario,
the slow rotational velocities of Class I stars therefore corroborates the
suggestion in Section 4.2 that these stars are as old as T Tauri stars.  
We speculate that if these Class I stars indeed represent the longest-lived
disk population of Class II stars (Section 4.6), this already slowly
rotating sample could produce the ``slow rotators'' observed at
intermediate pre-main sequence ages \citep[see][]{bouvier97}.

\subsection{The ``Luminosity Problem'' of Class I Stars, Revisited}

The mass infall rate during the main accretion phase can be approximated by
dividing the local Jeans mass by the free-fall timescale
\citep[$\dot{M}_{Infall} \sim M_{J}/\tau_{ff} = 5.4 \,c_s^3/G$;][]{sk04}.
For an isothermal sound 
speed, $c_s$, of 0.2 km/s characteristic of Taurus, this corresponds to
$\dot{M}_{Infall} = 1\times10^{-5}$ M$_\odot$/yr.  The effects of magnetic
support can produce smaller infall rates \citep[$2\times10^{-6}$
M$_\odot$/yr;][]{shu77} while the external compression in turbulent flows
can produce larger infall rates \citep[up to $\sim 10^{-4}$ M$_\odot$/yr,
at least initially;][]{sk04}.  Thus if Class I stars are in the main
accretion phase, they should have mass infall rates consistent with these
predictions.  In support of this, density profiles corresponding to infall 
rates of a few$\times 10^{-6}$ M$_\odot$/yr in envelope-only models
successfully fit the SEDs and scattered light images of many Class I stars
\citep{kenyon93a, kenyon93b, whitney97}.

As first pointed out by \citet{kenyon90}, if the infalling envelope
material is channeled onto the star via steady-state disk accretion at this
rate, the liberated accretion luminosity would be roughly 10 times that
emitted from the photosphere.  With accretion dominated luminosities,
Class I stars would have total luminosities roughly 10 times those of
similar mass T Tauri stars.  Direct tests of this prediction have been
inhibited, however, by the lack of measured stellar and accretion
luminosities of Class I stars.  Nevertheless, previous studies have tried
to assess the relative contribution of the accretion luminosity, under the
assumption of similar stellar luminosities, by comparing the bolometric 
luminosities of Class I stars and Class II/III T Tauri stars.  Bolometric
luminosities are calculated by integrating the entire energy distribution
of a star.  These comparisons show no statistically significant difference
between these Classes.  The discrepancy between the observed bolometric
luminosities and the predicted accretion dominated luminosities is often 
referred to as the ``Luminosity Problem'' for Class I stars
\citep{kenyon90, kenyon94}.

The spectroscopic measurements of 11 Class I stars observed here allow a
more direct measure of the relative stellar and accretion contributions to
the bolometric luminosity.  We first confirm that the stellar, accretion,
and disk emission can account for the observed bolometric luminosity.  The 
total luminosity, $L_{tot}$, is calculated as
\begin{equation}
L_{tot} = L_{star}+L_{acc,shock}+L_{acc,disk}+L_{rep,disk}, 
\end{equation}
where $L_{star}$ is the stellar luminosity calculated in Section 3.2.1
(Table 4), $L_{acc,shock}$ is the accretion luminosity calculated in
Section 3.3.1 ($L_{acc,shock} = G\dot{M}M_{star} [1/R_{star} - 1/R_{in}$]), 
$L_{acc,disk}$ is the viscously generated luminosity of the disk,
and $L_{rep,disk}$ is the reprocessed luminosity from the disk.  The
3rd and 4rth terms are calculated assuming a geometrically thin, optically
thick circumstellar disk with an inner radius ($R_{in}$) of 3 $R_{star}$
and infinite outer radius.  The luminosity viscously generated in the disk 
equals the change in potential energy of material moving from infinity to
the inner disk edge, and thus $L_{acc,disk} = G\dot{M}M_{star} (1/R_{in} -
1/R_\infty$), which simplifies to $L_{acc,disk} = 0.5 \times
L_{acc,shock}$.  Hence, $L_{acc,disk}$ can be determined from the 
inferred $L_{acc,shock}$ values.
The luminosity reprocessed by the disk originates from two 
sources, the central star and the bright accretion shock, and thus
$L_{rep,disk}$ = $0.25 (L_{star}+L_{acc,shock}) R_{star}/R_{in}$
\citep{as86}.  The total luminosity can then be simplified as
\begin{equation}
L_{tot} = 1.08L_{star}+1.58L_{acc,shock}.
\end{equation}

The total luminosities calculated following this prescription agree 
well with the bolometric luminosities for both Class I stars and Class II
stars (median Class I log($L_{tot}$/$L_{bol}$) $= -0.08$, $\sigma = 0.58$; 
median Class II log($L_{tot}$/$L_{bol}$) $= 0.19$, $\sigma = 0.43$).  
The agreement suggests that the dominant luminosity sources are accounted
for; the large scatter suggests that the absolute values
of the luminosity sources are very uncertain.  For the Class I stars, this
agreement suggests there is not significant contribution to the bolometric
luminosity generated from an infalling envelope accreting directly onto the
disk at large radii.

For both Class I and Class II stars, the contribution of disk accretion
($L_{acc,shock} + L_{acc,disk}$) to the bolometric luminosity can range
from a few percent to 50 percent.  However, in the typical (median)
case, only 25\% of the bolometric luminosity is generated through disk
accretion; the majority of the bolometric luminosity originates from the
star \citep[see also][]{muzerolle98}.  We note that the bolometric
luminosities of the 11 Class I stars studied here (median
log($L_{bol}/L_\odot) = -0.36, \sigma = 0.61$) are very similar to the  
bolometric luminosities of the remaining 15 Class I stars in Taurus
(median log($L_{bol}/L_\odot) = -0.17, \sigma = 0.56$).  Thus, if the
unobserved Class I stars have a similar distribution of stellar
luminosities, then they can not have disk accretion rates as high as
predicted by envelope infall models either.  These results strongly favor
one proposed resolution to the luminosity problem - Class I stars do not
have accretion dominated luminosities.  However, this resolution raises 2
additional questions: (1) are known Class I stars properly classified?, (2)
are Class I stars in the main accretion phase?  We address these 2 issues
in the next sections.

\subsection{The Effect of Orientation on Class Classification}

In the traditional classification scheme, Class I stars are true
protostars - stellar embryos surrounded by an infalling envelope.  However,
the observational criteria traditionally used to identify Class I stars
distinctly from Class II stars depend upon the disk/envelope orientation
relative to line of sight.  Radiative transfer models of still-forming
stars predict that edge-on systems will have optical, near- and
mid-infrared characteristics similar to less evolved, more embedded stars
\citep{kenyon93a, kenyon93b, yorke93, sonnhalter95, mh97, whitney03}.
As examples, Class II stars viewed edge-on will, in many cases, resemble
Class I stars, while Class I stars viewed at pole-on orientations will have
optical/infrared properties more characteristic of Class II stars.  The
predictions for pole-on orientations are less clear, however, since the
distribution and radiative properties of material from bipolar flows is not
included in most models.  We note that if Class II
stars with disk inclinations of $90\pm5^\circ$ appear ``edge-on'',
then 8.7\% of all disk systems should appear edge-on, assuming random
orientations (17.4\% if $i = 90\pm10^\circ$).  It is important to recognize
how these orientation effects could have introduced selection biases into
both the optically revealed Class I stars studied here, and the known
population of Class I stars in Taurus.

As discussed in Section 3.1.4, the emission-line profiles offer an indirect
constraint on the star/disk orientation.  The general trend of narrow
emission line profiles at the systemic velocity suggests a bias towards
edge-on orientations for the 11 Class I stars with measured photospheric
features.  Only 3 of these stars show high velocity ($> 60$
km/s) forbidden line emission, compared to more than half of Class II
stars \citep[e.g.][]{heg95}.  The H$\alpha$ 10\%-widths of these Class I
stars are systematically less than those of Class II stars (Section 3.1.4).
Only 1 of the 11 Class I stars (HL Tau) shows evidence of a strong blue
shifted absorption superimposed upon the H$\alpha$ emission profile.  In
comparison, roughly half of Class II stars show this absorption feature
\citep[e.g.][]{ab00}.  These combined results imply little or no high
velocity material along our line of sight for most Class I stars, as
expected for an edge-on orientation.  A orientation bias could also explain
the unusually large forbidden emission-line EWs of most Class I stars; 
the somewhat extended forbidden emission line region may be more directly
observable in this case than the obscured stellar photosphere.  Overall,
the emission line profiles 
of most optically revealed Class I stars favor nearly edge-on orientations.
One clear exception is HL Tau.  Its high velocity forbidden emission lines,
broad H$\alpha$ profile with superimposed absorption favor a less edge-on
orientation.  This orientation may explain why this is one of the most
optically bright environmentally young stars.

The evidence for edge-on orientations does not necessarily imply
that the observed sample of Class I stars are in fact Class II stars.
The presence of spatially extended envelope material, as determined from 
image morphology at infrared and millimeter wavelengths, offers a more
direct constraint on the evolutionary Class.  Based on criteria put forth
by \citet{ma01}, only 58\% (15/26) of the Class I stars listed in Table 1
are true protostars.  The remaining 42\% (11 stars) have envelope masses
$\lesssim 0.1$ M$_\odot$ and are spatially unresolved at 1.3-mm wavelengths 
\citep[referred to as ``unresolved Class I sources'' in][these stars are
  marked as Class I' stars in Table 1]{ma01}. 
\citet{ma01} suggest that these stars are more likely transitional Class
I/II stars or highly reddened Class II stars.  The complementary
near-infrared morphology survey by \citet{pk02} supports the claim that
these 11 stars are not bona fide Class I stars\footnote{Except for the 2
bona fide Class I stars with optically revealed companions (MHO 1/2 and GV 
Tau AB), which could bias $T_{bol}$ estimates towards hotter temperatures,
all bona fide Class I stars with $T_{bol}$ estimates (11/13) have $T_{bol} <
350$ K.  A $T_{bol} < 350$ K may be a better 
distinguishing value for Class I stars.  A comparison of the spectral
indices suggests no similar correction; the spectral indices of bona fide
and non-bona fide Class I stars overlap signficantly.}.  However, we note
that the morphological criteria used in these studies do not account for
the luminosity and mass of the central star.  IRAS 04158+2805 and IRAS
04489+3042 may appear more evolved and point-like because they are lower
luminosity Class I brown dwarfs with smaller disks and envelopes.  

We conclude that as many as one-half of the Class I sample listed in Table
1, and in particular 8 of 11 Class I stars studied here, could be
misclassified Class II stars.  The emission-line profiles and image
morphology suggests that in many cases this misclassification may have been
caused by a nearly edge-on orientation.  We nevertheless continue to refer
to all of these as Class I stars since the orientations and the mass
effects are generally not known.

\subsection{Are Class I Stars in the Main Accretion Phase?}

Although the absolute values of the mass accretion rates remain
considerably uncertain, typical Class I and Class II stars of K7-M1 
spectral type have $\dot{M}_{Acc} \approx 4\times10^{-8}$ M$_\odot$/yr
(Section 
3.3.1).  There is no difference between the disk accretion rates of Class I
stars and Class II stars.  \citet{muzerolle98} found similar results based
on Br$\gamma$ luminosity measurements and assumed stellar properties.
Although the new disk accretion rates we determine in Section 3.3.1 are
somewhat larger than previous estimates \citep[e.g.][]{gullbring98}, they
are still 2 orders of magnitude less than the envelope infall rates
inferred from SED modeling (few$\times 10^{-6}$ M$_\odot$/yr).

%

We explore possible explanations for reconciling this discrepancy.
One possibility suggested by \citet{kenyon90} is that the infalling
envelope material is not transferred to the star via disk accretion in a
steady-state fashion.  The accreting envelope mass accumulates in the
circumstellar disk until it becomes gravitationally unstable
\citep[e.g.][]{larson84} leading to a marked increase in the mass accretion
rate.  This process may be related to the FU Orionis outburst phenomenon,
which is currently understood as rapid ($\sim 10^{-5}$ M$_\odot$/yr), but
short lived ($\sim 100$ yr) accretion through a circumstellar disk onto a
central star \citep{hk87}.  At this rate of accretion, Class I stars would
only need to spend a 5-10\% percent of their lifetime in the high accretion
state to achieve typical T Tauri masses within 1 Myr.  The most luminous
Class I star in Taurus, L1551 IRS 5 \citep[20.9 $L_\odot$;][]{kh95} has
been proposed to be in an FU Ori-like outburst state, though a scenario in
which it is an embedded star of a few solar masses satisfies its observed
properties equally well.

If the envelope material of Class I stars is accumulating in their
circumstellar disks, they are expected to have more massive disks than
Class II stars.  One direct way to test this hypothesis is with 1.3-mm
continuum observation.  Since circumstellar dust is mostly optically thin
at these wavelengths, the emission should trace the total mass in
circumstellar dust.  Figure \ref{mm_age} shows the distribution of 1.3-mm
flux densities versus the evolutionary diagnostics $T_{bol}$
and $\alpha$ for all stars in Taurus with these values, including the
larger sample of Class I stars not spectroscopically studied here.  1.3-mm
measurements of Class II stars are taken from \citet{beckwith90},
\citet{ob95}, and \citet{ma01}, which all have beam sizes of $\sim 11-12$
arcseconds.  The emission in this beam is likely to be a good tracer of 
circumstellar disk mass without significant contamination from any possible 
envelope mass that would dominate on larger spatial scales \citep{ma01}.
Figure \ref{mm_age} also indicates (right hand ordinate) the corresponding
dust mass for a dust opacity per unit mass column density of 0.01
cm$^2$g$^{-1}$ and a dust temperature of 20 K for all
sources\footnote{Although there is some evidence that both the dust opacity
and temperature may change with evolutionary state, the expected changes 
are thought to be less than a factor of $\sim 2$ \citep{henning95}.}.

As Figure \ref{mm_age} clearly demonstrates, the distribution of
circumstellar disk masses for Class I and Class II stars are
indistinguishable.  On average, Class I stars do not have more massive 
disks than Class II stars.  One possible caveat is that if much of the
disk mass of Class I stars is confined to be very close to the star
($\lesssim 1$ AU), as would be expected immediately prior to an 
FU Ori outburst, this mass would likely be optically thick and thus 
un-revealed in our 1.3-mm comparison.  However, if the disks of Class I
stars undergo FU Ori outbursts at semi-regular intervals ($\sim$ 10\% of
the time), the optically thin outer disks would still need to be more
massive than those of Class II stars, on average, to sustain the larger 
time-averaged accretion rates.  This is inconsistent with the observations
and suggests Class I stars do not undergo FU Ori outbursts more often
than Class II stars.


An alternative possibility that we favor for reconciling the mass infall and
disk accretion rates is that most of the Class I stars in Taurus are not in
the main accretion phase.  First, the current populations of Class I and
Class II stars are likely biased because the available low spatial
resolution mid-infrared measurements (e.g. IRAS) used in Class
classification criteria (e.g. $\alpha$, T$_{bol}$) often include several
young stars (Section 2.1).  Some Class I and Class II stars may have been 
misclassified as a consequence.  Additionally, as discussed in Section 4.5,
42\% of stars classified as Class I stars do not appear to be bona fide
protostars (ie. stars with spatially extended envelope structures).  Many
of these are likely to be Class II or transitional Class I/II stars that
have been misclassified because of nearly edge-on disk orientations.  Stars
like IRAS 04016+2610 and IRAS 04302+2247, as examples, have morphologies
and kinematics that are better described by a rotating disk-like structure
\citep{hs00, boogert02, wolf03} than a collapsing envelope model
\citep{kenyon93a, whitney97}.  In the more general case, we suspect that
the envelope infall rates have been over-estimated, even for bona fide
Class I stars, by improperly accounting for the emission from a
circumstellar disk.  For an infalling envelope model, the ratio
$\dot{M}_{Infall}$/$M_\star^{1/2}$ sets the peak wavelength of the SED 
\citep{kenyon93a}; higher infall rates produce redder SEDs.  However, 
the addition of an optically thick circumstellar disk also shifts the peak
of the SED toward redder wavelengths \citep[cf.][]{kenyon93a, wolf03}.
Thus, if most Class I stars have circumstellar disks as expected, then
evelope-only models (ie. no circumstellar disk) will systematically
over-estimate the mass infall rate in order to account for their redder
SED.

We emphasize that even if, as we propose, most Class I stars are not in the
main accretion phase, they nevertheless must still have circumstellar
envelopes.  Disks alone are insufficient to explain the high extinctions,
scattered light images, 1.3-mm morphologies, and mid- to far-infrared SEDs.
Our proposal is simply that there is less mass in the envelope than has
been presumed for many of the Class I stars in Taurus.  Consequently, the
mass infall rates have been over-estimated.  Proper determination of the
envelope mass and infall rate is especially challenging for the low
luminosity Class I stars in Taurus, whose envelope emission can be confused
with diffuse cloud emission \citep[e.g. MHO 1/2, IRAS 04361+2547, IRAS
  04381+2540;][]{ma01}.  Additionally, the large fraction of companion
stars with $\sim 10^3$ AU separations \citep{duchene04} can give the
impression of spatially extended structures.  A convincing case for a
massive infalling envelope can only be established by spatially mapping
molecular line profiles 
and accounting for the effects of outflows and rotations \citep{evans99}.
Of the stars listed in Table 1, only the Class 0 star IRAS 04368+2557
(L1527) retains a massive extended envelope and shows unambiguous evidence
for infall \citep{gregersen97}.

If Class I stars are in fact as old as most T Tauri stars ($\sim 1$ Myr;
Section 4.1), the presence of envelope material in these systems requires a
long envelope dispersal timescale (ie. comparable to T Tauri ages).  A large
dispersion in the \textit{disk} dispersal timescale is thought to explain
the apparently coeval populations of Class II and Class III stars.
Extrapolating this idea to include high-latitude envelope material, many
Class I stars may represent a T Tauri star sub-sample with the longest
envelope/disk dispersal timescale.  The reason for this large dispersion is
unclear, but may involve variations in the initial properties of the cloud
core, dynamical and radiative effects from the presence of close stellar
companions, or the onset of planet formation.

The combined properties of Class I stars in Taurus support a scenario in
which most are not in the main accretion phase.  For these 
systems, the stellar (or substellar) masses must have accumulated prior to
the Class I phase, during the Class 0 phase.  The flat density profiles of
pre-stellar cores \citep{ward94} suggest initial conditions that favor more
rapid accretion during the Class 0 phase \citep{henriksen97}.  The stronger
mass outflows of Class 0 stars relative to Class I stars supports this
interpretation \citep{bontemps96}, if powered by mass accretion as
suspected.  A possible caveat to this suggestion is that if the Class 0
phase is the main phase of accretion, the small number of known Class 0
stars in Taurus suggests a very rapid formation timescale ($\sim 10^4$ yr)
and consequently, a very high mass accretion rate ($\sim 10^{-4}$
M$_\odot$/yr).  These predictions are inconsistent with the relatively low
luminosities of known Class 0 stars (assuming they too are properly
classified); Class 0 stars would have their own luminosity problem.  However,
the IRAS satelite may not have been sensitive enough to detect most Class 0
stars in Taurus \citep[e.g. IRAM 04191+1522;][]{andre99}.  Indeed, early
\textit{Spitzer Space Telescope} results show that some ``starless cores''
may not be not be starless \citep{young04}.  The 44 starless cores in
Taurus \citep{onishi02} may contain a yet hidden population of Class 0
stars.  Thus, concerns about the implied statistical ages and infall rates
of Class 0 stars in Taurus may be premature.

\subsection{Herbig-Haro Energy Sources}

Many of the Class I and Class II stars studied here have been identified as
the energy sources of HH objects (Section 2.1).  Some studies suggest HH
stars have higher mass accretion rates and larger circumstellar
reservoirs than the average T Tauri star \citep{reipurth93, chini97}.  It
has also been suggested that HH stars are more likely to be in binary
systems \citep{reipurth00}.  The stellar and circumstellar properties
inferred here may be used to identify possible differences between stars
which power HH flows and those which do not.

The radial velocities of the HH stars newly measured here are consistent
with the mean of Taurus, with the possible exception of HV Tau C (Section
3.1.1), which suggests that they are not more likely to be close
(spectroscopic) binaries than non-HH stars.  The locations of HH stars and
non-HH stars on an H-R diagram are similar (Figure \ref{sdf00}); HH stars
do not appear to be systematically younger.  We do find, however, a
decreasing frequency of HH objects associated with the lowest mass
stars and brown dwarfs (Sections 3.1.2 and 3.2.2).  The frequency may be
even less, as none of the recently identified accreting T Tauri brown
dwarfs \citep[e.g.][]{muzerolle03a} are known to power HH flows.  Although
the strong forbidden line emission confirms that many low mass stars/BDs
indeed power jets, the decreased frequency of spatially resolved regions of 
shocked emission suggests that their jets may be less powerful.  This
result is consistent with the mass dependent mass outflow rates and
mechanical luminosities determined in Section 3.3.2.  If the velocity
imparted to the outflow also decreases with stellar mass (we have assumed a
constant value in our calculations), this would lead to an even stronger
mass dependence on the mechanical energy of the outflow.  Although the
$v$sin$i$ values of HH stars are somewhat larger than those of non-HH
stars (Section 3.1.1), this difference disappears if the values are
normalized by their break-up velocity (Section 4.1).  HH stars and non-HH
stars have similar distirbutions of angular momentum.

In comparison to non-HH stars, HH stars have slightly higher mass
accretion rates ($\times$2.5 in the median), and much larger mass outflow
rates ($\times$20 in the median).  It is unclear if the apparent difference
in the mass outflow rate is a real time-averaged difference, or a
consequence of recent accretion/outflow history.  The similar line profiles
of most HH stars and non-HH stars (Class I stars being the exception)
suggests that edge-on orientation effects likely play less of a role in
increasing the apparent forbidden line EWs (Section 3.1.4), which would
translate into larger mass outflow rates.

The distributions of millimeter fluxes, a direct tracer of circumstellar
mass, for HH stars and non-HH stars are indistinguishable 
(Figure \ref{mm_age}).  This result is in contrast with the study of
\citet{reipurth93} who found that HH energy sources generally have, on
average, more than an order of magnitude 1.3 millimeter-wave flux densities
than T Tauri stars in Taurus do.  However, the \citet{reipurth93}
comparison is biased in that the majority of HH stars in that study are
much more luminous ($1 - 10^3$ L$_\odot$) than the Taurus T Tauri stars
compared to ($\sim 1$ L$_\odot$); as noted by \citet{reipurth93}, the 1.3
millimeter flux shows a strong dependence on the bolometric luminosity.
Our results suggest that when HH stars and non-HH stars of similar
luminosities are compared, their millimeter fluxes are likewise similar.
The lower luminosities of Taurus HH stars may stem from current
identification criteria \citep[spatially resolved optical
  jet;][]{reipurth99}.  Less energetic jets and lower luminosity sources
are easier to identify in Taurus given it's close proximity and low
extinction.  It is unclear if the more luminous HH stars in other star
forming regions are at an earlier evolutionary stage, or simply more
massive.

Finally, we note that since HH objects are more often associated with Class
I stars than Class II stars, the visual extinction of HH stars is
systematically larger than for non-HH stars.  Overall, however, with the
exception of apparently larger mass outflow rates, the stellar and
circumstellar properties of HH stars and non-HH stars of similar mass are
similar.

\section{Summary}

Using the W. M. Keck I telescope, we obtained high dispersion (R $\sim$
34,000) optical (6330 - 8750 \AA) spectra of 15 Class I stars and 21 Class II
stars in the nearby Taurus star forming region.  Targets were selected
based on evidence for either an infrared dominated luminosity as quantified
by the SED diagnostics $T_{bol}$ and $\alpha$, or the presence of a
spatially resolved optical jet (Section 2.1).  The optical emission from
these environmentally young stars is spatially extended scattered light in
some cases, and faint point-like in others.  For 28 of these 36 stars, our 
measurements are the first high dispersion optical spectra ever obtained.
Photospheric features are detected in 11 of the Class I stars (42\% of
known Taurus Class I stars) and in all 21 of the Class II stars; strong
emission lines (e.g. H$\alpha$) are detected in the spectra of all stars
observed, even Class I stars not visible on the POSS-II red plates.
Complementary $I_c$-band images were obtained for the majority of Class I
stars observed spectroscopically.

Radial and rotational velocities are determined via cross-correlation with
rotationally broadened spectral standards.  All stars have radial
velocities that are consistent (within 3$\sigma$) with the mean
of Taurus.  No spectroscopic binaries are identified, though HV Tau C and
HK Tau B are noted as possible single-lined spectroscopic binaries.  All
stars are slowly rotating ($v$sin$i < 35$ km/s).  Spectral types and
continuum excesses at 6500 \AA\, and 8400 \AA\, are determined from the best
fit rotationally broadened and veiled dwarf spectra.  The inferred spectral
types range from G8 to M6, with typical uncertainties of 1 spectral
subclass.  The new spectral types, in combination with $J,H,K_s$ photometry
from the 2MASS database, are used to estimate visual extinctions and
stellar luminosities.  Masses and ages are determined from comparison with
the \citet{siess00} evolutionary models.  Surface gravity signatures in all
spectra appear either dwarf-like or intermediate between those of dwarfs
and giants, consistent with the modestly less than dwarf surface gravities
expected from their H-R diagram positions.  Emission features associated
with pressure sensitive lines inhibits direct determination of surface
gravities from line profile analyses.

The measured continuum excesses range from 0.0 to 5.7 times that of the
photosphere, with one source (IRAS 04303+2240) showing variations by a
factor of $\sim 3$ between 2 observational epochs.  For all stars, the
continuum excesses appear to exhibit the trend of retaining constant flux
over the 6500 - 8400 \AA\, wavelength interval.  Although this has been
observed previously \citep{bb90}, current hot spot models
\citep[e.g.][]{kenyon94, cg98} can not account for it.  We interpret 
this emission as a cooler component of the shock, although emission from
the inner disk can not be completely excluded.  Mass accretion
rates are determined from the excess emission at 6500 \AA\, under the
assumption of a magnetically channeled accretion flow.  Mass accretion
rates for K7-M1 spectral types span $\sim 2$ orders of magnitude, with a
median value of $4 \times 10^{-8}$ M$_\odot$/yr.  This median value is
larger than previous estimates \citep[e.g.][]{gullbring98} determined from 
excess emission at shorter wavelengths $\lesssim$ 0.5 $\mu$m and based
on a model that does not account for the observed excesses at red optical
wavelengths.  Until the continuum excess spectrum can be more accurately
modeled, this will contribute a factor of 3 systematic uncertainty in mass
accretion rates.

H$\alpha$ emission is detected from all sources, and in most cases a wealth
of permitted (Fe II, Ca II) and forbidden ([OI], [NII], [SII]) emission
lines are seen.  Based on the relatively distinct emission line profiles of
3 optically veiled (and presumably accreting) edge-on disk systems (HH 30,
HV Tau, HK Tau B), we suggest that CoKu Tau 1, IRAS 04260+2642, and ZZ Tau
IRS also have nearly edge-on orientations.  We confirm previous results
that find larger forbidden-line emission associated with Class I stars than
Class II stars.  However, we attribute this to an orientation bias that
allows a more direct view of the forbidden emission line region than the
stellar photosphere, and not to larger mass outflow rates.  Mass outflow
rates are determined from the strength of [SII] 6731 \AA\, emission, under
the assumption of a bipolar jet.  Excluding edge-on disk systems, which may
bias the measured forbidden-line EWs to larger values, the mass outflow
rates for K7-M1 spectral types span $\sim 3$ orders of magnitude, with a
median value of $2 \times 10^{-9}$ M$_\odot$/yr.  The ratios of
$\dot{M}_{outflow}/\dot{M}_{inflow}$ span $\sim 2$ orders of magnitude,
with a median value of 0.05.

The inferred stellar and circumstellar properties are used to conduct
statistical comparisons of Class I and Class II stars, as well as HH stars 
and non-HH stars.  A summary of the median statistics is compiled in
Table 5, with brown dwarfs excluded.  The distribution of stellar masses of
Class I stars is similar to that of Class II stars, ranging from 
substellar to several solar masses.  Of particular interest are IRAS
04158+2805, IRAS 04248+2612, and IRAS 04489+3042 which have substellar
masses. These are the first spectroscopically confirmed Class I brown
dwarfs.  Brown dwarfs as low in mass as 0.05 M$_\odot$ likely form via
dynamical collapse of a cloud core, experience an embedded Class I phase
of evolution and are capable of powering molecular and HH flows.
Stellar luminosities of optically revealed Class I stars suggest ages of
$\sim 10^6$ yr, consistent with the ages of Class II and Class III T Tauri
stars but 
inconsistent with ages implied by stellar birthline predictions and 
relative number statistics ($\sim 10^5$ yr).  In light of the wide range of
disk dispersal time scales among T Tauri stars, we speculate that many
Class I stars represent a T Tauri subsample with the longest envelope/disk 
dispersal time scale.

For both Class I and Class II stars, approximately 25\% of the bolometric
luminosity is generated through disk accretion.  The majority of the
bolometric luminosity originates from the star.  This result strongly
supports one proposed resolution of the ``luminosity problem'' for Class I
stars - they do not have accretion dominated luminosities.  
We further propose that most Class I stars in Taurus are past their main
accretion phase.  In some cases, Class I stars may actually be Class II
stars that have been misclassified because of nearly edge-on orientations,
or because of a biased SED caused by spatially unresolved 25 $\mu$m
measurements.  In the more general case, however, we suggest that the
envelope infall rates, which are roughly 2 orders of magnitude larger than
the observed disk accretion rates, have been over-estimated by not properly
accounting for the emission from a circumstellar disk.

While the proposed scenario for Class I stars would explain the similar
stellar properties (masses, ages, rotation rates) and circumstellar
properties (mass accretion rates, mass outflow rates) of Class I and Class
II stars, there are still several issues that challenge this
interpretation.  The positions of Class I stars are correlated with 
the positions of dense cores \citep{hartmann02} suggesting they are still
close to their birth site.  It is somewhat surprising that they are not
more dispersed, like Class II stars, if Class I stars are indeed as old as
Class II stars.  The statistical ages of the few known Class 0 stars
suggest a very rapid formation timescale, and large mass accretion rates;
many Class 0 stars will have an even more severe luminosity problem
(e.g. IRAS 04368+2557).  Moreover, their rarity implies that there is very
little ongoing star formation in Taurus.  We emphasize, however, that the
limited sensitivity of current surveys make it difficult to assess the
validity of these so-called statistical problems.  Observations with the
\textit{Spitzer Space Telescope} will likely yield a more comprehensive 
picture of star formation in the Taurus molecular cloud.

Many of the Class I and Class II stars studied here have been identified as 
the energy sources of HH objects.  The primary difference between HH stars
and non-HH stars is that HH stars have stronger forbidden line emission,
which translates into larger mass outflow rates by a factor of $\sim 20$ in
the median.  Whether this is a real time-averaged difference or a
consequence of recent accretion/outflow history is unclear.  We also find 
that HH flows are less commonly associated with very low mass stars
($\lesssim 0.2$ M$_\odot$) and brown dwarfs than with more massive stars.
One remarkable exception is the Class I brown dwarf IRAS 04248+2612 which 
powers both an HH and a molecular flow.  Overall, with the exception of
larger mass outflow rates, the stellar and circumstellar properties of HH
stars and non-HH stars of similar mass are generally indistinguishable.

On a concluding and reflective note, as we anticipate the exciting young
star discoveries likely to be provided by the sensitive \textit{Spitzer
Space Telescope}, it is perhaps fitting that we are now determining 
fundamental stellar and circumstellar properties of environmentally young
stars identified 20 years ago by \textit{Spitzer's} predecessor, the IRAS
satellite.  The extensive surveys of star forming regions being conducted
with \textit{Spitzer} may revolutionize our generally accepted ideas
regarding star formation in the same way that the enlightening discoveries
of the IRAS satellite did.  However, proper interpretation of newly
discovered red, faint, and potentially very young objects will likely
depend critically on understanding the relation between the stellar,
accretion, and outflow properties, often best studied short-ward of $\sim 1
\mu$m, and the disk+envelope properties \textit{Spitzer} will study at
longer wavelengths.  Observations of stars in the Taurus star forming
regions, with its close proximity and the low cloud extinction, will likely
continue to play an important role in this endeavor.

\acknowledgements

We are grateful to G. Doppmann, S. Edwards, J. Eisner, P. Hartigan,
L. Hartmann, M. Liu and P. Williams for helpful discussions. This
publication makes use of data products from the Two Micron All Sky Survey,
which is a joint 
project of the University of Massachusetts and the Infrared Processing and
Analysis Center/California Institute of Technology, funded by the National 
Aeronautics and Space Administration and the National Science Foundation. 
Finally, we recognize and acknowledge the very significant cultural role
and reverence that the summit of Mauna Kea has always had within the
indigenous Hawaiian community.  We are most fortunate to have the
opportunity to conduct observations from this mountain.

\clearpage

\begin{figure}
\epsscale{0.9}
\plotone{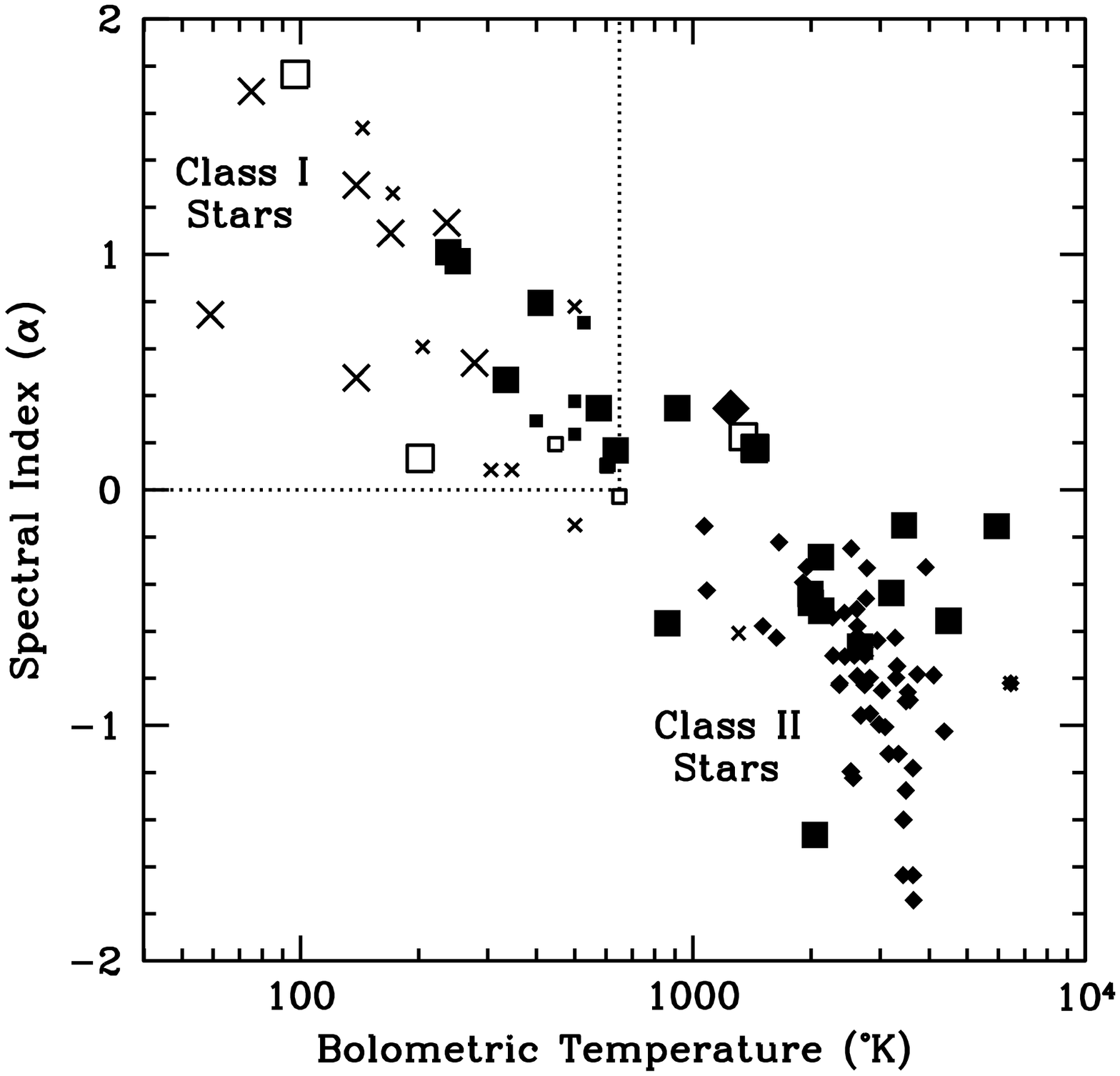}
\caption{Spectral index versus bolometric temperature for stars in Taurus.
  A total of 44 stars have either T$_{bol} < 650^\circ$ K or $\alpha > 0.0$
  or power an HH flow.  We consider these to be environmentally young stars
  (Table 1).  The subset of Class I stars have both T$_{bol} < 650^\circ$ K
  and $\alpha > 0.0$ (when both values are available; see text).
  \textit{Large symbols} indicate
  stars that power HH flows while \textit{small symbols} represent stars
  that do not.  \textit{Diamonds} indicate stars that have been observed
  spectroscopically previously \citep{bb90, heg95, wb03} and
  \textit{squares} indicate stars observed 
  in this study, 15 of which are within the Class I regime.  Of the 36
  stars observed here, 6 have no measurable continuum (\textit{unfilled 
  squares}; only 5 are visibly plotted since MHO-1 and MHO-2 overlap) while
  30 (3 of which are binary star components with overlapping positions)
  show a stellar continuum (\textit{filled squares}) which we use to study 
  the stellar and accretion properties.  The X's indicate stars that have
  never been observed at high dispersion.  
\label{index_tbol}}
\end{figure}
\clearpage

\begin{figure}
\epsscale{0.8}
\plotone{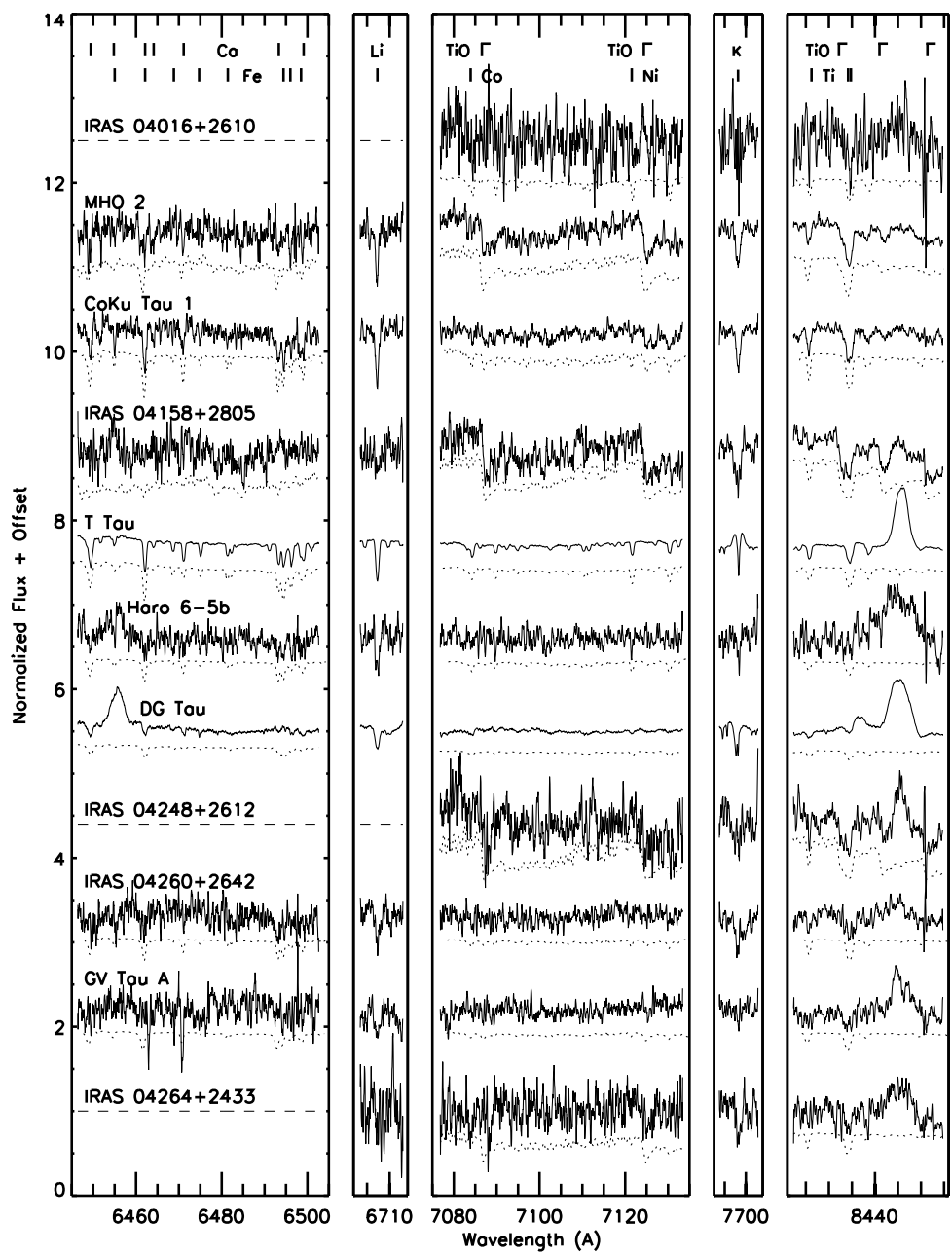}
\caption{Spectral segments obtained with Keck/HIRES of environmentally
  young stars.  Only stars with S/N $>$ 2 are plotted; portions of spectra
  with S/N $<$ 2 are shown as dashed lines.  The 6445-6500 \AA, 7060-7130
  \AA, 8420-8455 \AA\, regions are the primary ones used to determine the
  spectral types and veiling; the best fit dwarf spectral standards,
  rotationally broadened and veiled, are shown for each star.  The youth
  diagnostic Li I 6708 \AA\, and the gravity sensitive feature KI 7699
  \AA\, are also shown; these are typically the strongest and most easily
  identifiable lines in low S/N optically veiled spectra.  Several
  optically veiled, accreting stars show Fe II 6456 \AA\, and OI 8446 \AA\,
  emission.
\label{spec1}}
\end{figure} 
\clearpage

\begin{figure}
\epsscale{0.8}
\plotone{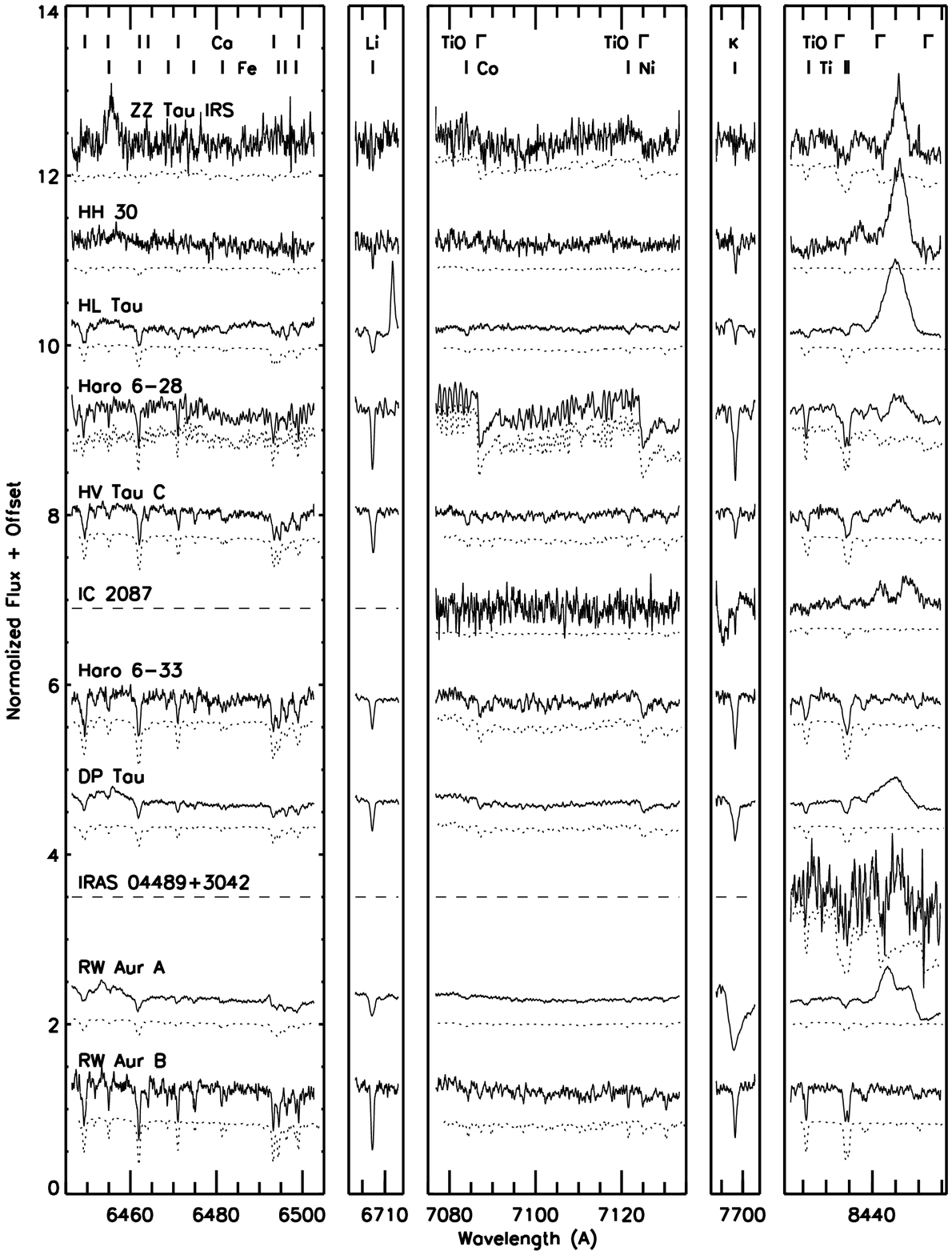}
\caption{Same as Figure \ref{spec1}.
  \label{spec2}}
\end{figure} 
\clearpage

\begin{figure}
\epsscale{0.8}
\plotone{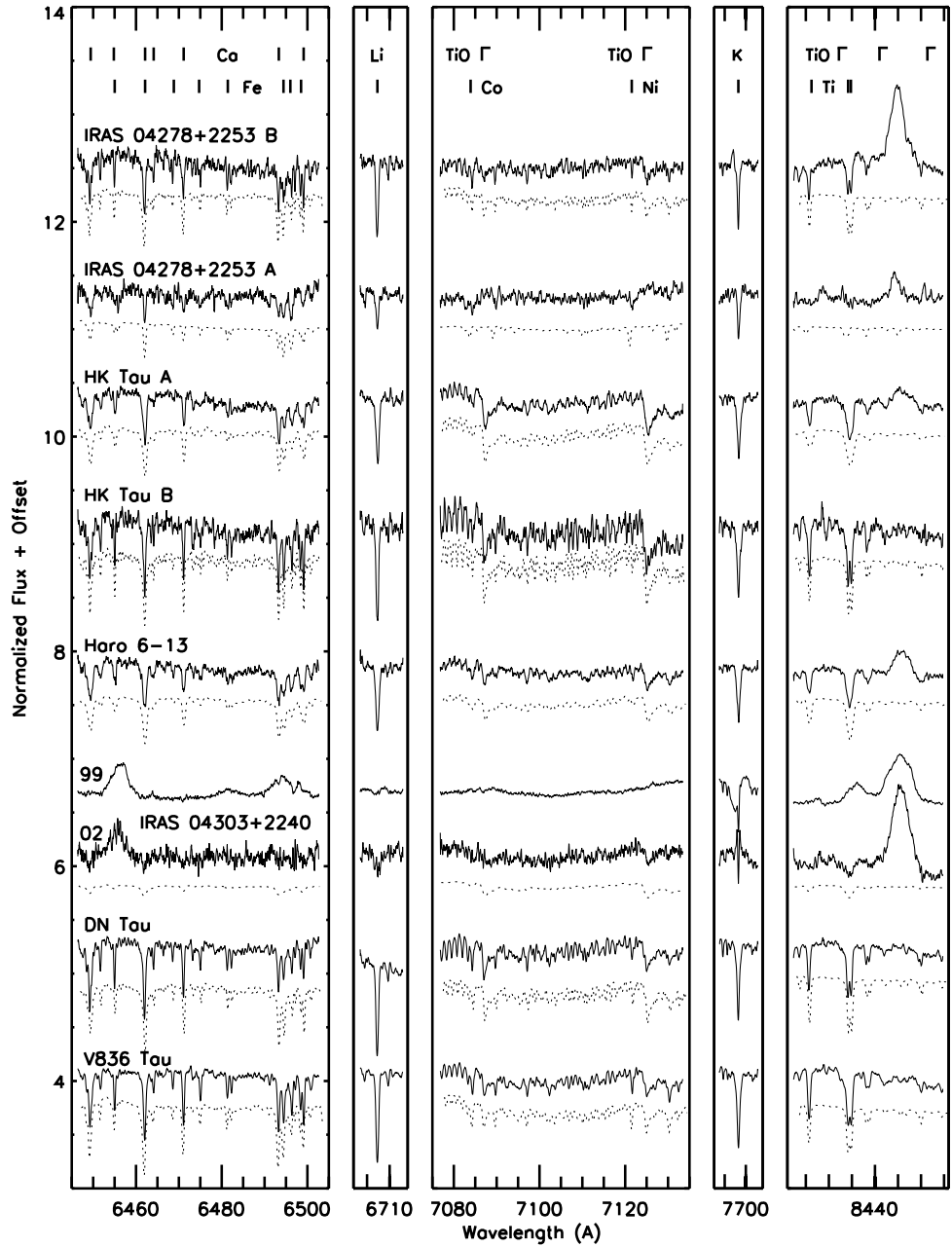}
\caption{Spectral segments obtained with Keck/HIRES of additional T Tauri
  stars not meeting the environmentally young criteria.  Features are as in 
  Figure \ref{spec1}.  Spectra from 2 epochs are shown for IRAS 04303+2240
  (see Table 2).  This star was both brighter and more optically
  veiled during the first epoch.
  \label{spec3}}
\end{figure} 
\clearpage

\begin{figure}
\epsscale{0.8}
\plotone{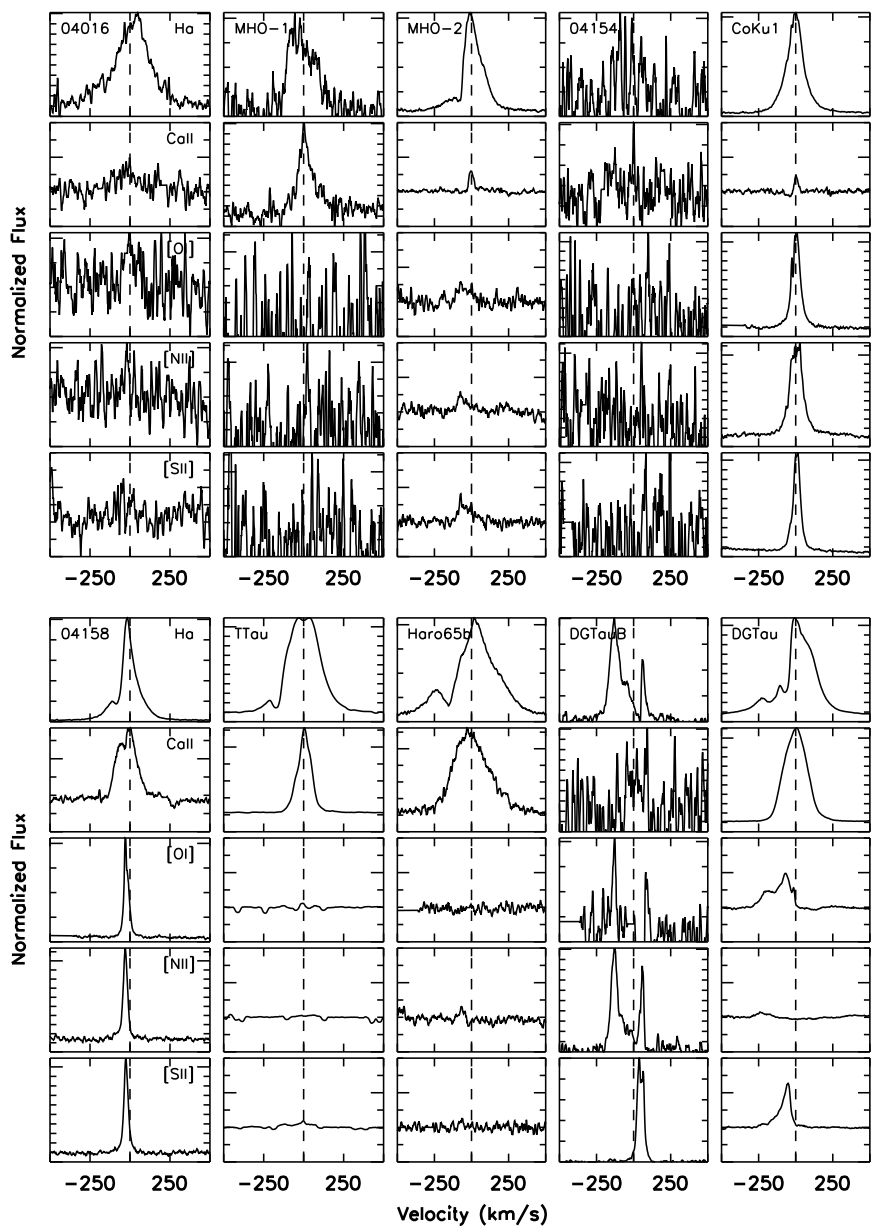}
\caption{H$\alpha$, Ca II 8498\AA, [OI] 6364 \AA, [NII] 6583 \AA, [SII]
  6731 \AA\, emission lines of the observed sample (Table 2).  All spectra
  are shifted in wavelength to the mean systemic velocity of Taurus (18
  km/s; vertical \textit{dashed line}) and normalized to unity in the
  continuum.  The vertical scale runs from 0.0 to 3.0, except when the
  peak to continuum ratio exceeds 3.0, in which case the scale runs from
  0.0 to the peak value.  \label{emission1}}
\end{figure} 
\clearpage

\begin{figure}
\epsscale{0.8}
\plotone{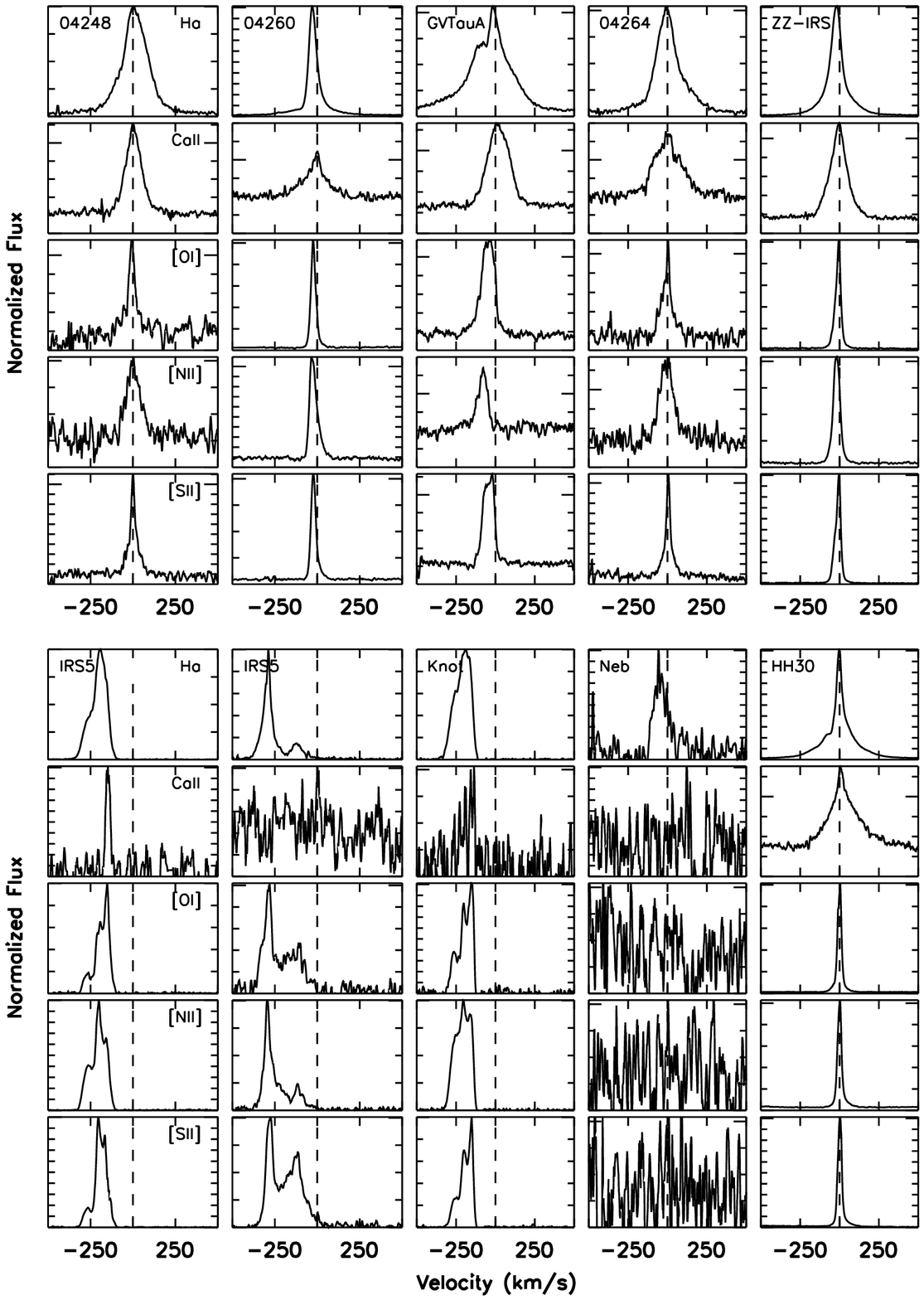}
\caption{Same as Figure \ref{emission1}.
  \label{emission2}}
\end{figure} 
\clearpage

\begin{figure}
\epsscale{0.8}
\plotone{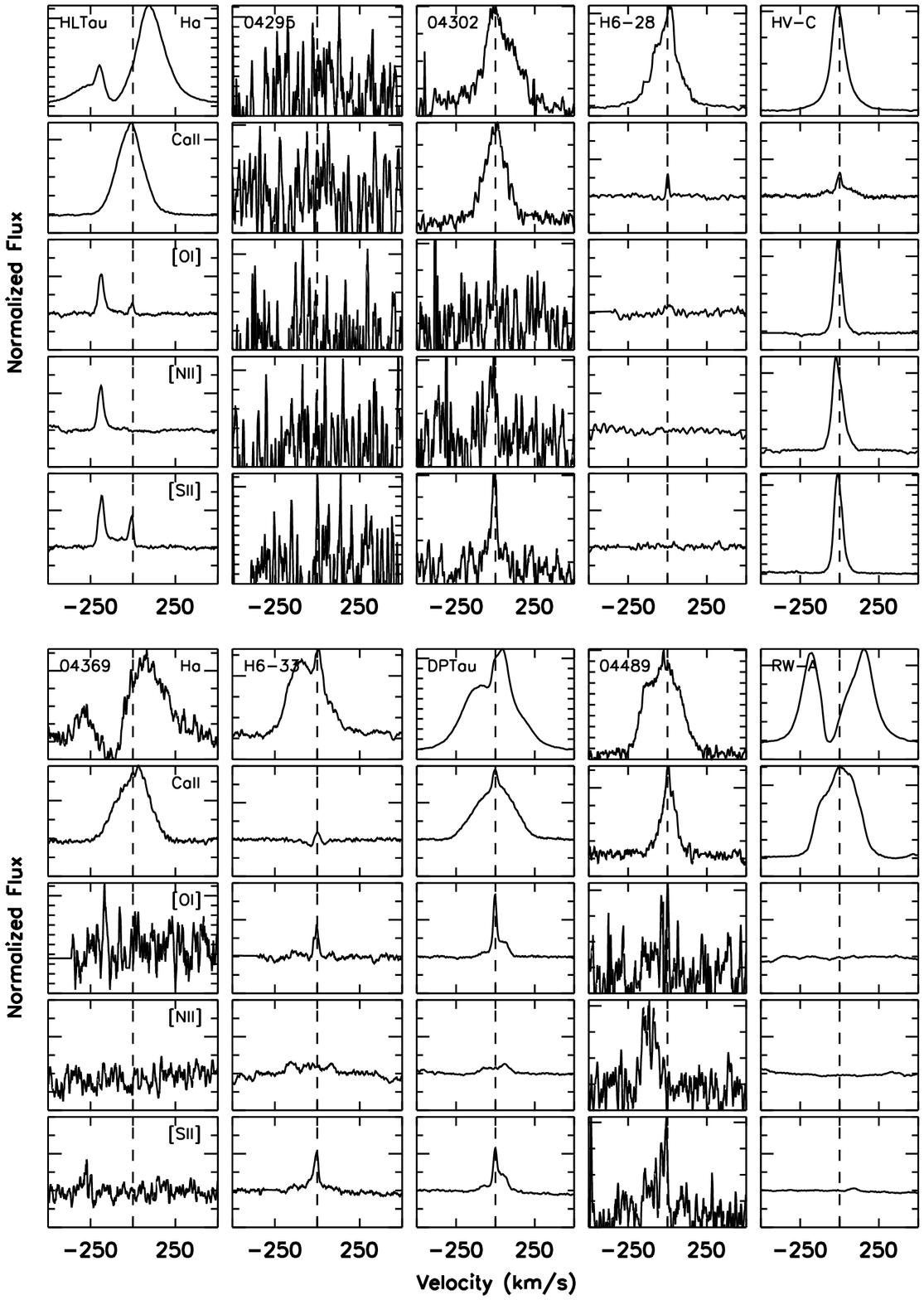}
\caption{Same as Figure \ref{emission1}.
  \label{emission3}}
\end{figure} 
\clearpage

\begin{figure}
\epsscale{0.8}
\plotone{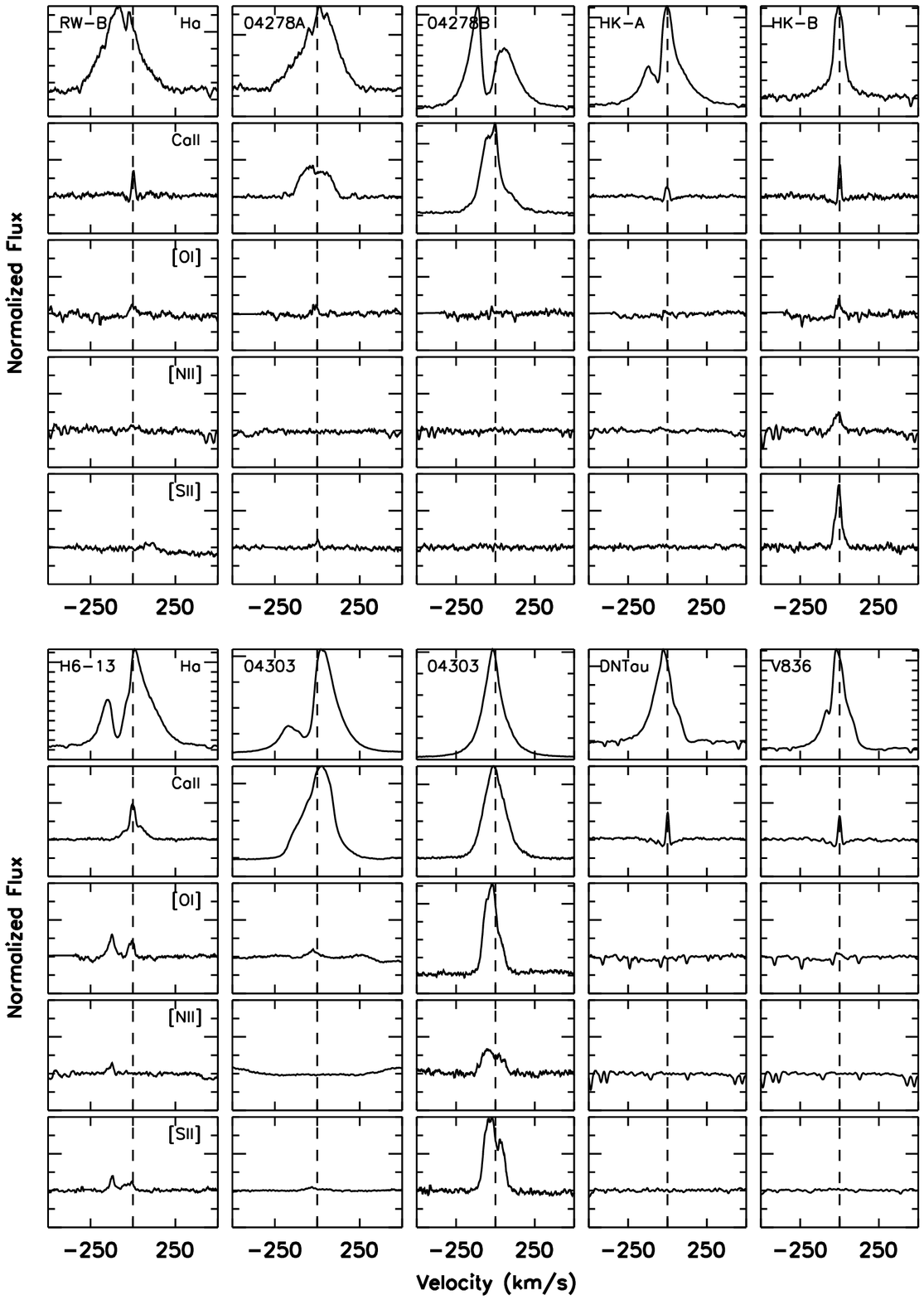}
\caption{Same as Figure \ref{emission1}.
  \label{emission4}}
\end{figure} 
\clearpage

\begin{figure}
\epsscale{0.8}
\plotone{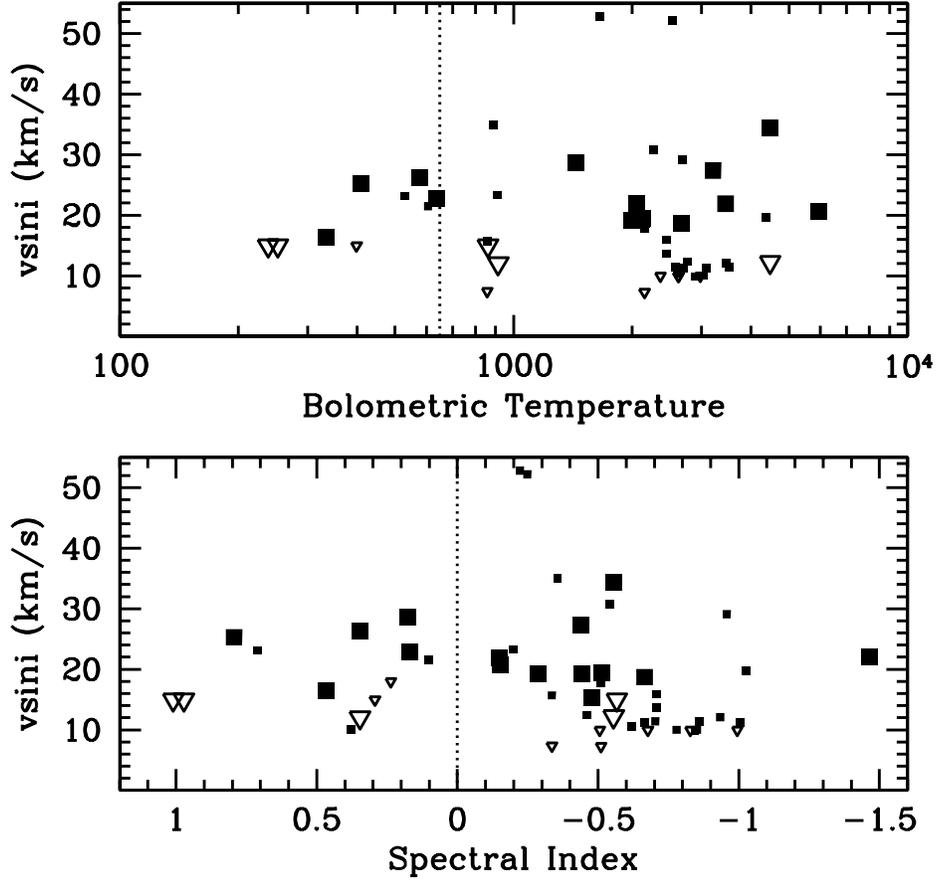}
\caption{Projected rotational velocities ($v$sin$i$) versus bolometric
  temperature (\textit{top panel}) and spectral index (\textit{bottom
  panel}).  The vertical \textit{dotted lines} are proposed values for
  separating Class I and Class II stars (see Section 2.1).  \textit{Large
  symbols} indicate stars that power HH flows (ie. HH stars).  \textit{Open 
  symbols} are $v$sin$i$ upper limits.  The stellar rotational velocities
  do not evolve significantly with either evolutionary diagnostic.  HH
  stars rotate slightly faster than non-HH stars, on average.
  \label{rot_age}}
\end{figure} 
\clearpage

\begin{figure}
\epsscale{1.0}
\plotone{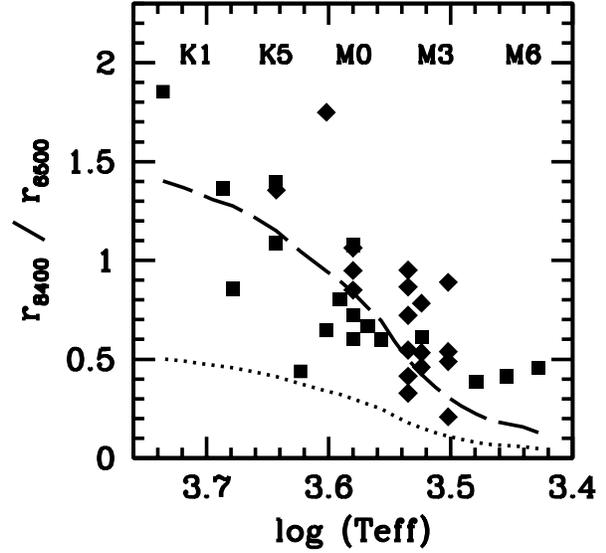}
\caption{Ratio of optical veiling at 8400 \AA\, to that at 6500 \AA\,
  versus stellar temperature.  \textit{Squares} are ratios determined
  from values listed in Table \ref{tab_obs} and from \citet{wb03};
  \textit{Diamonds} are determined from the values in
  \citet[][$r_{6110}$ to $r_{8115}$ for these points]{hk03}.  The
  observed veiling ratios are more consistent with a continuum excess of
  constant flux ($F_{ex} = C$; \textit{dashed line}) over this wavelength
  interval, than with a Rayleigh-Jeans approximation ($F_{ex}
  \propto \lambda^{-4}$; \textit{dotted line}) as would be expected if the
  excess originated from a spot much hotter than the stellar photosphere.
  We interpret this as evidence for a second, cooler component to the
  accretion generated excess emission.  \label{veil_ratio}}
\end{figure}
\clearpage

\begin{figure}
\epsscale{0.8}
\plotone{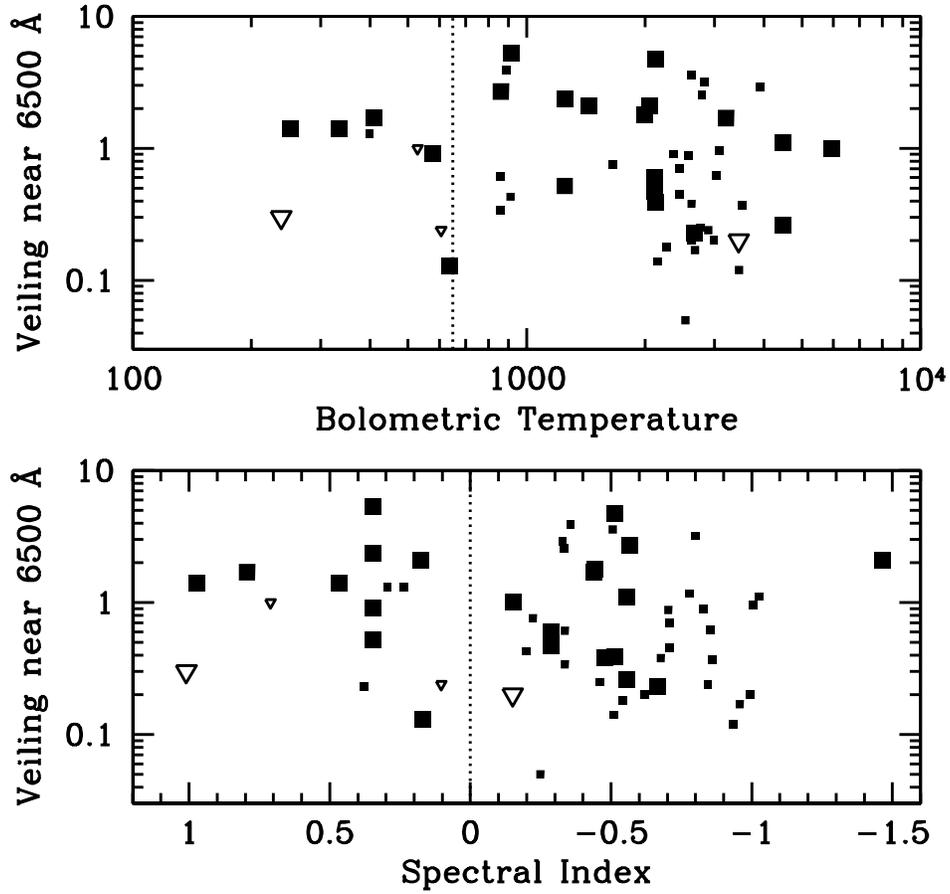}
\caption{Optical veiling near 6500 \AA\, ($r_{6500}$ in Table 2) versus
  bolometric temperature (\textit{top panel}) and spectral index
  (\textit{bottom panel}).  The symbols are the same as in
  Figure \ref{rot_age}.  The amount of excess emission does not evolve
  significantly with either evolutionary diagnostic.  HH stars and non-HH
  stars have similar excesses.
\label{veil_age}}
\end{figure} 
\clearpage

\begin{figure}
\epsscale{0.8}
\plotone{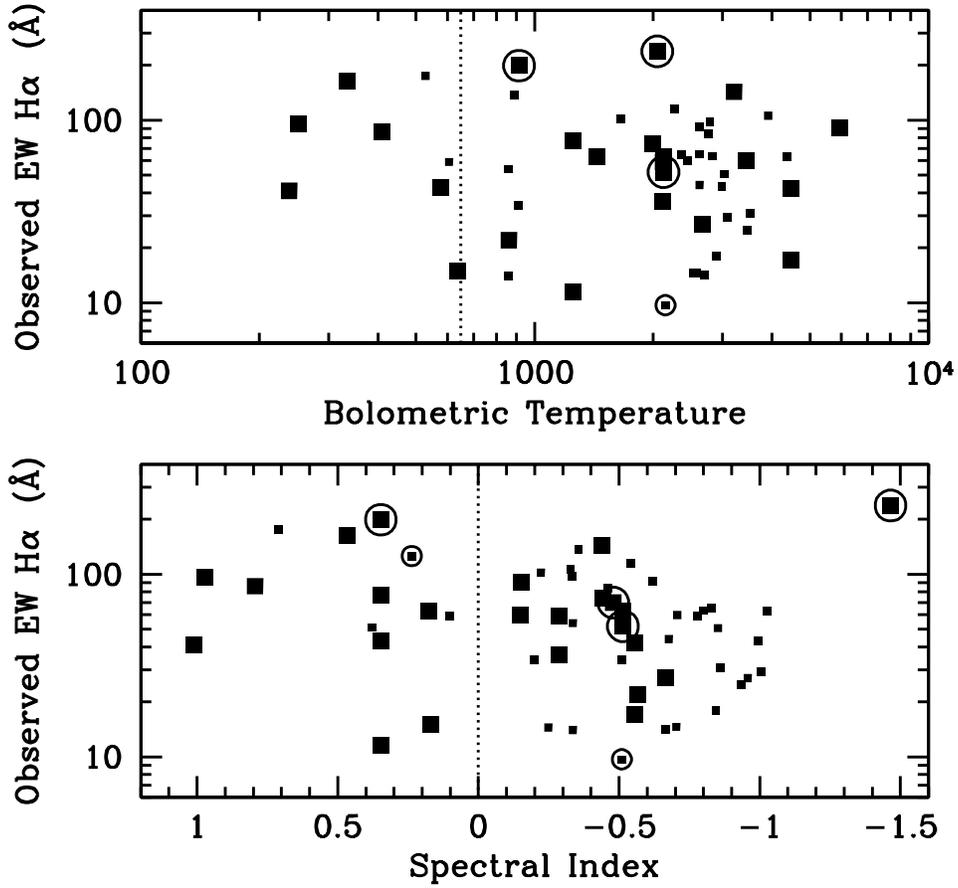}
\caption{Observed EW[H$\alpha$] versus bolometric temperature
  (\textit{top panel}) and spectral index (\textit{bottom panel}).  The
  symbols are the same as in Figure \ref{rot_age}; \textit{circled symbols}
  indicate stars with known or likely edge-on disk systems (see Section
  3.1.4).  The EW[H$\alpha$] does not evolve significantly with either
  evolutionary diagnostic.  HH stars and non-HH stars have similar
  EW[H$\alpha$] values.  \label{ha_age}}
\end{figure} 
\clearpage

\begin{figure}
\epsscale{0.8}
\plotone{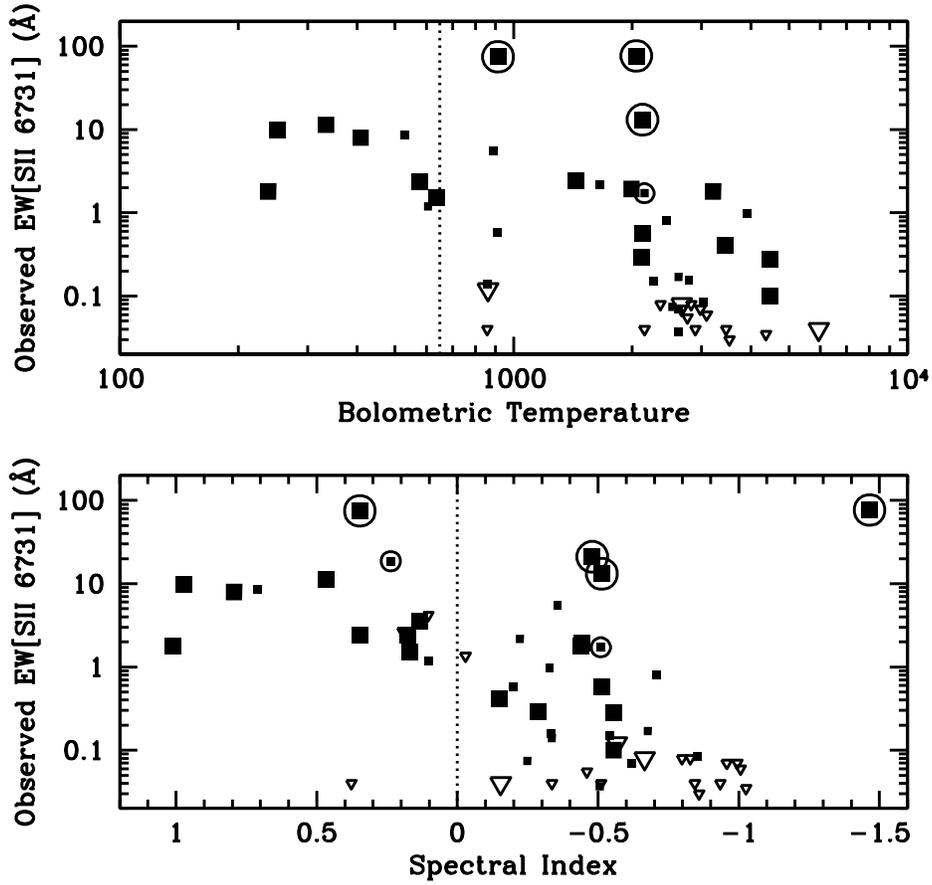}
\caption{Observed EW [SII 6731] emission versus bolometric temperature
  (\textit{top panel}) and spectral index (\textit{bottom panel}).  The
  dashed lines and symbols are the same as in Figure \ref{rot_age}; 
  \textit{circled symbols} indicate stars with known or likely edge-on disk
  systems.  The average EW[SII] decreases toward hotter
  bolometric temperatures and negative spectral indices.  HH stars have
  systematically larger EW[SII]s than non-HH stars.  Stars with edge-on
  orientation have EW[SII]s biased towards artificially large values
  (Section 3.1.4). \label{sii_age}}
\end{figure} 
\clearpage

\begin{figure}
\epsscale{0.8}
\plotone{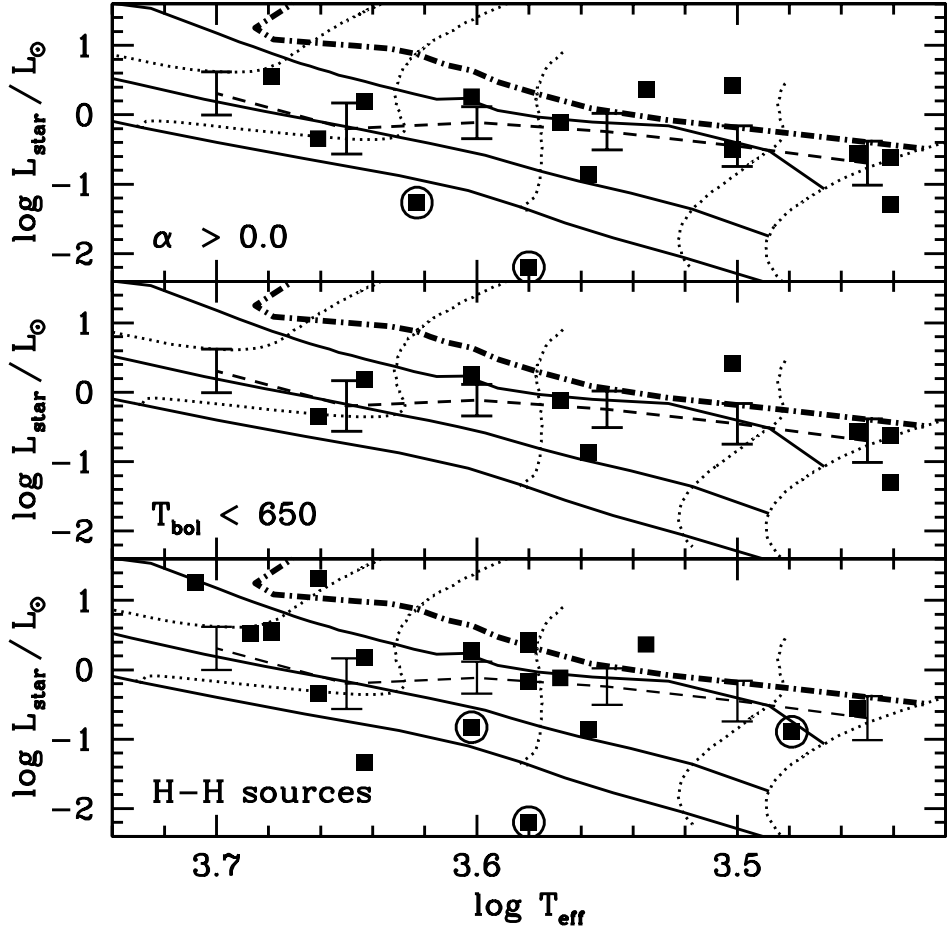}
\caption{H-R diagrams for the young stars in Taurus selected according to 3
    evolutionary diagnostics: spectral index $> 0.0$, T$_{Bol} \le 650$,
    and stars that power HH flows.  Values are from Table 3; 
    \textit{circled squares} indicate stars with known or likely edge-on
    disk systems.  Typical uncertainties are 0.4 in LogL and 0.02 in logT.
    The \textit{dashed 
    line} in each plot indicates the mean stellar luminosities of optically
    visible Taurus T Tauri stars and error bars indicate the dispersion
    in these luminosities.  The evolutionary models of Siess et al. (2000)
    are shown with isochrones at $10^6$, $10^7$, $10^8$ years
    (\textit{solid lines}), and mass tracks at 0.2, 0.5, 1.0, 2.0, 4.0
    M$_\odot$ (\textit{dotted lines}).  The \textit{thick dot-dashed line}
    indicates the ``stellar birthline'' for the case of $\dot{M}_{Acc} =
    10^{-5}$ M$_\odot$/yr \citep{fs94}.  The environmentally young stars
    appear to have ages of $\sim 10^6$ years, similar to the mean ages of
    environmentally older Taurus T Tauri stars.
    \label{sdf00}}
\end{figure} 
\clearpage

\begin{figure}
\epsscale{0.8}
\plotone{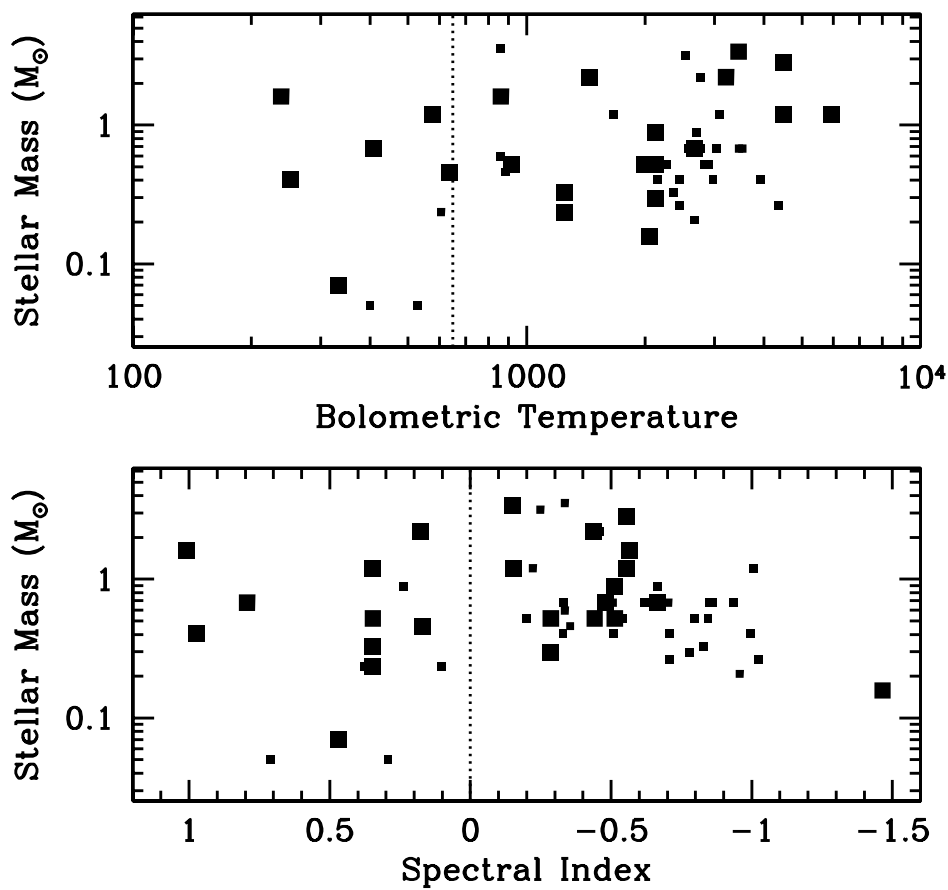}
\caption{Stellar mass versus bolometric temperature (\textit{top panel})
  and spectral index (\textit{bottom panel}).  The symbols are the same as
  in Figure \ref{rot_age}.  Class I stars have masses similar to Class II
  stars, ranging from substellar masses to several solar masses.  Note that
  although many Class II brown dwarfs are known, they are not plotted here
  since they do not have bolometric temperature or spectral index values
  because of insufficient infrared and sub-millimeter measurements.
  \label{mass_age}}
\end{figure} 
\clearpage

\begin{figure}
\epsscale{0.8}
\plotone{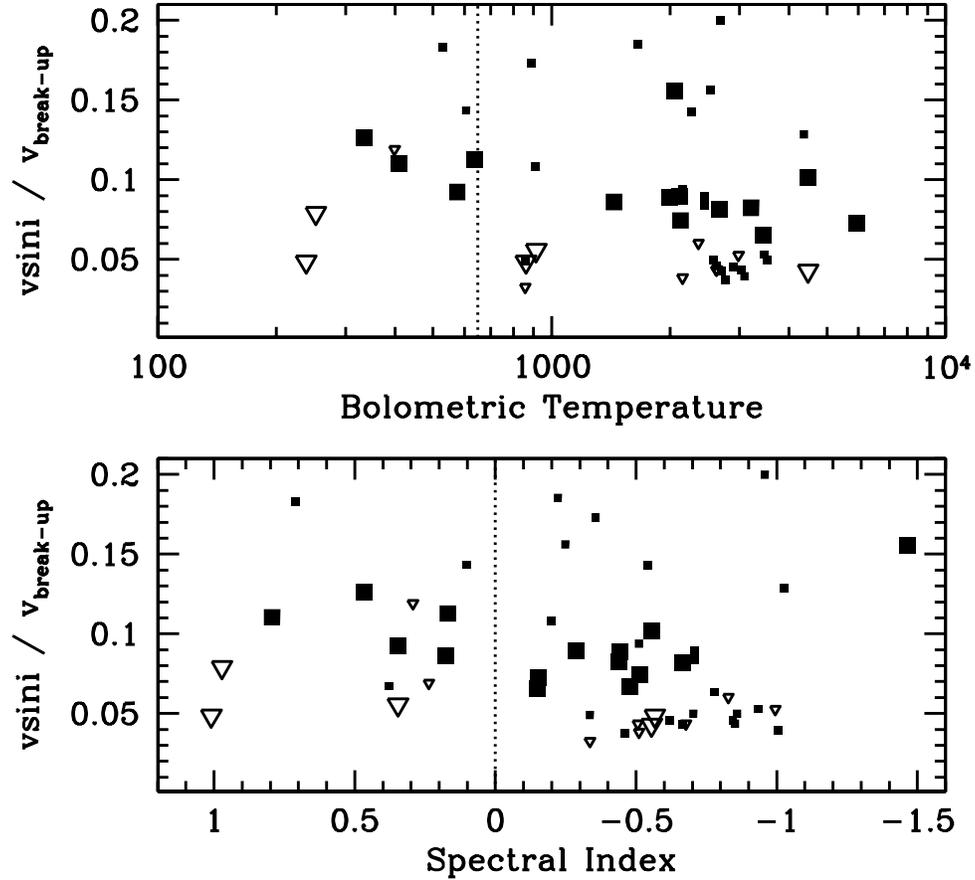}
\caption{Projected rotational velocities ($v$sin$i$) normalized by the
  break-up velocity versus bolometric temperature (\textit{top panel}) and
  spectral index (\textit{bottom panel}).  The symbols are the same as in
  Figure \ref{rot_age}.  Angular momentum does not evolve significantly
  with either evolutionary diagnostic.  The angular momentum of HH stars
  and non-HH stars are similar. \label{break_age}}
\end{figure} 
\clearpage

\begin{figure}
\epsscale{1.0}
\plotone{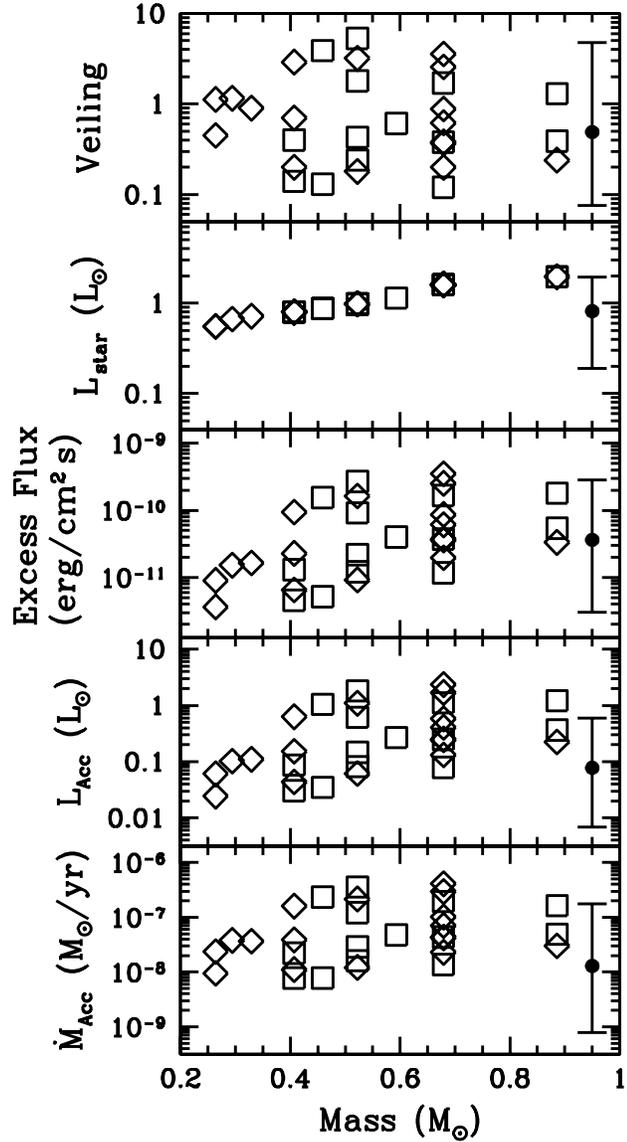}
\caption{Quantities used in the conversion of optical veiling values at
  6500 \AA\, to mass accretion rates (veiling, stellar luminosity, excess
  flux in the 6000 - 6500 \AA\, passband, total accretion luminosity, mass
  accretion rate) versus stellar mass.  \textit{Squares} are new veiling
  measurements and \textit{diamonds} are veiling measurements from the
  literature (see Section 3.3.1).  The vertical error bar in each panel
  indicates the range of values determined by \citet{gullbring98} from a
  broader but bluer wavelength interval (3200 - 5300 \AA), constrained to
  the same mass range shown here.  Surprisingly, the excess flux levels
  measured over the 6000 - 6500 \AA\, passband are similar to those
  measured by \citet{gullbring98}; the standard hot spot model predicts
  they should be only one-tenth.  A factor of 2 can be attributed larger
  stellar luminosities, from which the excess fluxes are determined
  relative to, but the red excess fluxes are still a factor of $\sim 5$
  greater than predicted.  A reduced bolometric correction is adopted
  to convert the excess flux levels to total accretion luminosity, which
  leads to accretion luminosities only $\sim 3$ times larger than those of
  \citep{gullbring98}.  This, in combination with a different geometric
  assumption, leads to mass accretion rates that are, on average, $\sim 4$
  times larger than those inferred by
  \citet{gullbring98}. \label{macc_mass}}
\end{figure}
\clearpage

\begin{figure}
\epsscale{0.8}
\plotone{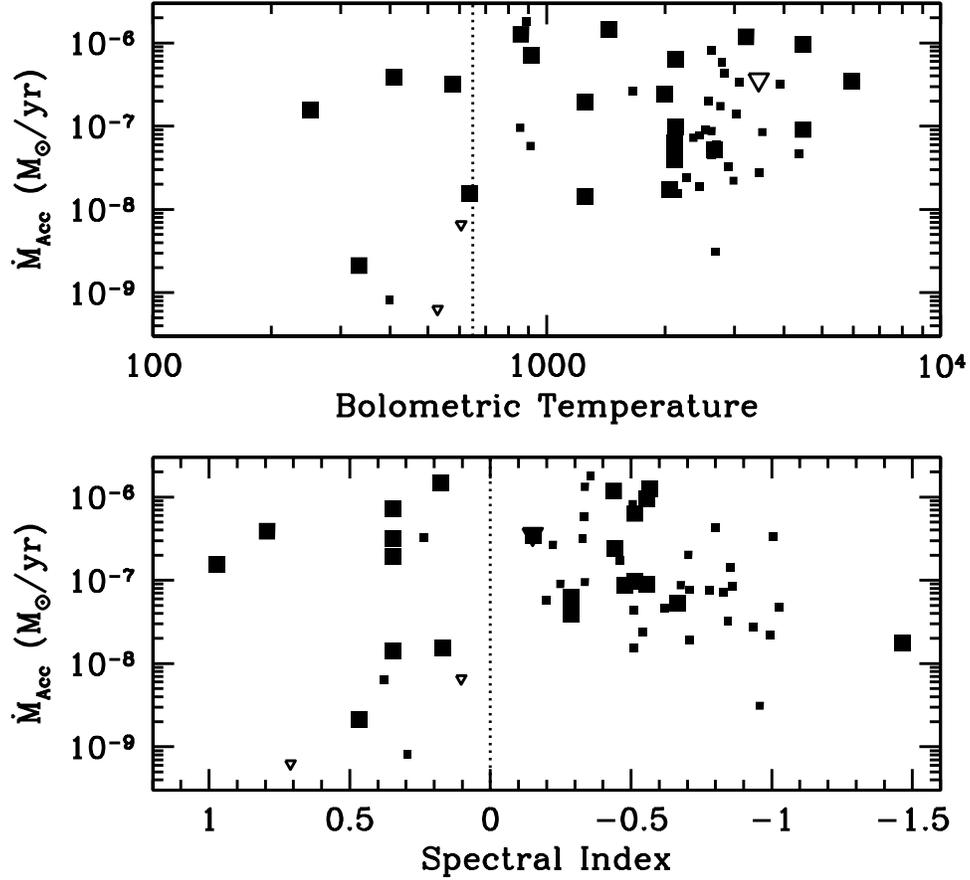}
\caption{Mass accretion rate versus bolometric temperature (\textit{top
  panel}) and spectral index (\textit{bottom panel}).  The symbols are the
  same as in Figure \ref{rot_age}.  Although the mass accretion rates for
  Class I stars extend to lower values than for Class II stars, this is
  primarily a consequence of lower masses.  Over a similar stellar mass
  range, the mass accretion rates do not evolve significantly with either
  evolutionary diagnostic.  HH stars have mass accretion rates that are
  similar to those of non-HH stars, although larger by a factor of 
  $\sim 2.5$ on average.  \label{macc_age}}
\end{figure}
\clearpage

\begin{figure}
\epsscale{0.8}
\plotone{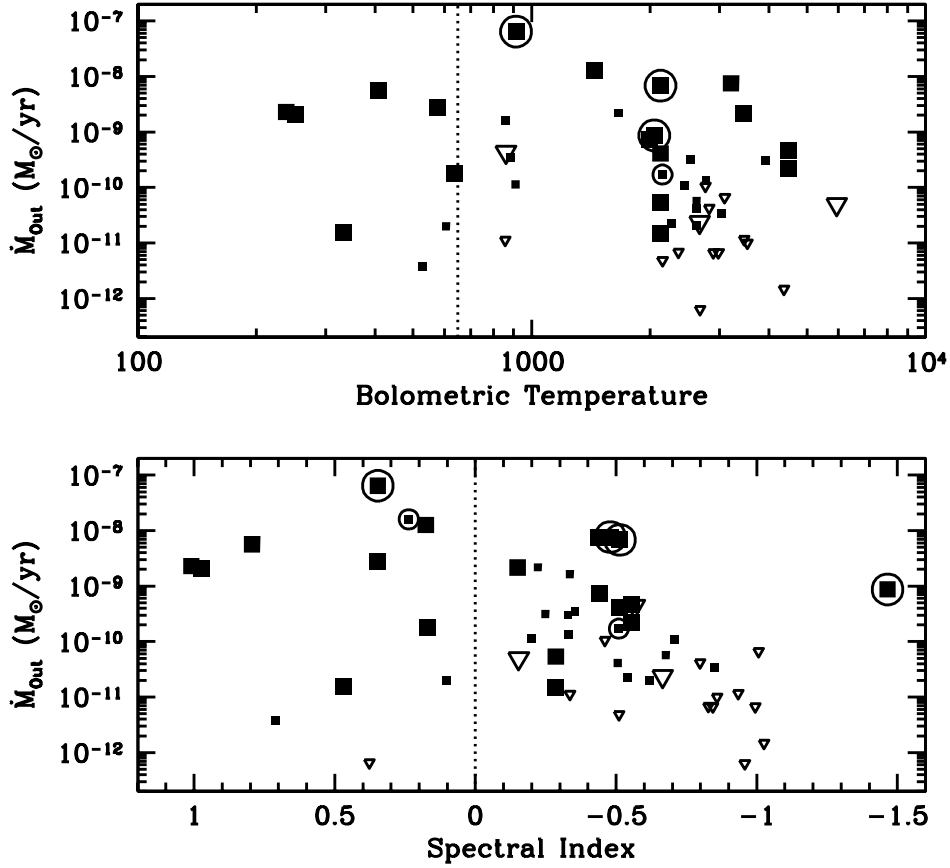}
\caption{Mass outflow rate versus bolometric temperature (\textit{top
  panel}) and spectral index (\textit{bottom panel}).  The symbols are the
  same as in Figure \ref{rot_age}; \textit{circled symbols} indicate stars
  with known or likely edge-on disk systems (see Section 3.1.4).  
  Over a similar stellar mass range, the mass outflow rates appears to
  decline by a median factor of $\sim 20$ with either bolometric 
  temperature or spectral index.  Similarlly, HH stars have mass outflow
  rates that are, in the median, $\sim 20$ times that of non-HH stars.
  These difference may be a consequence of biases in the observed EW[SII] 
  emission, as is the case for known edge-on disk systems \label{mout_age}}
\end{figure}
\clearpage

\begin{figure}
\epsscale{0.8}
\plotone{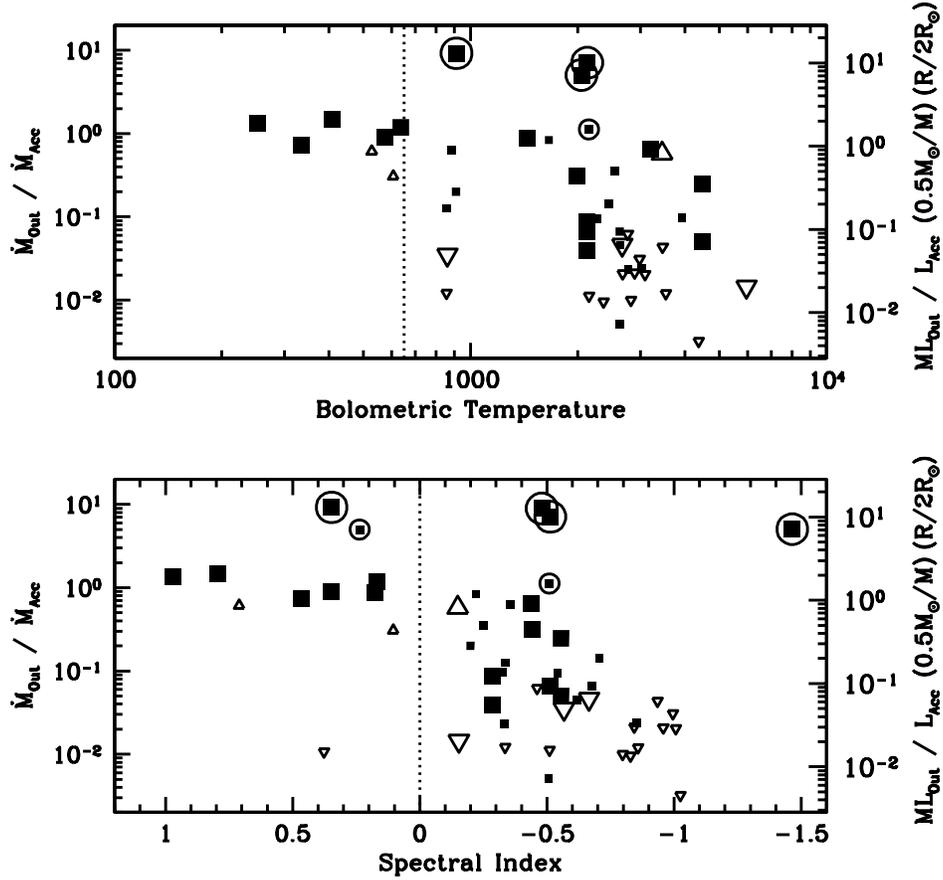}
\caption{Ratio of mass outflow to mass accretion rate versus bolometric 
  temperature (\textit{top panel}) and spectral index (\textit{bottom
  panel}).  The symbols are as in Figure \ref{rot_age}; \textit{upperward
  pointing triangles} are lower limits and \textit{downward pointing
  triangles} are upper limits.  \textit{Circled symbols} indicate stars
  with known or likely edge-on disk systems.  The right hand ordinate shows
  the corresponding ratio of outflow mechanical lumionsity to accretion
  luminosity, assuming a 0.5 M$_\odot$ star with a radius of 2.0
  R$_\odot$.  Class I stars and HH stars have systematically larger ratios
  than Class II stars or non-HH stars.  These difference may be a
  consequence of biases in the observed EW[SII] emission, as is the case
  for known edge-on disk systems. \label{mflow_age}}
\end{figure}
\clearpage

\begin{figure}
\epsscale{0.8}
\plotone{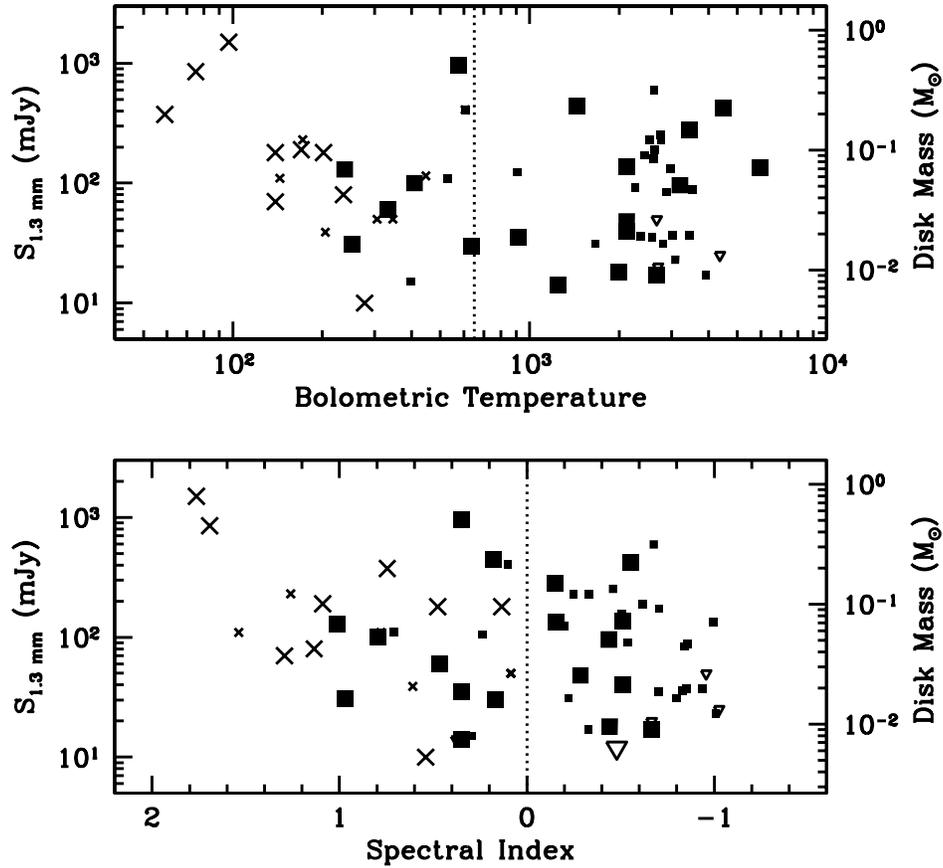}
\caption{1.3-mm Flux density versus bolometric 
  temperature (\textit{top panel}) and spectral index (\textit{bottom
  panel}).  The right hand ordinate shows the conversion to disk mass, as
  described in the text (Section 4.6).  The symbols are as in Figure
  \ref{rot_age}; the \textit{crosses} indicate Class I stars without
  spectroscopically determined stellar properties.  The disk mass does not 
  evolve significantly with either evolutionary diagnostic.  HH stars and 
  non-HH stars have similar disk masses. \label{mm_age}}
\end{figure}
\clearpage


\begin{landscape}
\begin{deluxetable}{llccccccccccc}
\tabletypesize{\scriptsize}
\tablecaption{A Sample of Young Stars in Taurus-Auriga
  \label{tab_sample}}
\tablewidth{0pt}
\tablehead{\colhead{Common}
& \colhead{Associated}
& \colhead{}
& \colhead{}
& \colhead{}
& \colhead{Spectral}
& \colhead{T$_{bol}$$^c$}
& \colhead{Evol}
& \colhead{HH$^e$}
& \colhead{visible$^f$}
& \colhead{$I_c$}
& \colhead{$J^g$}
& \colhead{$I_c^h$} \\
\colhead{Name}
& \colhead{IRAS Source}
& \colhead{Obs?$^a$}
& \colhead{RA(2000)$^b$}
& \colhead{DEC(2000)$^b$}
& \colhead{Index$^c$}
& \colhead{($^\circ K$)}
& \colhead{Class$^d$}
& \colhead{source?}
& \colhead{in DSS?}
& \colhead{($mag$)}
& \colhead{($mag$)}
& \colhead{Ref}	}
\startdata
\multicolumn{13}{c}{Environmentally Young Stars in Taurus-Auriga} \\
\nodata			& 04016+2610	&new& 04 04 43.27	& +26 18 53.6	& +1.01	& 238	& I'	&yes& yes	& 17.94 & 12.72 & kh95 \\
\nodata			& 04108+2803 B	&\nodata& 04 13 54.72	& +28 11 32.9	&(+0.61)& (205)	& I'	&\nodata& no	& 22.1: & 16.48 & wh \\
CW Tau			& 04112+2803	&prev& 04 14 17.00	& +28 10 57.8	& -0.44 & 3207	& II	& yes	& yes	& 11.42	&  9.54 & kh95\\
MHO 2/1$^\dag$		& 04113+2758 AB	&new& 04 14 26.40	& +28 05 59.7	&(+0.10)& (606)	& I	&\nodata& yes	&\nodata& 11.14 & wh \\
\nodata			& 04154+2823	&new& 04 18 32.03	& +28 31 15.4	& -0.03	& 650	& II	&\nodata& no	&\nodata& 15.19 & wh \\
CoKu Tau 1 AB		& 04157+2813	&new& 04 18 51.48	& +28 20 26.5	&(-0.48)&\nodata& II	& yes	& yes	&\nodata& 12.87 & wh \\
\nodata			& 04158+2805	&new& 04 18 58.14	& +28 12 23.5	& +0.71	& 528	& I'	&\nodata& yes	&\nodata& 13.78 & wh \\
\nodata			& 04166+2706	&\nodata& 04 19 42.5	& +27 13 40	& +0.48 & 139	& I	& yes	& no	& $>$24:&\nodata& wh \\
\nodata			& 04169+2702	&\nodata& 04 19 58.45	& +27 09 57.1	& +1.09 & 170	& I	& yes	& no	& 20.1: & 17.19 & wh \\
\nodata			& 04181+2655	&\nodata& 04 21 07.95	& +27 02 20.4	& +0.54 & 278	& I	&yes& no	& 22.5: & 13.86 & wh \\
\nodata			& 04181+2654 A	&\nodata& 04 21 11.47	& +27 01 09.4	&(+0.09)& (346)	& I	&\nodata& no	& 19.9: & 16.22 & wh \\
\nodata			& 04181+2654 B	&\nodata& 04 21 10.39	& +27 01 37.3	&(+0.09)& (306)	& I	&\nodata& no	& $>$24:& 18.89 & wh \\
T Tau AB		& 04190+1924	&new& 04 21 59.43	& +19 32 06.4	&(-0.15)&(3452/501)&II	& yes	& yes	& 8.50  &  7.24 & kh95 \\
Haro 6-5 B		& 04189+2650	&new& 04 22 00.70	& +26 57 32.5	&(-0.15)&(5948)	& II	& yes	& yes	&\nodata& 15.08 & wh \\
\nodata			& 04239+2436	&\nodata& 04 26 56.30	& +24 43 35.3	&(+1.13)& (236)	& I	& yes	& no	& 19.4: & 15.75 & wh \\
DG Tau B		& 04240+2559	&new& 04 27 02.67	& +26 05 30.4	&(+0.18)& (1440)& II	& yes	& yes	&\nodata& 14.72 & wh \\
DG Tau			& 04240+2559	&new& 04 27 04.70	& +26 06 16.3	&(+0.18)& (1440)& II	& yes	& yes	& 10.54	& 8.69 & kh95\\
HH31 IRS 2 AB		& 04248+2612	&new& 04 27 57.33	& +26 19 18.1	&(+0.47)& (334)	& I 	& yes	& yes	& 16.9: & 11.62 & wh \\
\nodata			& 04260+2642	&new& 04 29 04.99	& +26 49 07.3	& +0.24 &\nodata& I'	&\nodata& yes	&\nodata& 14.68 & wh \\
GV Tau AB       	& 04263+2426	&new& 04 29 23.73	& +24 33 00.3	&(+0.79)& (409)	& I	& yes	& yes	& 12.80 & 11.54 & kh95\\
HH 414 IRS AB		& 04264+2433	&new& 04 29 30.08	& +24 39 55.1	&(+0.97)& (252)	& I'	& yes	& yes	& 18.5: & 13.73 & wh \\
ZZ Tau IRS		& 04278+2435	&new& 04 30 51.71	& +24 41 47.5	&(-1.46)& (2048)& II	& yes	& yes	&\nodata& 12.84 & wh \\
L1551 IRS 5		& 04287+1801	&new& 04 31 34.08	& +18 08 04.9	& +1.76	& 97	& I	& yes	& yes	& 18.6: & 13.71 & wh \\
HH 30 IRS		& 04287+1807	&new& 04 31 37.47	& +18 12 24.5	&(+0.35)& (913) & II	& yes	& yes	& 17.3  & 15.18 & mf83 \\
HL Tau			& 04287+1807	&new& 04 31 38.44	& +18 13 57.7	&(+0.35)& (576)	& I'	& yes	& yes	& 12.56	& 10.62 & kh95 \\
XZ Tau AB		& 04287+1807	&prev& 04 31 40.07	& +18 13 57.2	&(+0.35)& (1250)& II	& yes	& yes	& 12.00	&  9.36 & kh95 \\
L1551 NE		& 04287+1801	&\nodata& 04 31 44.45	& +18 08 31.5	& +1.69 & 75	& I	& yes	& no	&\nodata& 16.61 & wh \\
\nodata			& 04295+2251	&new& 04 32 32.05	& +22 57 26.7	& +0.20 & 447	& I'	&\nodata& no	& 20.0: & 14.89 & wh \\
\nodata			& 04302+2247	&new& 04 33 16.50	& +22 53 20.4	& +0.13	& 202	& I'	& yes	& yes	&\nodata& 14.68 & wh \\
GK Tau			& 04305+2414	&prev& 04 33 34.56	& +24 21 05.9	&(-0.67)&(2667)	& II	& yes	& yes	& 10.62	&  9.05 & kh95\\
\nodata			& 04325+2402 ABC&\nodata& 04 35 35.39	& +24 08 19.4	&(+0.78)&\nodata& I	&\nodata& no	& 21.5: & 17.13 & wh \\
Haro 6-28 AB		& 04328+2248	&new& 04 35 56.84	& +22 54 36.0	&(+0.38)&\nodata& I:	&\nodata& yes	&\nodata& 11.14 & wh \\
DO Tau			& 04353+2604	&prev& 04 38 28.58	& +26 10 49.4	&(-0.51)& (2125)& II	& yes	& yes	& 11.17	& 9.47 & kh95\\
HV Tau C		& 04353+2604	&new& 04 38 35.48	& +26 10 41.5	&(-0.51)& (2125)& II	& yes	& yes	& 14.93	&\nodata& mm94\\
\nodata			& 04361+2547 AB	&\nodata& 04 39 13.89	& +25 53 20.9	&(+1.54)& (144)	& I	&\nodata& no	& 22.3: & 16.44 & wh \\
\nodata			& 04365+2535	&\nodata& 04 39 35.19	& +25 41 44.7	& +1.26	& 172	& I	&\nodata& no	& 23.5: & 16.91 & wh \\
\nodata			& 04368+2557	&\nodata& 04 39 53.8	& +26 03 06	& +0.74	& 59	& 0	&yes& no	& 19.0: &\nodata& wh \\
IC 2087	        	& 04369+2539	&new& 04 39 55.75	& +25 45 02.0	& -0.57 & 860	& II	& yes	& yes	&\nodata& 10.66 & wh \\
\nodata			& 04381+2540	&\nodata& 04 41 12.68	& +25 46 35.4	& +1.29	& 139	& I	&yes& no	& 21.1: & 17.15 & wh \\
Haro 6-33		& 04385+2550	&new& 04 41 38.82	& +25 56 26.8	& +0.17	& 636	& I'	& yes	& yes	& 15.58 & 11.85 & kh95\\
DP Tau			& 04395+2509	&new& 04 42 37.70	& +25 15 37.5	& -0.44	& 1991	& II	& yes	& yes	& 11.95	& 11.00 & kh95\\
UY Aur AB		& 04486+3042	&prev& 04 51 47.38	& +30 47 13.5	&(-0.29)& (2120)& II	& yes	& yes	& 10.83	&  9.13 & kh95\\
\nodata			& 04489+3042	&new& 04 52 06.68	& +30 47 17.5	& +0.29 & 399	& I'	&\nodata& no	& 20.3: & 14.43 & wh \\
RW Aur AB		& 05046+3020	&new& 05 07 49.54	& +30 24 05.1	&(-0.56)& (4478)& II	& yes	& yes	& 9.34	&  8.38 & kh95\\
\multicolumn{13}{c}{Additional T Tauri Stars in Taurus-Auriga} \\
\nodata		& 04278+2253	&new	& 04 30 50.28		& +23 00 08.8	&(-0.34)& (856)	& II	&\nodata& yes	&\nodata&  8.78 & wh \\
HK Tau  		& 04288+2417	&new& 04 31 50.57	& +24 24 18.1	&(-0.51)& (2148)& II	&\nodata& yes	& 12.37	& 10.45 & kh95\\
Haro 6-13		& 04292+2422	&new& 04 32 15.41	& +24 28 59.7	& -0.20	& 910	& II	&\nodata& yes	&\nodata& 11.24 & wh \\
\nodata		& 04303+2240	&new	& 04 33 19.07		& +22 46 34.2	& -0.35	& 886	& II	&\nodata& yes	& 15.73	& 11.10 & kh95\\
DN Tau		& 04324+2408	&		new& 04 35 27.37	& +24 14 58.9	& -0.84	& 2890	& II	& no	& yes	& 10.49	& 9.14	& kh95\\
V836 Tau	& \nodata	&		new& 05 03 06.60	& +25 23 19.7	& -0.93	& 3462	& II	& no	& yes	& 11.19	& 9.91	& kh95\\
\enddata
\tablenotetext{a}{Indicates optical observations at high dispersion from
  this study ('new') or previous work ('prev').}
\tablenotetext{b}{Positions are from the 2MASS Point Source Catalogue 
if available.  The positions for 3 of the remaining 6 stars are from the
2MASS Extended Source Catalogue (IRAS 04016+2610, DG Tau B, and IRAS
04248+2612).  The positions for IRAS 04166+2706 and IRAS 04368+2557, which
are too faint to be detected by the 2MASS survey, are from \citet{pk02} and
\citet{kenyon90}, respectively.  The position of HV Tau C is determined
from the 2MASS measurement of HV Tau AB offset by the separation and
position angle measurements of \citet{wl98}.  The position listed for the
binary star MHO 1/2 is that of the primary, MHO 2.}
\tablenotetext{c}{Spectral indices are determined from the K-[25] colors
  listed in \citet{kh95}.  Bolometric temperatures are from \citet{chen95}.
  Values listed in parentheses have been determined from colors or spectral
  energy distribution that include contributions from more than 1 star.}
\tablenotetext{d}{The Class type (0, I or II) is set according to the
  bolometric temperature and spectral index criteria discussed in Section
  2.1.  Class I stars which are unresolved (ie. point-like) at 1.3
  millimeters are marked with an apostrophe \citep{ma01}; Haro 6-28, which
  is marked with a colon, has no spatial millimeter information.}
\tablenotetext{e}{Indicates if the star powers a Herbig-Haro flow
  \citep{gomez97, reipurth99}.}
\tablenotetext{f}{Indicates if the object is visible in the Second
Generation POSS-II Red plates (ref).}
\tablenotetext{g}{$J$ magnitudes are from the 2MASS Point Source Catalogue except for 
IRAS 04016+2610, DG Tau B, and IRAS 04248+2612, which are from the 2MASS Extended 
Source Catalogue.}
\tablenotetext{h}{kh95 = \citet{kh95}; wh = this work; mm94 = \citet{mm94};
  mf83 = \citet{mf83} }
\end{deluxetable}
\end{landscape}

\begin{landscape}
\begin{deluxetable}{llrccllllll}
\tabletypesize{\scriptsize}
\tablecaption{Observational Summary and Measured Photospheric Properties
  \label{tab_obs}}
\tablewidth{0pt}
\tablehead{ \colhead{}
& \colhead{}
& \colhead{Exp.}
& \colhead{}
& \colhead{} 
& \colhead{}
& \colhead{Radial} 
& \colhead{}
& \colhead{}
& \colhead{}
& \colhead{}
 \\
\colhead{}
& \colhead{Obs}
& \colhead{Time}
& \colhead{S/N}
& \colhead{S/N} 
& \colhead{EW[Li\,I]}
& \colhead{Velocity} 
& \colhead{$v$sin$i$}
& \colhead{}
& \colhead{}
& \colhead{}
 \\
\colhead{Star}
& \colhead{Date}
& \colhead{($min$)}
& \colhead{6700}
& \colhead{8450}
& \colhead{(\AA)}
& \colhead{(km/s)}
& \colhead{(km/s)}
& \colhead{SpT}
& \colhead{$r_{6500}$}
& \colhead{$r_{8400}$}
}
\startdata
\multicolumn{11}{c}{Environmentally Young Stars in Taurus-Auriga} \\
IRAS 04016+2610	&99 Dec 5,6& 215 & $1.6$& $2.1$	& $\le 0.5$	&$19\pm7:$	& $\le 15:$ 	&K:		&$\le 0.3:$	&$\le 0.37$ \\
MHO 1		&03 Feb 17 & 10 & $0.1$	& $1.3$	& $\le 4.2$	&\nodata	&\nodata	&\nodata	&\nodata	&\nodata \\
MHO 2		&03 Feb 17 & 30 & $5.0$	& $18$	& $0.39 \pm0.06$&$14.8\pm3.0$	& $21.5\pm4.2$	&M3.5$\pm$1	&$<0.24$	&$<0.05$ \\
IRAS 04154+2823	&03 Feb 17 & 60 & $0.2$	& $0.9$	& $\le 1.4$	&\nodata	&\nodata	&\nodata	&\nodata	&\nodata \\
CoKu Tau 1	&02 Dec 13 & 30 & $8.5$	& $15$	& $0.49\pm0.03$	& $15.0\pm0.8$	& $15.3\pm1.7$	&K7$\pm 1+$M	&$0.38\pm0.18$	&$\le 0.19$ \\
IRAS 04158+2805	&02 Dec 13 & 30 & $4.0$	& $10$	& $0.21\pm0.07$	& $18.1\pm5.4$	& $23.1\pm4.5$	&M6$\pm$1	&$\le 1.0$    &$0.58\pm0.30$ \\
T Tau		&03 Feb 18 & 1.5& $144$	& $119$	& $0.36\pm0.02$	& $20.2\pm0.4$	& $20.9\pm1.0$	&K0$\pm$2	&$\le 0.20$	&$\le 0.19$ \\
Haro 6-5 B	&02 Dec 13 & 20 & $7.0$	& $5.1$ & $0.45\pm0.04$ & $18.6\pm3.6$	& $20.7\pm6.4$	&K5$\pm$2	&$1.0\pm0.2$	&$1.4\pm1.2$ \\
DG Tau B	&02 Dec 13 & 30 & $0.6$	& $0.4$ & $\le0.5$	&\nodata	&\nodata	&\nodata	&\nodata	&\nodata \\
DG Tau		&99 Dec  6 & 15 & $59$	& $65$	& $0.24\pm0.02$	& $19.3\pm2.7$	& $28.6\pm5.1$	&K3$\pm$2	&$2.1\pm0.5$	&$1.8\pm0.5$ \\
IRAS 04248+2612	&99 Dec  6 & 180 & $1.3$& $4.4$ & $\le 0.2$	& $14.4\pm2.5$	& $16.4\pm4.2$  &M5.5$\pm$1	&$1.4:$  	&$0.60\pm0.21$ \\
IRAS 04260+2642	&02 Dec 13 & 60 & $5.3$	& $8.2$ & $0.43\pm0.05$	& $22.3\pm3.6$	& $\le 18$	&K6$\pm$2	&$1.3\pm0.5$	&$0.57\pm0.27$ \\
GV Tau		&99 Dec  6 & 30 & $5.4$	& $7.1$ & $0.25\pm0.05$ & $13.4\pm3.8$	& $25.3\pm7.0$	&K7$\pm$2	&$1.7\pm0.4$	&$1.1\pm0.4$ \\
IRAS 04264+2433	&99 Dec  5 & 60 & $2.0$	& $5.1$ & $0.21\pm0.14$ & $11\pm8:$	& $\le 15:$	&M$1\pm$2     	&$1.4:$  	&$0.99\pm0.49$ \\
ZZ Tau IRS	&03 Feb 18 & 60 & $4.7$	& $8.4$	& $0.15\pm0.06$ & $18.4\pm3.7$	& $22\pm12$	&M4.5$\pm$2	&$2.1\pm0.7$	&$0.81\pm0.29$ \\
L1551 IRS 5	&99 Dec  5 & 60 & $0.7$	& $0.1$ & $\le 0.4$	&\nodata	&\nodata	&\nodata	&\nodata	&\nodata \\
	        &03 Feb 17 & 30 & $0.6$	& $1.4$	& $\le 0.5$	&\nodata	&\nodata	&\nodata	&\nodata	&\nodata \\
L1551 IRS 5 HH  &03 Feb 17 & 30 & $0.6$	& $0.1$	& $\le 0.5$	&\nodata	&\nodata	&\nodata	&\nodata	&\nodata \\
L1551 IRS 5 Neb	&03 Feb 18 & 60 & $1.0$	& $0.3$	& $\le 0.8$	&\nodata	&\nodata	&\nodata	&\nodata	&\nodata \\
HH 30 IRS	&02 Dec 13 & 60 & $9.6$	& $8.5$ & $0.17\pm0.03$	& $20.3\pm3.5$	&$\le 12$      &M0$\pm2$	&$5.3\pm1.0$  	&$5.7\pm0.9$ \\
HL Tau		&99 Dec  5 & 20 & $29$	& $33$	& $0.26\pm0.02$ & $18.3\pm2.6$	& $26.3\pm5.2$	&K5$\pm$1	&$0.91\pm0.07$	&$0.99\pm0.15$ \\
IRAS 04295+2251	&99 Dec  5 & 60 & $0.1$	& $0.8$ & $\le 2.7$	&\nodata	&\nodata	&\nodata	&\nodata	&\nodata \\	
IRAS 04302+2247	&99 Dec  5 & 120& $0.8$	& $1.8$	& $\le 0.4$	&\nodata	&\nodata	&\nodata	&\nodata	&\nodata \\
Haro 6-28	&02 Dec 13 & 20 & $28$	& $47$	& $0.46\pm0.02$ & $16.1\pm0.3$	& $10.1\pm1.0$	&M3.5$\pm$0.5	&$0.23\pm0.05$	&$\le 0.07$ \\
HV Tau C	&02 Dec 13 & 40 & $21$	& $19$	& $0.47\pm0.02$ & $23.1\pm0.6$	& $19.4\pm1.9$	&K6$\pm$1	&$0.39\pm0.07$	&$\le 0.19$ \\
IC 2087 	&02 Dec 13 & 20 & $2.5$	& $14$	& $\le 0.12$	& $22\pm8:$     & $\le 15:$ 	&K:	        &$2.7:$ 	&$3.2\pm0.6$ \\
Haro 6-33	&02 Dec 13 & 20 & $14$	& $23$	& $0.59\pm0.02$ & $15.8\pm0.6$	& $22.8\pm1.6$	&M0.5$\pm$0.5	&$0.13\pm0.06$	&$\le 0.04$ \\
DP Tau		&03 Feb 17 &  5 & $47$	& $48$	& $0.28\pm0.02$ & $16.8\pm0.6$	& $19.2\pm1.3$	&M0$\pm$1	&$1.8\pm0.2$	&$1.3\pm0.3$ \\
IRAS 04489+3042	&99 Dec  5 & 60 & $0.5$	& $2.5$	& $\le 0.6$	& $15\pm5:$	& $\le 15:$ 	&M6$\pm$2	&$1.3:$ 	&$0.56\pm0.20$ \\
RW Aur A	&03 Feb 17 &  1 & $67$	& $49$	& $0.29\pm0.02$ & $16.0\pm1.7$	& $34.4\pm5.7$	&K2$\pm$2	&$1.1\pm0.3$	&$1.5\pm0.6$ \\
RW Aur B	&03 Feb 17 &  1 & $14$	& $16$	& $0.50\pm0.02$ & $15.9\pm0.5$	& $12.2\pm1.6$	&K6$\pm$1	&$0.26\pm0.12$	&$0.17\pm0.13$ \\
\multicolumn{11}{c}{Additional T Tauri Stars in Taurus-Auriga} \\
IRAS 04278+2253b&02 Dec 13 & 15 & $14$	& $24$	& $0.42\pm0.02$	& $15.7\pm0.4$	& $\le7.4$	&K7/M0$\pm$1	&$0.61\pm0.10$	&$0.49\pm0.11$ \\
IRAS 04278+2253a&02 Dec 13 & 15 & $15$	& $27$	& $0.25\pm0.02$	& $19.2\pm2.7$	& $15.7\pm4.9$	&G8$\pm$2	&$0.34\pm0.22$	&$0.63\pm0.33$ \\
HK Tau A	&02 Dec 13 & 8  & $23$	& $31$	& $0.50\pm0.02$ & $21.6\pm0.5$	& $17.8\pm1.1$	&M1$\pm$0.5	&$0.40\pm0.10$	&$0.24\pm0.06$ \\
HK Tau B	&02 Dec 13 & 20 & $12$	& $16$	& $0.58\pm0.02$ & $17.4\pm0.3$	& $\le 7.3$	&M1$\pm$0.5	&$0.14\pm0.06$	&$< 0.12$ \\
Haro 6-13	&02 Dec 13 & 20 & $24$	& $40$	& $0.50\pm0.02$ & $18.6\pm0.8$	& $23.3\pm1.1$	&M0$\pm$0.5	&$0.43\pm0.08$	&$0.26\pm0.08$\\
IRAS 04303+2240	&99 Dec  6 & 30 & $53$	& $69$	& $\le 0.05$	&\nodata	&\nodata	&\nodata	&$14.9:$      	&$7.9:$ \\
		&03 Feb 17 & 20 & $10.3$& $18$	& $0.25\pm0.03$	& $20.8\pm3.6$	& $35\pm16$	&M0.5$\pm1$	&$3.9\pm0.4$	&$2.6\pm1.1$ \\
DN Tau		&99 Dec  6 & 20 & $49$	& $55$	& $0.53\pm0.02$ & $16.9\pm0.3$ & $9.8\pm0.9$	&M0$\pm$0.5	&$0.24\pm0.07$	&$< 0.07$ \\
V836 Tau	&99 Dec  6 & 20 & $71$	& $71$	& $0.55\pm0.02$ & $18.5\pm0.3$ & $12.1\pm1.0$	&K7$\pm1+$M	&$0.12\pm0.10$	&$< 0.16$ \\
\multicolumn{11}{c}{T Tauri stars in the TW Hydrae Association} \\
TWA 8a		&03 Feb 18 & 20 & $195$	& $231$	& $0.\pm0.02$	& $8.7\pm0.2$&  $\le 5.5$	&M3$\pm$0.5	&$\le 0.08$	&$\le 0.08$ \\
TWA 8b		&03 Feb 18 & 20 & $44$	& $88$	& $0.\pm0.02$	& $8.2\pm0.2$ & $10.8\pm1.0$	&M5.5$\pm$0.5	&$\le 0.06$	&$\le 0.17$ \\
TWA 9a		&03 Feb 18 & 18 & $239$	& $190$	& $0.\pm0.02$	& $10.9\pm0.3$ & $11.1\pm1.0$	&K7$\pm$1	&$\le 0.09$	&$\le 0.10$ \\
TWA 9b		&03 Feb 18 & 18 & $63$	& $88$	& $0.\pm0.02$	& $12.2\pm0.4$ & $9.0\pm1.0$	&M3.5$\pm$0.5	&$\le 0.20$	&$\le 0.15$ \\
\enddata
\end{deluxetable}
\end{landscape}

\begin{deluxetable}{llclclclclclclc}
\tabletypesize{\scriptsize}
\tablecaption{Summary of Median Properties, Substellar Objects Excluded
  \label{tab_median} }
\tablewidth{0pt}
\tablehead{\colhead{}
& \colhead{}
& \colhead{}
& \colhead{}
& \colhead{}
& \colhead{}
& \colhead{}
& \colhead{H$\alpha$ 10\%}
& \colhead{}
& \colhead{}
& \colhead{}
& \colhead{}
& \colhead{}
& \colhead{} 
& \colhead{} \\
& \colhead{$v$sin$i$}
& \colhead{}
& \colhead{}
& \colhead{}
& \colhead{EW[H$\alpha$]}
& \colhead{}
& \colhead{width}
& \colhead{}
& \colhead{EW[SII]}
& \colhead{}
& \colhead{log$\dot{M}_{Acc}$}
& \colhead{}
& \colhead{log$\dot{M}_{Out}$} 
& \colhead{} \\
\colhead{type}
& \colhead{$(km/s)$}
& \colhead{N}
& \colhead{$r_{6500}$}
& \colhead{N}
& \colhead{(\AA)}
& \colhead{N}
& \colhead{(km/s)}
& \colhead{N}
& \colhead{(\AA)}
& \colhead{N}
& \colhead{(M$_\odot$/yr)}
& \colhead{N}
& \colhead{(M$_\odot$/yr)}
& \colhead{N} }
\startdata
Class Is& 19.8$\pm$5.7 & 8	& 0.6$\pm$0.6 & 8	& $-51\pm$33 & 13
	& $344\pm131$ & 13	& $-2.4\pm$5.1 &13	
	& $-7.1\pm0.8$ & 8	& $-7.0\pm1.5$ & 8 \\
Class IIs& 13.7$\pm$11.0 & 41	& 0.6$\pm$1.3 & 50	& $-59\pm$48 & 49
	& $429\pm108$ & 46	& $-0.1\pm$17 &42
	& $-7.3\pm0.7$ & 51	& $-8.2\pm1.1$ & 42 \\
\hline
HH	& 19.4$\pm$6.0  & 19 & 1.0$\pm$1.4 & 23		& $-61\pm$69 & 24 
	& $409\pm120$ & 21	& $-1.8\pm$22 & 21
	& $-7.0\pm0.6$ & 23	& $-7.2\pm0.9$ & 20 \\
non-HH	& 11.4$\pm$11.9 & 31 & 0.5$\pm$1.1 & 34		& $-58\pm$37 & 37 
	& $414\pm116$ & 38	& $-0.1\pm$3.4 &33
	& $-7.4\pm0.7$ & 35	& $-8.7\pm1.0$ & 30 \\
\enddata
\end{deluxetable}

\begin{deluxetable}{lrcrrrrrr}
\tabletypesize{\scriptsize}
\tablecaption{Permitted and Forbidden Emission-Line Properties
 \label{tab_emission}}
\tablewidth{0pt}
\tablehead{ \colhead{}
& \colhead{EW\,H$\alpha$}
& \colhead{H$\alpha$}
& \colhead{EW\,CaII}
& \colhead{EW\,CaII}
& \colhead{EW\,[O\,I]}
& \colhead{EW\,[N\,II]} 
& \colhead{EW\,[S\,II]}
& \colhead{EW\,[S\,II]} \\
\colhead{}
& \colhead{6563}
& \colhead{10\%-width}
& \colhead{8498}
& \colhead{8662}
& \colhead{6364}
& \colhead{6583}
& \colhead{6716}
& \colhead{6731} \\
\colhead{Star}
& \colhead{(\AA)}
& \colhead{(km/s)}
& \colhead{(\AA)}
& \colhead{(\AA)}
& \colhead{(\AA)}
& \colhead{(\AA)}
& \colhead{(\AA)}
& \colhead{(\AA)}
}
\startdata
\multicolumn{9}{c}{Environmentally Young Stars in Taurus-Auriga} \\
IRAS 04016+2610	& 41	& 550	& 4.4	& 1.9	& 0.6	& 2.1	& 1.1	& 1.8	\\
MHO 1		& 58	& 253	& 17	& 15	&$<4.$	&$<4.$	&$<4.$	&$<4.$	\\
MHO 2		& 59	& 249	& 0.61	& 0.46	& 1.0	& 0.91	& 0.7	& 1.2	\\
IRAS 04154+2823	&$>17$	& 340	&\nodata&\nodata&$<1.4$	&$<1.4$	&$<1.4$	&$<1.4$	\\
CoKu Tau 1	& 70	& 285	& 0.37	& 0.50	& 16	& 19	& 12	& 21	\\
IRAS 04158+2805	& 175	& 287	& 9.3	& 6.8	& 13	& 6.1	& 4.5	& 8.6	\\
T Tau		& 60	& 380	& 12	& 12	&$>0.1$	& 0.11	&$>0.05$& 0.41	\\
Haro 6-5 B	& 91	& 481	& 31	& 32	&$<0.1$	& 0.37	&$<0.04$&$<0.04$\\
DG Tau B	&$>276$	& 330	&$>10$	&$>12$	&$>36$	&$>96$	&$>298$	&$>426$	\\
DG Tau		& 63	& 517	& 49	& 41	& 3.0	& 0.28	& 1.1	& 2.4	\\
IRAS 04248+2612	& 163	& 410	& 14	& 11	& 8.5	& 4.5	& 6.8	& 11	\\
IRAS 04260+2642	& 126	& 139	& 4.0	& 3.7	& 19	& 13	& 8.9	& 19	\\
GV Tau		& 86	& 552	& 15	& 13	& 13	& 3.1	& 5.3	& 8.0	\\
IRAS 04264+2433	& 96	& 367	& 8.3	& 7.5	& 8.9	& 4.9	& 6.4	& 9.8	\\
ZZ Tau IRS	& 238	& 268	& 23	& 19	& 51	& 30	& 40	& 77	\\
L1551 IRS 5	&$>999$	& 196	&$>32$	&$>44$	&$>611$	&$>999$	&$>669$	&$>935$	\\
L1551 IRS 5	&$>412$	& 290	&$<0.8$	&$<0.9$	&$>281$	&$>295$	&$>127$	&$>225$	\\
L1551 IRS 5 HH  &$>818$	& 196	&$>15$	&$>18$	&$>163$	&$>685$	&$>530$	&$>739$ \\
L1551 IRS 5 Neb	& 44	& 275	&\nodata&\nodata&$<1.7$	&$<1.7$	&$<1.7$	&$<1.7$ \\
HH 30 IRS	& 199	& 293	& 13	& 11	& 49	& 27	& 58	& 75	\\
HL Tau		& 43	& 390	& 26	& 22	& 2.4	& 1.2	& 1.6	& 2.4	\\
IRAS 04295+2251	&$> 11$	& 160	&$>3.3$	&$>2.8$	&$<3.$	&$<3.$	&$<3.$	&$<3.$	\\
IRAS 04302+2247	& 67	& 485	& 36	& 43	&$<2.$	&$<3.$	& 2.8	& 3.5	\\
Haro 6-28	& 51	& 344	& 0.35	& 0.24	& 0.25	&$<0.1$	&$<0.04$&$<0.04$\\
HV Tau C	& 52	& 261	& 1.5	& 1.0	& 7.3	& 10.3	& 7.3	& 13	\\
IC 2087 	& 22	& 610	& 14	& 9.0	&$<0.4$	&$<0.3$	&$<0.1$	&$<0.1$ \\
Haro 6-33	& 15	& 377	& 0.48	& 0.31	& 0.62	& 1.0	&$>0.6$	& 1.5	\\
DP Tau		& 74	& 601	& 13	& 10	& 1.8	& 0.8	& 0.85	& 1.9	\\
IRAS 04489+3042	&$>96$	& 415	& 11	& 7.9	&$>7.2$	&$>7.0$	&$>2.7$	&$>4.7$	\\
RW Aur A	& 42	& 365	& 36	& 29	&$<0.13$&$<0.1$	&$>0.03$& 0.10	\\
RW Aur B	& 17	& 466	& 5.9	& 3.5	& 0.11	& 0.35	& 0.16	& 0.28	\\
\multicolumn{9}{c}{Additional T Tauri Stars in Taurus-Auriga} \\
IRAS 04278+2253 B& 54	& 498	& 17	& 15	& 0.18	& 0.12	&$<0.04$&$<0.04$\\
IRAS 04278+2253 A& 14	& 434	& 4.9	& 3.8	& 0.21	& 0.10	& 0.06	& 0.14	\\
HK Tau A	& 34	& 435	& 0.36	& 0.42	&$<0.1$	&$<0.1$	&$<0.04$&$<0.04$\\
HK Tau B	& 9.7	& 196	& 0.68	& 0.59	&$>0.6$	& 0.53	& 1.0	& 1.7	\\
Haro 6-13	& 34	& 321	& 1.8	& 1.2	& 1.1	& 0.18	&$>0.15$& 0.58	\\
IRAS 04303+2240	& 67	& 487	& 30	& 26	& 0.35	& 0.05	& 0.14	& 0.20	\\
IRAS 04303+2240	& 137	& 337	& 24	& 22	& 9.5	& 1.9	& 2.4	& 5.5	\\
DN Tau		& 18	& 291	& 0.64	& 0.54	&$<0.4$	&$<0.4$	&$<0.4$ &$<0.4$	\\
V836 Tau	& 25	& 270	& 0.69	& 0.47	& 0.18	&$<0.4$	&$<0.4$	&$<0.4$ \\
\enddata
\end{deluxetable}

\begin{deluxetable}{llllllrcrrc}
\tabletypesize{\scriptsize}
\tablecaption{Inferred Stellar, Accretion and Outflow Properties
  \label{tab_prop}}
\tablewidth{0pt}
\tablehead{ \colhead{}
& \colhead{Adopt}
& \colhead{}
& \colhead{A$_V$}
& \colhead{L$_{star}$}
& \colhead{L$_{star}^{1Myr}$}
& \colhead{L$_{bol}$}
& \colhead{1 Myr Mass}
& \colhead{log($\dot{M}_{Acc}$)}
& \colhead{log($\dot{M}_{Out}$)}
& \colhead{log} \\
\colhead{Star}
& \colhead{SpT}
& \colhead{T$_{eff}$}
& \colhead{($mag$)}
& \colhead{(L$_\odot$)}
& \colhead{(L$_\odot$)}
& \colhead{(L$_\odot$)}
& \colhead{(M$_\odot$)}
& \colhead{(M$_\odot$/yr)$^a$}
& \colhead{(M$_\odot$/yr)}
& \colhead{($\dot{M}_{Out}$/$\dot{M}_{Acc}$)$^a$}
}
\startdata
\multicolumn{11}{c}{Environmentally Young Stars in Taurus-Auriga} \\	
IRAS 04016+2610	& K4	& 4580	& 10.2	& 0.45	& 4.63	& 3.70	& 1.61	&$-7.15$	& $-6.93$	& $+0.22$\\
MHO 1		&\nodata&\nodata&\nodata&\nodata&\nodata&($>1.60$)&\nodata&\nodata	&\nodata	&\nodata \\
MHO 2		& M3.5	& 3180	& 18.1	& 2.61	& 0.44	&($>1.60$)& 0.24&$<-8.48$	& $-8.99$	& $>-0.51$ \\
IRAS 04154+2823	&\nodata&\nodata&\nodata&\nodata&\nodata&$>0.25$&\nodata&\nodata	&\nodata	&\nodata \\
CoKu Tau 1	& K7	& 4000	& 6.80	& 0.15	& 1.60	&$>0.29$& 0.68	& $-7.36$	& $-6.41$:	& $+0.95$: \\
IRAS 04158+2805	& M6	& 2760	& 8.63	& 0.050	& 0.022	&$>0.44$& 0.05	& $<-9.50$	& $-9.72$	& $>-0.22$ \\
T Tau		& K0	& 5235	& 4.24	& 19.9	& 29.6	& 15.5	& 3.38	& $<-6.74$	& $-6.96$	& $>-0.21$ \\
Haro 6-5 B	& K5	& 4395	& 9.96	& 0.047	& 3.0	&$>0.02$& 1.20	& $-6.76$	& $<-8.60$	&$<-1.84$ \\
DG Tau B	&\nodata&\nodata&\nodata&\nodata&\nodata&$>0.02$&\nodata&\nodata	&\nodata	&\nodata \\
DG Tau		& K3	& 4775	& 3.32	& 3.62	& 7.7	& 6.36	& 2.22	& $-6.13$	& $-6.19$	& $-0.06$ \\
IRAS 04248+2612	& M5.5	& 2845	& 7.02	& 0.27	& 0.044	& 0.36	& 0.07	& $-8.97$	& $-9.11$	& $-0.13$ \\
IRAS 04260+2642	& K6	& 4200	& 9.24	& 0.054	& 1.95	& 0.09	& 0.89	& $-6.79$	& $-6.09$:	& $+0.70$: \\
GV Tau		& K7	& 4000	& 12.1	& 1.82	& 1.60	& 6.98	& 0.68	& $-6.71$	& $-6.54$	& $+0.17$ \\
IRAS 04264+2433	& M1	& 3605	& 10.4	& 0.14	& 0.80	& 0.37	& 0.41	& $-7.11$	& $-6.98$	& $+0.13$ \\
ZZ Tau IRS	& M4.5	& 3015	& 7.60	& 0.13	& 0.21	&\nodata& 0.16	& $-8.06$	& $-7.35$:	& $+0.70$: \\
L1551 IRS 5	&\nodata&\nodata&\nodata&\nodata&\nodata&$>21.9$&\nodata&\nodata	&\nodata	&\nodata\\
HH 30 IRS	& M0	& 3800	& 2.96	& 0.0064& 0.98	&\nodata& 0.52	& $-6.45$	& $-5.48$:	& $+0.97$: \\
HL Tau		& K5	& 4395	& 7.43	& 1.53	& 3.0	& 6.60	& 1.20	& $-6.80$	& $-6.84$	& $-0.04$ \\
IRAS 04295+2251	&\nodata&\nodata&\nodata&\nodata&\nodata& 0.44	&\nodata&\nodata	&\nodata	&\nodata \\
IRAS 04302+2247	&\nodata&\nodata&\nodata&\nodata&\nodata& 0.34	&\nodata&\nodata	&\nodata	&\nodata \\
Haro 6-28	& M3.5	& 3175	& 4.38	& 0.31	& 0.44	& 0.96	& 0.24	& $-8.50$	& $<-10.47$	& $<-1.97$ \\
HV Tau C	& K6	& 4200	&\nodata&\nodata& 1.95	&\nodata& 0.89	& $-7.31$	& $-6.46$:	& $+0.85$: \\
IC 2087 	& K4	& 4580	& 18.1	& 20.7	& 4.6	& 3.80	& 1.61	& $-6.20$	& $<-7.65$	& $<-1.46$ \\
Haro 6-33	& M0.5	& 3700	& 10.2	& 0.76	& 0.87	&$>0.36$& 0.46	& $-8.11$	& $-8.03$	& $+0.08$ \\
DP Tau		& M0	& 3800	& 6.31	& 0.68	& 0.98	& 0.70	& 0.52	& $-6.92$	& $-7.42$	& $-0.51$ \\
IRAS 04489+3042	& M6	& 2760	& 17.5	& 0.24	& 0.022	& 0.30	& 0.05	& $-9.39$	& $>-9.92$	& $<-0.53$ \\
RW Aur A	& K2	& 4955	& 1.58	& 3.37	& 12.9	&(3.20)	& 2.84	& $-6.32$	& $-7.63$	& $-1.31$ \\
RW Aur B	& K6	& 4200	&\nodata&\nodata& 3.0	&(3.20)	& 1.20	& $-7.34$	& $-7.95$	& $-0.61$ \\
\multicolumn{11}{c}{Additional T Tauri Stars in Taurus-Auriga} \\
IRAS 04278+2253 B& K8	& 3900	&\nodata&\nodata& 1.14	&($>4.50$)& 0.59& $-7.32$	& $<-9.24$	& $<-1.92$ \\
IRAS 04278+2253 A& G8	& 5445	& 11.4	& 29.6	& 41.9	&($>4.50$)& 3.54& $-5.18$	& $-7.08$	& $-0.90$ \\
HK Tau A	& M1	& 3605	& 5.42	& 0.84	& 0.80	& 0.81	& 0.41	& $-7.65$	& $<-9.60$	& $-1.95$ \\
HK Tau B	& M1	& 3605	&\nodata&\nodata& 0.80	&$>0.01$& 0.41	& $-8.11$	& $-8.06$:	& $+0.05$: \\
Haro 6-13	& M0	& 3800	& 11.9	& 2.11	& 0.98	& 1.30	& 0.52	& $-7.54$	& $-8.24$	& $-0.70$ \\
IRAS 04303+2240	& M0.5	& 3700	& 11.7	& 2.20	& 0.87	&$>2.60$& 0.46	& $-6.05$	& $-7.76$	& $-1.71$ \\
IRAS 04303+2240	& M0.5	& 3700	& 11.7	& 2.20	& 0.87	&$>2.60$& 0.46	& $-6.63$	& $-6.83$	& $-0.20$ \\
DN Tau		& M0	& 3800	& 1.89	& 0.68	& 0.98	& 1.00	& 0.52	& $-7.79$	& $<-9.46$	& $<-1.67$ \\
V836 Tau	& K7	& 4000	& 1.68	& 1.21	& 1.60	& 0.51	& 0.68	& $-7.86$	& $<-9.22$	& $<-1.36$ \\
\enddata
\tablenotetext{a}{Values marked with a colon indicate edge-on disk systems.
  The [SII] emission in these systems appears to be biased towards large
  values, leading to overestimates of $\dot{M}_{Out}$ and
  log($\dot{M}_{Out}$/$\dot{M}_{Acc})$. }
\end{deluxetable}

\end{document}